\documentclass[12pt]{iopart}
\usepackage{multirow}
\usepackage{setstack}
\usepackage{csquotes}
\usepackage{epstopdf}
\usepackage{amsfonts}
\usepackage{cases}
\usepackage{graphicx}
\usepackage{xcolor}
\usepackage{cite}

\newcommand\restr[2]{{
  \left.\kern-\nulldelimiterspace 
  #1 
  \right|_{#2} 
  }}

\newcommand{\cc}[1]{{#1}} 
\newcommand{\eqref}[1]{(\ref{#1})}

\usepackage{iopams}  

\begin{document}

\title[Diffusion NMR in periodic media]{Diffusion NMR in periodic media: efficient computation and spectral properties}

\author{Nicolas Moutal$^1$, Antoine Moutal$^1$, Denis Grebenkov$^1$}


\address{$^1$ Laboratoire de Physique de la Mati{\`e}re Condens{\'e}e, Ecole Polytechnique, CNRS, IP Paris,
91128 Palaiseau, France}
\ead{nicolas.moutal@polytechnique.edu,~denis.grebenkov@polytechnique.edu}

\begin{abstract}
The Bloch-Torrey equation governs the evolution of the transverse magnetization in diffusion magnetic resonance imaging, where two mechanisms are at play: diffusion of spins (Laplacian term) and their precession in a magnetic field gradient (imaginary potential term). In this paper, we study this equation in a periodic medium: a unit cell repeated over the nodes of a lattice. Although the gradient term of the equation is not invariant by lattice translations, the equation can be analyzed within a single unit cell by replacing a continuous-time gradient profile by narrow pulses. In this approximation, the effects of  precession and diffusion are separated and the problem is reduced to the study of a sequence of diffusion equations with pseudo-periodic boundary conditions. This representation allows for efficient numerical computations as well as new theoretical insights into the formation of the signal in periodic media. In particular, we study the eigenmodes and eigenvalues of the Bloch-Torrey operator. We show how the localization of eigenmodes is related to branching points in the spectrum and we discuss low- and high-gradient asymptotic behaviors.
The range of validity of the approximation is discussed; interestingly the method turns out to be more accurate and efficient at high gradient, being thus an important complementary tool to conventional numerical methods that are most accurate at low gradients.
%
%

\end{abstract}

%
%
%
%
%
\section{Introduction}
Diffusion magnetic resonance imaging (dMRI) is an experimental technique that probes the random diffusive motion of spin-bearing particles through the dephasing they acquire under an applied magnetic field gradient (i.e., spin precession rate gradient), and allows one to indirectly probe the microstructure of a sample well below the spatial resolution of conventional MRI \cite{Callaghan1991a,Grebenkov2007a,Price2009a,Kiselev2017a}. For example, this technique has been applied in the brain, where the anisotropy of the diffusive motion of water molecules can be related to the orientation of neuronal fibers. In this way, the neuronal fiber pathways can be reconstructed \cite{Pfeuffer1999a,Wedeen2012a}. This example is emblematic of dMRI in the sense that the sample ($\sim 10~\mathrm{cm}$) is much larger than the voxels ($\sim 1~\mathrm{mm}$) in which individual measurements are performed, that are in turn much larger than the microstructural details ($\sim 1~\mathrm{\mu m}$) probed by diffusion.

The basic principle of dMRI is as follows. A population of spins initially aligned with a magnetic field is flipped to the transverse plane by a short $90^\circ$ radio-frequency (rf) pulse, forming a uniform initial transverse magnetization within a limited region of interest called a voxel. These spins rotate at the Larmor frequency that is proportional to the strength of the magnetic field. Inhomogeneities in the magnetic field (especially the applied magnetic field gradient) lead to a fast defocusing of spins and a subsequent decay of the total magnetization of the sample. By applying a $180^\circ$ rf pulse at some moment $T/2$, all dephasing are reversed and spins are refocused at a later time $T$, forming a measurable macroscopic signal \cc{(``spin echo'')} \cite{Hahn1950a}. The diffusive motion of spin-bearing particles makes this refocusing imperfect, allowing for a possibility to recover the diffusive properties of the medium and its microstructure from the decay of the magnetization at echo time $T$ \cc{(which is often denoted by $T_E$ or $TE$ in the literature)}.
The evolution of $m$ in  $\Omega$ is governed by the Bloch-Torrey equation \cite{Torrey1956a} that describes diffusion and precession in a magnetic field gradient:
\begin{eqnarray}
&\frac{\partial m}{\partial t} = D_0\nabla^2 m + i (g_x(t) x + g_y(t) y + g_z(t) z) m
\label{eq:BT_equation}
\\
&\restr{\mathbf{n}\cdot D_0\nabla m + \kappa m}{\partial \Omega} = 0\;,
\label{eq:robin_boundary}\\
&m(t=0,x,y,z)=1\;,
\label{eq:initial_condition}
\end{eqnarray}
where $D_0$ is the diffusion coefficient of the spin-bearing particles in the medium and $g_x,g_y,g_z$ are the components of the Larmor frequency gradient \cc{(with this definition, the gyromagnetic ratio $\gamma$ is included in the magnetic field gradient $\mathbf G$ such that $\mathbf g=\gamma \mathbf G$). Precession caused by the constant polarizing field $B_0$ was implicitly taken into account by considering the magnetization in the frame rotating at Larmor frequency $\omega_0=\gamma B_0$.}
Furthermore, $\partial \Omega$ denotes the boundary of the domain, $\mathbf{n}$ is the normal vector at the boundary (pointing outward the domain), and $\kappa$ is the surface relaxivity of the boundary.
In this paper, we focus on the Robin boundary condition \eqref{eq:robin_boundary}, but the case of permeable boundaries can be treated in a similar way, see for example \cite{Grebenkov2014b,Grebenkov2014a,Moutal2019a} and references therein. \cc{The Bloch-Torrey equation \eqref{eq:BT_equation} was written under the hypothesis that bulk relaxation effects are neglected. If the bulk relaxation rate $1/T_2$ is homogeneous in the medium, then it simply contributes through a factor $e^{-t/T_2}$ to the magnetization. The case of inhomogeneous bulk relaxation is not considered in this paper but can be included in the theoretical frame with minor changes.
}
The application of the $180^\circ$ refocusing rf pulse is equivalent to imposing
\begin{equation}
\int_0^T g_j(t)\,\mathrm{d}t=0\;, \quad j=x,y,z\;.
\label{eq:refocusing}
\end{equation}
This implies that the phase encoding of the spins depends only on their motion during the gradient sequence ($0\leq t\leq T$) and not on their mean positions. Throughout the text, we implicitly take into account the refocusing pulse in the sign of the gradient, so that $(g_x,g_y,g_z)$ is an ``effective'' gradient.
 Due to lack of spatial resolution, the complex-valued transverse magnetization $m(t,x,y,z)$ is not accessible experimentally, and the measurable quantity is the macroscopic signal formed by the nuclei of the whole voxel $\Omega$ (a prescribed spatial region):
\begin{equation}
S=\int_{\Omega} m(T,x,y,z)\,\mathrm{d}x\,\mathrm{d}y\,\mathrm{d}z\;.
\end{equation}
Varying the time $T$ and/or the applied gradient profile $(g_x(t),g_y(t),g_z(t))$ and measuring the signal $S$, one aims at recovering the microstructure inside the voxel that affected diffusion and thus the signal. This is a formidable inverse problem that has not been fully solved in spite of many decades of intensive research \cite{Callaghan1991a,Grebenkov2007a,Price2009a,Kiselev2017a}.

\cc{dMRI has been broadly applied in material sciences, neurosciences and medicine as a very powerful technique for probing microstructure in a noninvasive way. Most former theoretical advances and applications relied on using low or moderate magnetic field gradients. In turn, larger gradients open new modalities for detecting finer details of the microstructure but suffer from two major limitations:} (i) the consequent weak signal-to-noise ratios that limit the range of experimental parameters and the observation of subtle effects; (ii) mathematical difficulties hinder the understanding of the signal formation at large gradients and/or in complex microstructures. In particular, the right-hand side of Eq. \eqref{eq:BT_equation} is not Hermitian that \cc{leads} to unusual mathematical properties. For instance, in the case of a constant gradient, the spectrum of the BT operator
\begin{equation}
\mathcal{B}=-D_0\nabla^2 - i(g_x x+g_y y+g_z z)
\label{eq:BT_operator}
\end{equation}
is empty in the free space $\mathbb{R}^3$ and becomes discrete with the addition of an obstacle, even if the domain remains unbounded \cite{Grebenkov2017a,Grebenkov2018b,Almog2018a,Almog2019a}. This is in sharp contrast with the spectrum of the Laplace operator that is continuous in unbounded domains, including $\mathbb{R}^3$.

Numerous theoretical, numerical, and experimental works have been devoted to studying the BT operator and the dMRI signal in bounded domains (intracellular space, isolated pores)
\cite{Stejskal1965a,Wayne1966a,Robertson1966a,Tanner1968a,Neuman1974a,Brownstein1979a,Coy1994a,Swiet1994a,Callaghan1995a,Froehlich2006a}.
On theoretical side, the linear potential in Eq. \eqref{eq:BT_operator} is a bounded perturbation of the (unbounded) Laplace operator, which has a discrete spectrum in bounded domains. As a consequence, the BT operator also has a discrete spectrum, and its spectral properties can be analyzed by rather standard mathematical tools \cite{Huerlimann1995a,Swiet1994a,Grebenkov2017a,Grebenkov2018b,
Stoller1991a,Grebenkov2014b,Herberthson2017a}. At low gradient strength, perturbation methods are applicable and we shall discuss their limitations when we present bifurcation points in the spectrum.
On the numerical side, different computational techniques for dMRI have been developed, including finite difference/finite elements PDE solvers \cite{Hwang2003a,Li2014a,Nguyen2014a,Novikov1998a,Ziener2019a}, Monte-Carlo simulations \cite{Burcaw2015a,Carl2007a,Fieremans2010a,Grebenkov2011a}, and spectral methods (matrix formalism) \cite{Barzykin1998a,Callaghan1997a,Grebenkov2007a,Grebenkov2008b}. However, all of these techniques are numerically challenging at high gradients because of the fine spatial scales involved in the signal formation, as well as the weak signal. This requires a fine mesh (for PDE solvers), a fine diffusion step and a large number of particles (for Monte Carlo algorithms), and a large number of Laplace eigenmodes (for spectral methods).

In contrast, unbounded domains (that can model extracellular space or connected porous media, for example) are much harder to study both theoretically and numerically. In fact, numerical simulations in unbounded domains require adding a virtual outer boundary to the domain with convenient boundary conditions (e.g., Dirichlet boundary condition). To ensure that the effect of this boundary is negligible, the boundary should be sufficiently far away from the area of interest so that very few particles can diffuse from one to the other. As such, the computational domain can be much bigger than the area of interest, especially in long diffusion time simulations, which makes the technique inefficient. Mathematically, the unbounded gradient term prevents the use of perturbation techniques even at vanishingly small gradient strength, which leads to singular spectral properties as mentioned above.

In this light, periodic media present a somewhat intermediate setting between bounded and unbounded domains, keeping the advantages of both: they can model macroscopic samples but computations can be performed in a single unit cell that dramatically reduces the size of the computational domain and the computation time.
Evidently, complex biological or mineral samples, on which dMRI experiments are usually performed, are not simple periodic structures. In a living tissue, one would most likely find very diverse cell shapes, sizes, and arrangements, in a given voxel. However, the microstructure is probed at the scale of the diffusion length traveled by spin-bearing particles, that is much smaller than the voxel size. 
Although two ``unit cells'' of the real structure are always different, they are often statistically similar at this mesoscopic scale and may lead to almost identical behavior of the signal. In that regard, a periodic medium may be the best compromise between simplicity and relevance.
To our knowledge, the spectral properties of the BT operator have not been studied at all in periodic domains. This paper aims therefore to provide the first results in this direction. One of the major challenges is that the gradient term in Eq. \eqref{eq:BT_operator} is not periodic, so that standard methods of the quantum theory of solids \cite{Bloch1929a,Kittel2004a}, in which potentials are typically periodic, are not applicable here. 
To overcome this problem, we will approximate the constant gradient in Eq. \eqref{eq:BT_operator} or, more generally, the continuous-time gradient profile in the BT equation \eqref{eq:BT_equation}, by a sequence of infinitely narrow gradient pulses. In this approximation, the effects of \cc{the} gradient term and of the Laplace operator are separated, and the problem can be reduced to that of the Laplace operator in a single unit cell with pseudo-periodic boundary conditions. This representation will allow us to develop efficient numerical computations and to investigate the spectral properties of the BT operator. In particular, we will show how the localization of eigenmodes is related to branching points in the spectrum. We will also discuss the validity of this approximation.

The paper is organized as follows. In Sec. \ref{section:theory}, we present the theoretical basis of our numerical technique. We show that the BT equation cannot be straightforwardly reduced to a single unit cell and how to overcome this difficulty. The numerical implementation and results are described in Sec. \ref{section:numerical_results}. As the gradient strength increases, the magnetization localizes sharply around obstacles in the medium, at points where the boundary is orthogonal to the gradient direction. This behavior can be interpreted in terms of localized eigenmodes of the BT operator. We investigate these eigenmodes and the corresponding eigenvalues in Sec. \ref{section:eigenmodes}.
Finally, Sec. \ref{section:conclusion} summarizes our results and concludes the paper.

\section{Theoretical ground}
\label{section:theory}

For pedagogical reasons, the presentation of our technique is split into different steps of increasing generality. First we consider the case of a medium that is periodic along one axis (\cc{and} bounded along the other two). To be concise we call it a 1D-periodic medium, although the medium itself is not one-dimensional. 
The gradient is initially aligned with the periodicity axis, then we show how to take into account a general gradient direction. Finally the \cc{general} case of periodicity along several axes is discussed.

\subsection{Bloch-Torrey equation adapted to a 1D-periodic medium}

\begin{figure}[t]
\centering
\includegraphics[width=0.6\linewidth]{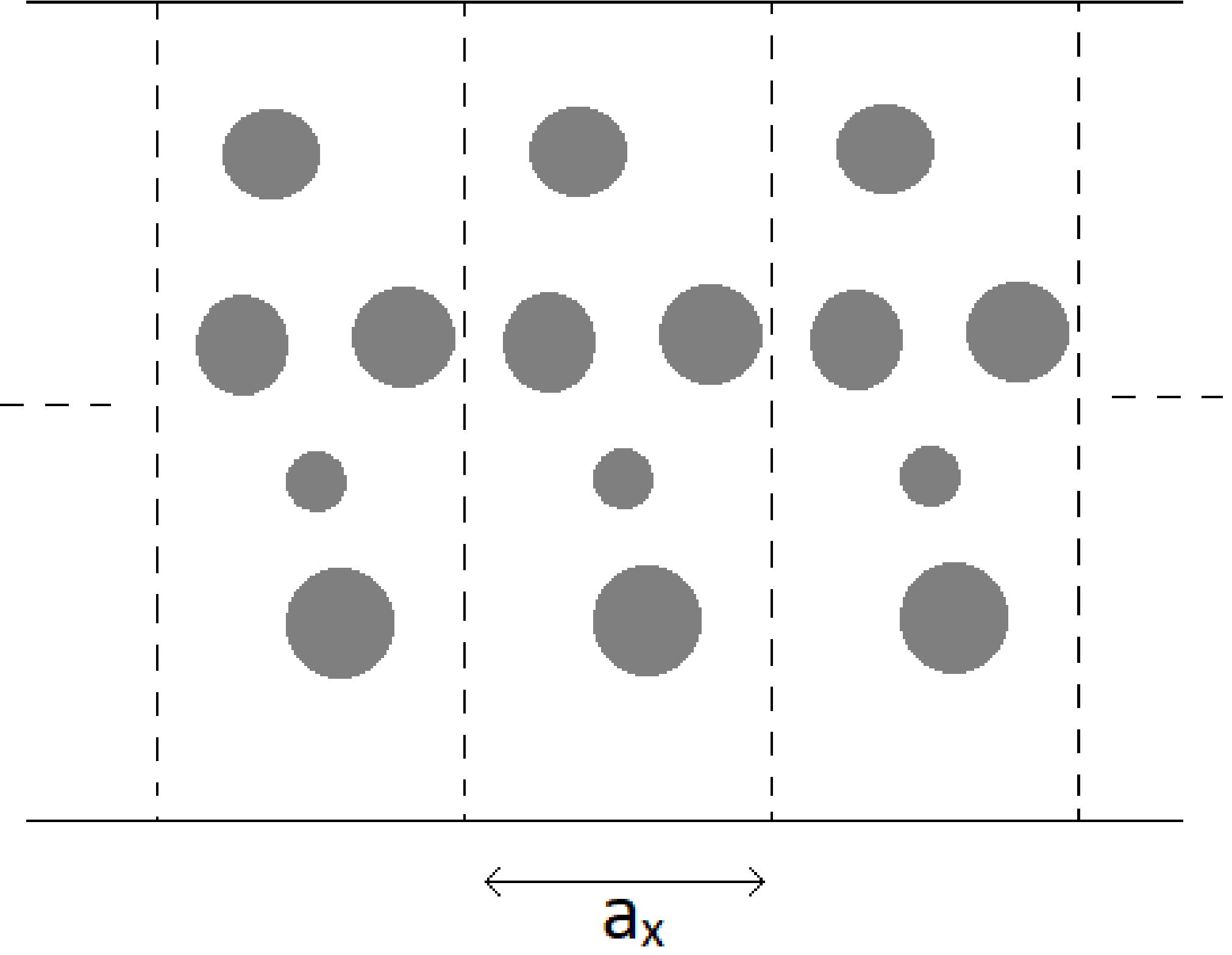}
\caption{A schematic example of a 1D-periodic medium, where $a_x$ is the spatial period. The dashed lines help to visualize a unit cell. The gray regions represent obstacles. Diffusion can occur either only in white region (so that boundaries of gray regions are impermeable), or in both gray and white regions (in which case boundaries of gray regions are permeable). In this paper, we focus on the former setting but the method can be generalized to the latter one.}
\label{fig:periodic_1d}
\end{figure}

Let us first consider a medium $\Omega$ which is periodic along a given direction, say $x$. In other words, the medium is invariant by the translation $x \to x+a_x$, where $a_x$ is the spatial period of the medium along $x$. A natural idea is to reduce the study of the whole medium to the study of a single unit cell, that is to a slab $\Omega_1= \left\{(x,y,z) \;|\; -a_x/2\leq x \leq a_x/2\right\}$, with appropriate boundary conditions, and then to expand the results to the whole medium. Note that this ``slab'' \cc{may contain} microstructural features, as illustrated on Fig. \ref{fig:periodic_1d}. 

A simple case in which the reduction is straightforward is when the transverse magnetization $m(t,x,y,z)$ is at all times periodic along $x$; in that case, one can study the magnetization and related quantities on $\Omega_1$ with periodic boundary conditions. However, although the initial condition \eqref{eq:initial_condition} is uniform (hence periodic), the BT equation \eqref{eq:BT_equation} that governs the time-evolution of the transverse magnetization is not periodic unless $g_x=0$.
Therefore, let us consider the case of $g_x \neq 0$ where one cannot directly reduce the BT equation to a single unit cell with periodic boundary conditions. For clarity we assume that the gradient is along $x$, in other words $g_y=g_z=0$, and the general case will be presented later. We introduce
\begin{equation}
q_x(t)=\int_0^t g_x(t')\,\mathrm{d}t'\;.
\end{equation}
From Eq \eqref{eq:BT_equation}, one can see that the magnetization at the position $x+a_x$ evolves in the same way as the magnetization at $x$, except for an accumulated phase:
\begin{equation}
m(t,x+a_x,y,z)=e^{i q_x(t) a_x} m(t,x,y,z)\;.
\label{eq:time_boundary_conditions}
\end{equation}
Thus, in principle one can reduce the BT equation to a single unit cell, with the \emph{time-dependent} boundary condition $m(t,a_x/2,y,z)=e^{i q_x(t) a_x} m(t,-a_x/2,y,z)$. This time-dependent boundary condition makes the problem impractical from both theoretical and numerical points of view\cc{, except in the particular case of a constant $q_x(t)$, which would result from two opposite narrow-gradient pulses at $t=0$ and $t=T$. This observation is the basis of our numerical technique described in the next subsection. In particular, if $q_x(t)a_x$ is a multiple of $2\pi$, then Eq. \eqref{eq:time_boundary_conditions} simply expresses a periodic boundary condition. In general, } an often-employed trick to discard the phase $e^{i q_x(t) a_x}$ and \cc{to} reduce the problem to simple periodic boundary conditions is to \cc{define a new function} \cite{Xu2007a,Nguyen2014a}:
\begin{eqnarray}
m_{\rm per}(t,x,y,z)= e^{-iq_x(t)x} m(t,x,y,z) \;,
\end{eqnarray}
so that Eq. \eqref{eq:time_boundary_conditions} becomes
\begin{equation}
m_{\rm per}(t,x+a_x,y,z) = m_{\rm per}(t,x,y,z)\;.
\end{equation}
\cc{We emphasize that this periodicity property is valid at all times and any point of the domain.}
Moreover, $q_x(t)=0$ at $t=0$ and the refocusing condition \eqref{eq:refocusing} implies that $q_x(t)=0$ at the end of the gradient sequence, so that $m$ and $m_{\rm per}$ coincide before and after the gradient sequence.
The BT equation \eqref{eq:BT_equation} and boundary condition \eqref{eq:robin_boundary} on $m$ become new equations on $m_{\rm per}$ in the unit cell $\Omega_1$:
\begin{eqnarray}
&\frac{\partial m_{\rm per}}{\partial t} = D_0\nabla^2 m_{\rm per} + 2D_0 iq_x(t)\frac{\partial m_{\rm per}}{\partial x} - D_0 q_x^2(t) m_{\rm per}
\label{eq:BT_new}
\\
&\restr{\mathbf{n}\cdot D_0\nabla m_{\rm per}+ iD_0 q_x(t) n_x m_{\rm per}+ \kappa m_{\rm per}}{\partial \Omega_1} = 0\;,
\label{eq:robin_new}\\
&m_{\rm per}(t,a_x/2,y,z)=m_{\rm per}(t,-a_x/2,y,z)\;,
\label{eq:perio_perio}
\end{eqnarray}
with $m_{\rm per}(t=0,x,y,z)=1$.
As expected, the non-periodic $i g_x x$ term in Eq. \eqref{eq:BT_equation} has been replaced by new, periodic terms. Note that the boundary $\partial \Omega_1$ does not include frontiers between neighboring unit cells (here, the sections $x=-a_x/2$ and $x=a_x/2$), since these are taken into account by the periodic boundary condition \cc{\eqref{eq:perio_perio}}.

The modified BT equation \eqref{eq:BT_new} now has time-dependent coefficients and the new boundary condition \eqref{eq:robin_new} is complex-valued and time-dependent.
 These features prevent the use of spectral methods that were very efficient to solve the BT equation in bounded domains.
In the next section we show how one can reformulate the BT equation in a different way, in order to reduce the problem to a single unit cell while allowing the use of spectral methods.

\subsection{Periodic boundary conditions}

We still assume that the gradient is along the $x$-axis, in other words $g_y=g_z=0$. The main idea of the method is to replace the continuous-time gradient profile by a series of infinitely narrow gradient pulses: computing the magnetization is then reduced to solving a series of diffusion problems with different (pseudo-)periodic boundary conditions. Note that the idea of replacing a gradient profile by multiple narrow pulses was introduced and exploited in \cite{Caprihan1996a,Callaghan1997a,Sukstanskii2002a}  to compute the magnetization in bounded domains. One will see that the case of periodic domains is much more subtle.

\begin{figure*}[t]
\centering
\includegraphics[width=0.99\linewidth]{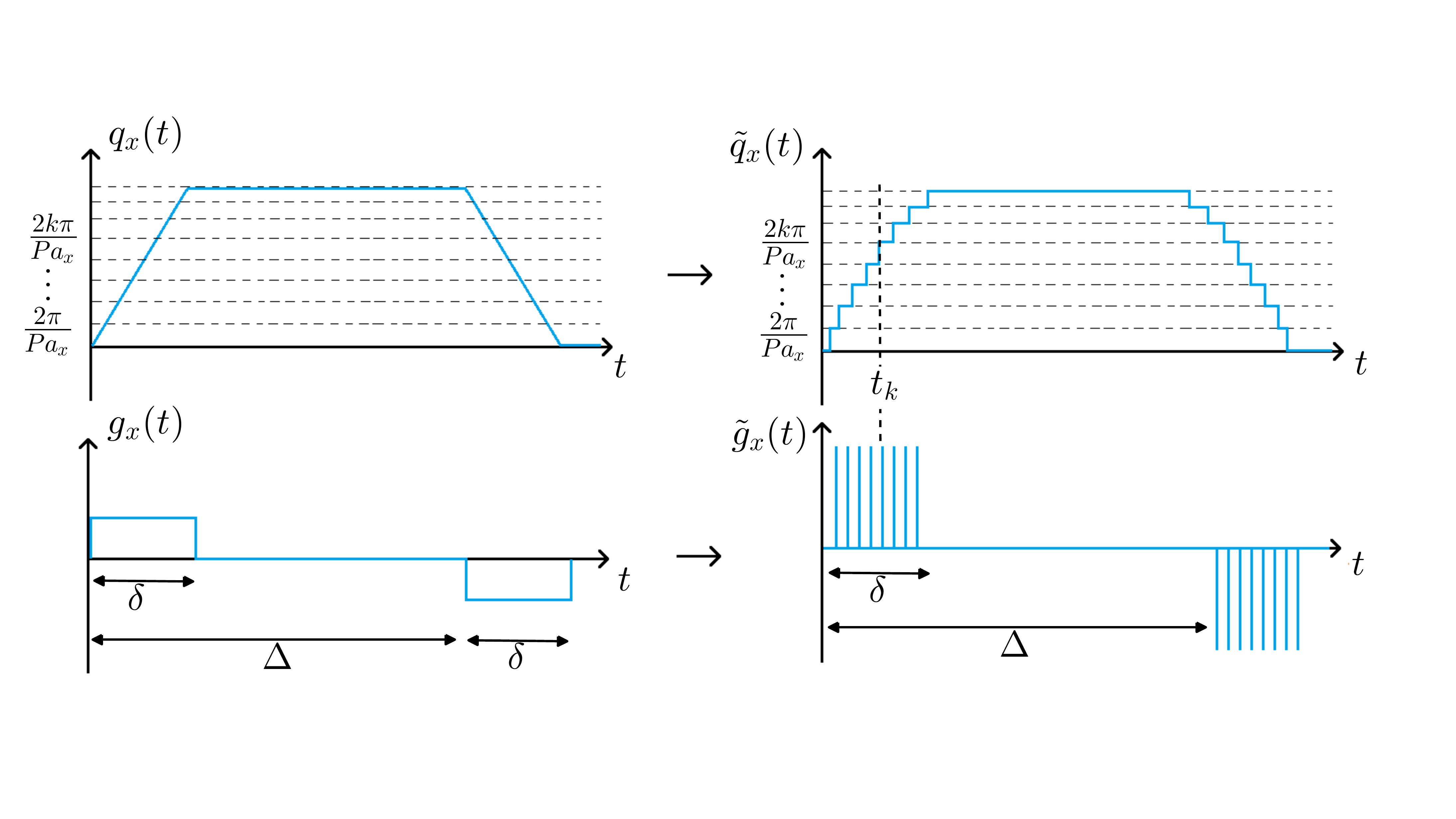}
\caption{The $q_x(t)$ function is sampled at multiples of $2\pi/(P a_x)$, the gradient $g_x(t)$ becomes a series of Dirac peaks at times $t_k$, $k=1, 2, \ldots$. Here, an example with two rectangular gradient pulses (``pulsed-gradient spin-echo sequence'') is shown but the technique is applicable to any gradient profile.}
\label{fig:sampling}
\end{figure*}

For the sake of clarity, let us first present the simplest case that involves only periodic boundary conditions. If we sample the function $q_x(t)$ at multiples of $2\pi/a_x$ and replace it by a step function $\tilde{q}_x(t)$, the gradient is then replaced by a series of Dirac peaks $\tilde{g}_x(t)$ with weights $\pm 2\pi/a_x$ (see Fig. \ref{fig:sampling} with $P=1$). In other words, a positive/negative gradient pulse effectively multiplies the magnetization by $\exp(\pm 2i\pi x/a_x)$.

If the initial magnetization is periodic along $x$, then it remains periodic at all times. Indeed, the gradient pulses and the diffusion steps both preserve the periodicity. Thus, one can project the magnetization on the eigenmodes of the Laplace operator on the slab $\Omega_1$ with a periodic boundary condition along $x$ coordinate:
\begin{eqnarray}
&\cc{-D_0 \nabla^2 u_{0,n}=\lambda_{0,n}u_{0,n}\nonumber }\\
&u_{0,n}(x=a_x/2,y,z)=u_{0,n}(x=-a_x/2,y,z)\nonumber
\\
&\restr{\mathbf{n}\cdot D_0 \nabla u_{0,n} + \kappa u_{0,n}}{\partial \Omega_1} = 0\;,\nonumber
\end{eqnarray}
where $n=0,1,\ldots$, and the eigenmodes $u_{0,n}$ are $L^2(\Omega_1)$-normalized. The reason for the index ``$0$'' will be clarified when we move to the general (pseudo-periodic) case. 

After projection on the eigenmode basis $u_{0,n}$, the magnetization is represented by a vector $M$:
\begin{eqnarray}
m(t,x,y,z)=\sum_n M_n(t) u_{0,n}(x,y,z)\;,\\
M_n(t)=\int_{\Omega_1} m(t,x,y,z)u_{0,n}^*(x,y,z)\,\mathrm{d}x\,\mathrm{d}y\,\mathrm{d}z\;.
\end{eqnarray}
The computation of the magnetization is then reduced to matrix multiplications. A diffusion step of duration $\tau$ corresponds to left-multiplication by the matrix \cc{$\exp(-\tau\Lambda_{0})$}, where $\Lambda_0$ is a diagonal matrix with elements $\lambda_{0,n}, n\geq 0$. A positive gradient pulse corresponds to left-multiplication by the matrix $G^x_0$, whose elements are
\begin{equation}
\left[G^{x}_{0}\right]_{n,n'}=\int_{\Omega_1} u_{0,n}^* \exp(2i\pi x/a_x) u_{0,n'} \,\mathrm{d}x\,\mathrm{d}y\,\mathrm{d}z\;,
\end{equation}
and a negative pulse corresponds to the matrix $(G^x_0)^\dagger$.

In summary, for a  given periodic medium, one first computes a sufficient number of eigenmodes, constructs the matrix $G^x_0$, discretizes the gradient profile, and then computes the magnetization and/or the normalized signal via a matrix product of the form:
\begin{eqnarray}
&\cc{M=\left(e^{-\tau_N\Lambda_0}\,(G^{x}_0)^\dagger \cdots  G^{x}_0 \,e^{-\tau_1\Lambda_0}\, G^{x}_0\right) M_0\;, }
\label{eq:matrix_magn}\\
&S=M_0^\dagger M/(M_0^\dagger M_0)\;,
\label{eq:matrix_signal}
\end{eqnarray}
where $M_0$ represents the initial condition \eqref{eq:initial_condition}:
\begin{equation}
\left[M_0\right]_n=\int_{\Omega_1} u_{0,n}^* \,\mathrm{d}x\,\mathrm{d}y\,\mathrm{d}z\;,
\label{eq:matrix_magn0}
\end{equation}
and the left multiplication by $M_0^\dagger$ represents the integration over a unit cell.
In Eq. \eqref{eq:matrix_magn}, $N$ is the number of narrow pulses, $\tau_1, \tau_2,\ldots \tau_N$ are the time intervals between adjacent narrow pulses, and one has $\tau_1+\tau_2+\dots+\tau_N = T$ the total duration of the gradient sequence (here we assume that the gradient sequence lasts up to the echo time $T$, at which the signal is measured).
With the notations of Fig. \ref{fig:sampling}, $\tau_k=t_{k+1}-t_k$ for $k=1,\ldots,N$ with the convention $t_{N+1}=T$.

\subsection{Pseudo-periodic boundary conditions}

In general, it may be restrictive to sample $q_x(t)$ at multiples of $2\pi/a_x$, especially at low gradient strength. The above method can be generalized to any sampling: for example one can sample $q_x(t)$ at every multiple of $2\pi/(P a_x)$, with a given integer $P$ (see again Fig. \ref{fig:sampling}). In that case, each gradient pulse multiplies the magnetization by $\exp(\pm 2i\pi x/(P a_x))$. Naturally, other sampling choices are possible.
Because of the sampling, at all times the magnetization obeys:
\begin{eqnarray}
m(t,a_x/2,y,z)=e^{ip a_x} m(t,-a_x/2,y,z)\;,
\label{eq:pseudo-periodic-boundary-conditions}\\
p=\tilde{q}_x(t) \quad (\mathrm{mod} \; 2\pi/a_x)\;.
\label{eq:def_p}
\end{eqnarray}
Throughout the text, we will call ``$p$-pseudo-periodic'' a function that obeys Eq. \eqref{eq:pseudo-periodic-boundary-conditions}, and $p$ is the wavenumber that defines the pseudo-periodicity condition. Note that, as the function $\tilde{q}_x(t)$ is piecewise constant, there is a finite number of different values of $p$ involved during the gradient sequence. For example, if one samples $q_x(t)$ at multiples of $2\pi/(Pa_x)$ as in Fig. \ref{fig:sampling}, there are only $P$ different values of $p$.

Every $p$-pseudo-periodic function can be projected onto the $p$-pseudo-periodic eigenmode basis of the Laplace operator on $\Omega_1$\cc{\cite{Kuchment1993a}}:
\begin{eqnarray}
&\cc{-D_0\nabla^2 u_{p,n}=\lambda_{p,n}u_{p,n}}  \\
&u_{p,n}(x=a_x/2,y,z)=e^{ipa_x}u_{p,n}(x=-a_x/2,y,z)\;
\\
&\restr{\mathbf{n}\cdot D_0\nabla u_{p,n} + \kappa u_{p,n}}{\partial \Omega_1} = 0\;.
\end{eqnarray}
The wavenumber $p=0$ corresponds to periodic eigenmodes, which is consistent with our previous notations. A diffusive step of duration $\tau$ translates then into left-multiplication by the matrix \cc{$\exp(-\tau \Lambda_p)$}, with $\Lambda_p$ being a diagonal matrix with elements $\lambda_{p,n}, n\geq0$. A narrow gradient pulse of weight $q_0$ corresponds to the left-multiplication by $G^x_{p\to p+q_0}$:
\begin{equation}
\left[G^x_{p\to p+q_0}\right]_{n,n'}=\int_{\Omega_1} u^*_{p+q_0,n} e^{iq_0 x} u_{p,n'} \,\mathrm{d}x\,\mathrm{d}y\,\mathrm{d}z\;,
\end{equation}
that is the projection of the $p$-pseudo-periodic basis onto the $(p+q_0)$-pseudo-periodic basis after multiplication by $\exp(iq_0 x)$. 
Note that by performing several pulses  in succession in order to cycle through a $2\pi$ phase difference between $x=-a_x/2$ and $x=a_x/2$, one gets the equivalent of one pulse of weight $2\pi/a_x$, in other words
\begin{equation*}
G^x_{0}=G^x_{p_{N-1}\to 2\pi/a_x}\,G^x_{p_{N-2}\to p_{N-1}}\dots \,G^x_{p_1\to p_2}\,G^x_{0\to p_1}\;,
\end{equation*}
where $0<p_1<p_2<\cdots<p_{N-1}<2\pi/a_x$. This algebraic relation is a direct consequence of the completude of the $p$-pseudo-periodic Laplacian eigenmode bases and shows how the $G^x_{p\to p+q_0}$ matrices generalize the $G^x_0$ matrix from the previous subsection.

Similarly to the periodic case presented above, one can compute the magnetization at all times by successively applying the matrix multiplications corresponding to the gradient sequence:
\begin{eqnarray}
\cc{M=\left(e^{-\tau_N\Lambda_0}\, G^x_{p_N\to 0}\cdots G^x_{p_1\to p_2}\, e^{-\tau_1 \Lambda_{p_1}}\,G^x_{0\to p_1}\right)M_0\;,}
\label{eq:matrix_magn_bis}
\end{eqnarray}
where $N$ is the number of narrow pulses, $\tau_1,\tau_2,\ldots,\tau_N$ are the time intervals between adjacent narrow pulses and satisfy $\tau_1+\tau_2+\dots+\tau_N=T$ the duration of the gradient sequence, and $p_1,p_2,\ldots,p_N$ are the sampled values of $q_x(t)$ modulo $2\pi/a_x$. With the notations of Fig. \ref{fig:sampling}, one has $p_k=\tilde{q}_x(t_k) \quad (\mathrm{mod} \; 2\pi/a_x)$ and $\tau_k = t_{k+1} - t_k$ for $k=1,\ldots,N$ with the convention $t_{N+1}=T$.
Due to the refocusing condition \eqref{eq:refocusing}, the magnetization at the end of the gradient sequence is periodic again, so that the wavenumber $p$ is equal to zero, hence the last gradient pulse matrix $G^x_{p_N\to 0}$ in Eq. \eqref{eq:matrix_magn_bis}. The initial condition $M_0$ is still given by Eq. \eqref{eq:matrix_magn0}, and the normalized signal can be computed with Eq. \eqref{eq:matrix_signal}.

%

\subsection{Relation with Bloch bands, diffusion-diffraction, and diffusion pore imaging}
\label{section:blabla}
The collection of all $p$-pseudo-periodic eigenvalues are exactly the Bloch bands of the periodic medium, a fundamental concept in condensed matter physics \cite{Bloch1929a,Kittel2004a}. The previous formulas potentially allow one to measure the Bloch bands of a periodic medium by performing a short-gradient pulses experiment and fitting the signal by a multi-exponential function of the diffusion time between two pulses (see Fig. \ref{fig:sampling}, with the pulse duration $\delta\to 0$ and variable inter-pulse duration $\Delta$) . Indeed, a short gradient pulse of weight $q_x$ allows one to select a given pseudo-periodicity wavenumber $p$, and the signal decays then according to the Laplacian eigenvalues corresponding to that wavenumber \cc{\cite{Sen1994a,Dunn1994a,Bergman1995a}}:
\begin{eqnarray}
\cc{S=\sum_{n=0}^{\infty} \left | C_{p,n}(q_x) \right|^2 \exp(-\lambda_{p,n} \Delta)\;, \label{eq:structure_factor_pseudo}}\\
C_{p,n}(q_x)=\frac{1}{\sqrt{\mathrm{vol}(\Omega_1)}} \int_{\Omega_1} e^{iq_x x} u^*_{p,n}(x,y,z)\,\mathrm{d}x\,\mathrm{d}y\,\mathrm{d}z \;, \nonumber
\\
p=q_x \quad (\mathrm{mod}\;2\pi/a_x)\;.\nonumber
\end{eqnarray}
The notation for $C_{p,n}(q_x)$ is somewhat redundant because $p$ is a function of $q_x$; its purpose is to present $C_{p,n}$ as a generalization of the form factor of bounded domains that corresponds to $C_{0,0}$.
As the eigenvalues $\lambda_{p,n}$ generally scale as \cc{$D_0/a_x^2$}, the signal $S$ typically exhibits a multi-exponential time-decay over the duration $\Delta\sim a_x^2/D_0$, and then becomes mono-exponential \cc{at longer times $\Delta$}. In contrast with diffusion in the free space $\mathbb{R}^3$ where the signal decays as $\exp(-q_x^2 D_0\Delta)$, the long-time decay of the signal in a periodic medium with microstructural features is controlled by $\lambda_{p,0}$ that is a bounded function of $q_x$.

The above formula generalizes the expression used by Callaghan \textit{et al.} in their seminal work \cite{Callaghan1991b}. In that work, a packing of monodisperse beads is treated as a collection of pores separated by a constant spacing along the gradient direction, i.e., a periodic lattice.
As they were interested in the long-time limit when water molecules could diffuse through multiple pores, their main formula is exactly the first term ($n=0$) of Eq. \eqref{eq:structure_factor_pseudo}. 
If one assumes zero surface relaxivity on the obstacles and pore boundaries, then $\lambda_{p,0}=0$ for $p=0 
$ and $\lambda_{p,0}>0$ otherwise.
Thus at long times, Eq. \eqref{eq:structure_factor_pseudo} displays relatively sharp maxima at $q_x = 2 k \pi / a_x$, $k=0,1,2,\ldots$. This important feature, called ``diffusion-diffraction pattern'', allowed Callaghan \textit{et al.} to recover the lattice step $a_x$ (i.e., pore spacing). Moreover, the value of the squared generalized form factor $\left|C_{p,0}\right|^2$ allowed them to extract geometrical features of the pores, in particular their diameter (assuming a spherical shape). 

Note that one could improve this last step by getting access to the phase information of the form factor (lost because of the absolute value). This possibility was shown by Laun \textit{et al.} in bounded \cc{and periodic} domains by using asymmetric gradient sequences (short and long pulses, double diffusion encoding, and others), thus opening the field of diffusion pore imaging \cc{\cite{Laun2011a,Laun2012a,Demberg2017a,Laun2016a}}.  Plots of the magnetization and signal for a short-gradient pulses sequence are presented in \ref{section:NPA}.

\subsection{Optimization and range of validity}
\subsubsection{Sampling scheme optimization}
\label{section:sampling_scheme}

One can sample $q_x(t)$ in different ways that lead to different step functions $\tilde{q}_x(t)$, which are more or less close to the original profile. In essence, this is similar to approximating integrals by Riemann sums or to rounding a decimal number.
In fact, there are at least 4 natural approximation schemes: (i) ``flooring'' scheme where $\tilde{q}_x(t)$ is equal to the sampled value immediately below $q_x(t)$; (ii) ``ceiling'' scheme would choose the value immediately above $q_x(t)$; (iii) ``rounding'' scheme would choose the value which is the closest to $q_x(t)$; (iv) ``midpoint'' scheme would be to place the gradient pulses (i.e. the jumps in $\tilde{q}_x(t)$) inbetween the pulses of the flooring scheme and those of the ceiling scheme. Although these 4 schemes are the most straigthforward ones, many others are possible. Note that the midpoint and rounding schemes give the same results if the gradient is constant.
If one considers the free diffusion case as a benchmark, the criterion for the sampling scheme is to reproduce the $b$-value, $b=\int_0^T q_x^2(t)\,\mathrm{d}t$, as accurately as possible. From the theory of Riemann sums, the most accurate sampling scheme among the four considered above would be the midpoint one, followed by the rounding one.

\subsubsection{Fine sampling}
\label{section:fine_sampling}
The second point to optimize is the size of the steps of $\tilde{q}_x(t)$. For simplicity, we assume that $q_x$ is sampled at multiples of $2\pi/(P a_x)$ as in Fig. \ref{fig:sampling}. The larger we choose $P$, the finer the sampling and the better the approximation. To have a more quantitative view on this question, one can again consider free diffusion as a benchmark and compare the effect of a finite pulse of strength $g$ and duration $\tau$ with a narrow pulse of weight $q_0=g\tau$ such that $q_0=2\pi/(Pa_x)$. Following the conclusion of the previous subsection, the narrow pulse is performed at $t=\tau/2$.

Because the $q$-value associated to both pulses is the same, the only difference is the decay of the magnetization during the pulse itself. This decay is simply expressed as $\exp(- g^2 \tau^3 D_0/3) = \exp(- q_0^2 D_0\tau/3)$ for the continuous pulse, and $\exp(-q_0^2 D_0\tau/2)$ for the narrow pulse, resulting in a ratio of $\exp( - q_0^2 D_0\tau/6)$. This additional decay accumulates over all pulses, so that if $T$ is the total time during which the gradient is turned on, one gets that the multiple narrow pulses create an additional attenuation factor $\exp(- q_0^2 D_0 T/6)$ compared to the continuous gradient. One obtains the same formula by directly comparing the continuous-time value $bD_0=D_0\int_0^T q_x^2(t)\,\mathrm{d}t$ with its discrete version.

Now, according to the sampling scheme detailed previously, one should replace $q_0$ by $2\pi /(P a_x)$ hence the relative error created by the sampling reads
\begin{equation}
\epsilon=1-\exp\left(-\frac{4\pi^2}{6 P^2}\frac{D_0 T}{a_x^2}\right) \approx \frac{7}{P^2} \frac{D_0 T}{a_x^2}\;.
\label{eq:error_estimation}
\end{equation}
This estimation allows one to control the quality of the approximation as a function of $P$. Since any microstructure on a much finer scale than the diffusion length $\sqrt{D_0 T}$ would be modeled via reduced (effective) medium diffusivity, it is reasonable to assume that the diffusion length is at most of the order of magnitude of the lattice step: $\sqrt{D_0 T} \lesssim a_x$. Thus it is possible to choose a value of $P$ to ensure a good compromise between accuracy and computation time.

It should be noted that many gradient sequences, especially the pulsed-gradient spin-echo (PGSE) sequence \cite{Stejskal1965a,Tanner1968a} (see Fig. \ref{fig:sampling}), contain a free diffusion step during which the gradient is off. This means that $q_x(t)$ (resp. $\tilde{q}_x(t)$) would take a constant value $q_{\rm off}$ (resp. $\tilde{q}_{\rm off}$) over a duration $t_{\rm off}$. In terms of $b$-value, the discrepancy between $q_{\rm off}$ and $\tilde{q}_{\rm off}$ would accumulate over the whole duration $t_{\rm off}$ and yield a difference in $b$-values equal to $t_{\rm off}(q^2_{\rm off}-\tilde{q}^2_{\rm off})$. Thus, if $t_{\rm off}$ is large, even a very fine sampling may lead to an important error. To prevent this, a simple solution is to add the constant value $q_{\rm off}$ explicitly in the sampling scheme.

\subsection{Extension to higher dimensions}

In the previous sections we dealt with a medium that is periodic along one direction, and the gradient was aligned with that direction. In this section, we show how to extend the results to an arbitrary gradient direction, as well as multi-dimensional periodic media.

\subsubsection{1D-periodic medium and arbitrary gradient}
\label{section:periodic_bounded}

Here we assume that the medium is still periodic along $x$ and bounded along $y$ and $z$. The gradient direction is arbitrary and may change over time as well.
Since the medium is bounded along $y$ and $z$, the effect of $g_y$ and $g_z$ can be implemented using standard spectral methods \cite{Barzykin1998a,Callaghan1997a,Grebenkov2007a,Grebenkov2008b}. Two main schemes were proposed in the literature, in which the gradient is either replaced by (i) a collection of narrow pulses \cite{Caprihan1996a,Callaghan1997a,Sukstanskii2002a} (similar to our method but without restrictions introduced by periodicity); or (ii) a stepwise function \cite{Barzykin1998a,Grebenkov2007a,Grebenkov2008b}.
For clarity and consistency of notations we show here how to implement the narrow pulse approach, and the extension to the stepwise gradient approach is detailed in \ref{section:spectral_method_bis}.

Between two narrow $g_x$ pulses, the magnetization is $p$-pseudo-periodic with a given wavenumber $p$ and one can compute the effect of narrow gradient pulses of weight $q_0$ along $y$ or $z$ with the following matrices:
\begin{eqnarray}
&\left[G^y_{p}\right]_{n,n'}=\int_{\Omega_1} u^*_{p,n} e^{iq_0y} u_{p,n'} \,\mathrm{d}x\,\mathrm{d}y\,\mathrm{d}z\;,
\label{eq:G_y}\\
&\left[G^z_{p}\right]_{n,n'}=\int_{\Omega_1} u^*_{p,n} e^{iq_0z} u_{p,n'} \,\mathrm{d}x\,\mathrm{d}y\,\mathrm{d}z\;.
\label{eq:G_z}
\end{eqnarray}
There are two main differences between the $G^y_p$, $G^z_p$ matrices and the $G^x_{p\to p+q_0}$ matrices presented above.
First, since the $y$ or $z$ pulses do not interfere with the pseudo-periodic boundary condition along $x$, there is no restriction on the sampling of $q_y(t)$ and $q_z(t)$ as it is the case with $q_x(t)$ where each new additional sampled value $p$ requires the computation of a family of eigenmodes $u_{p,n}$. Moreover, since the boundary condition along $y$ and $z$ does not evolve with the $g_y$ and $g_z$ gradient pulses, one needs to compute only one $G^y_p$ and $G^z_p$ matrix for each value of $p$. The only requirement is that the value of $q_0$ in Eqs. \eqref{eq:G_y} and \eqref{eq:G_z} is sufficiently small to provide a correct sampling of $q_y(t)$ and $q_z(t)$.

\subsubsection{2D/3D periodic medium and arbitrary gradient}
\label{section:multi-periodic}
If the medium is periodic along, say, $x$ and $y$, then one has to sample both $q_x(t)$ and $q_y(t)$ in order to apply the same numerical technique. This leads to two pseudo-periodicity wavenumbers $p_x$ and $p_y$, and two families of matrices $G^x_{p_x\to p_x+q_0,p_y}$, $G^y_{p_x,p_y\to p_y+q_0}$. If the medium is periodic along $x$, $y$ and $z$, one has three indices and three families of $G$ matrices.

\begin{figure}[t]
\centering
\includegraphics[width=0.29\linewidth]{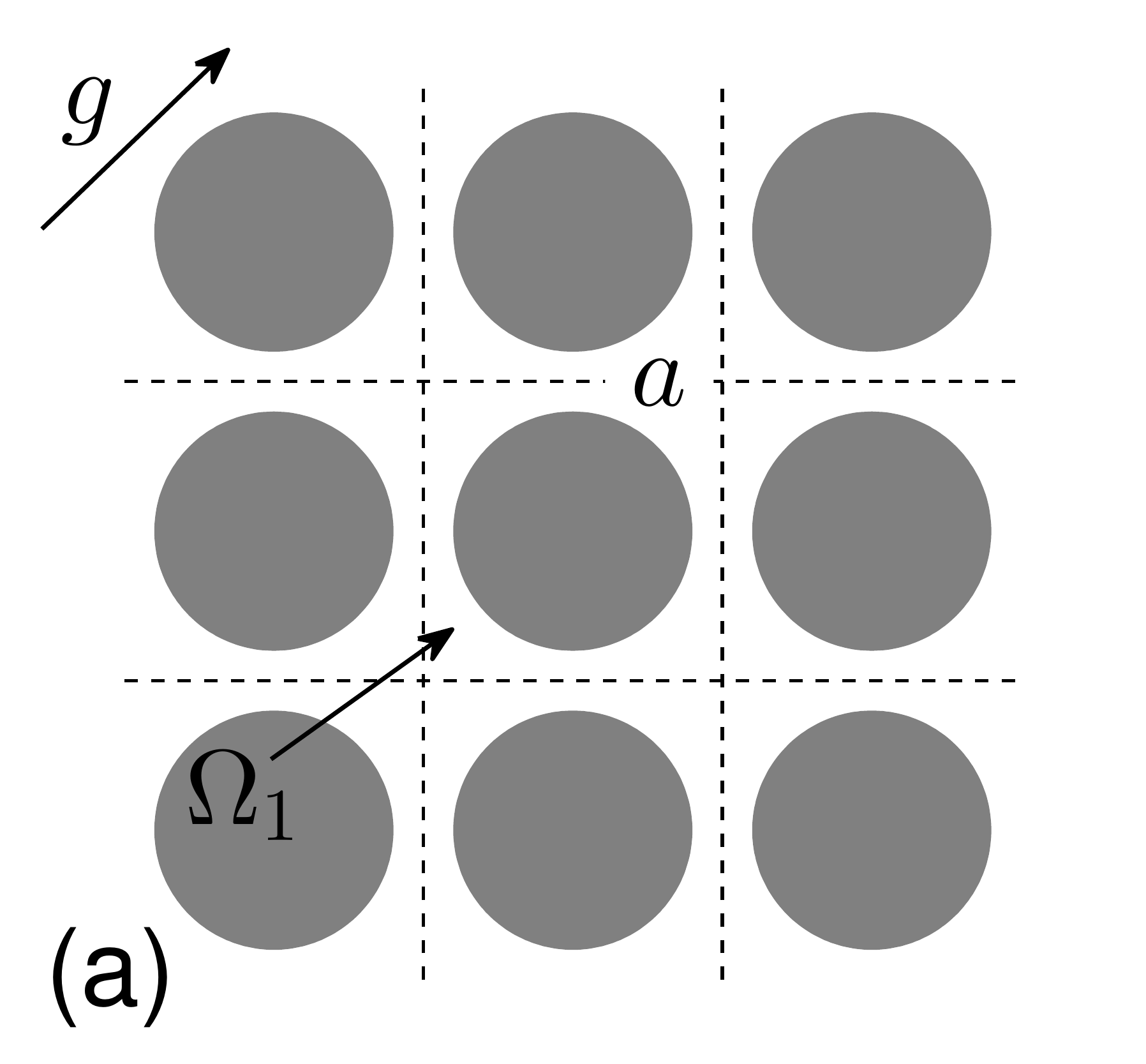}
\includegraphics[width=0.34\linewidth]{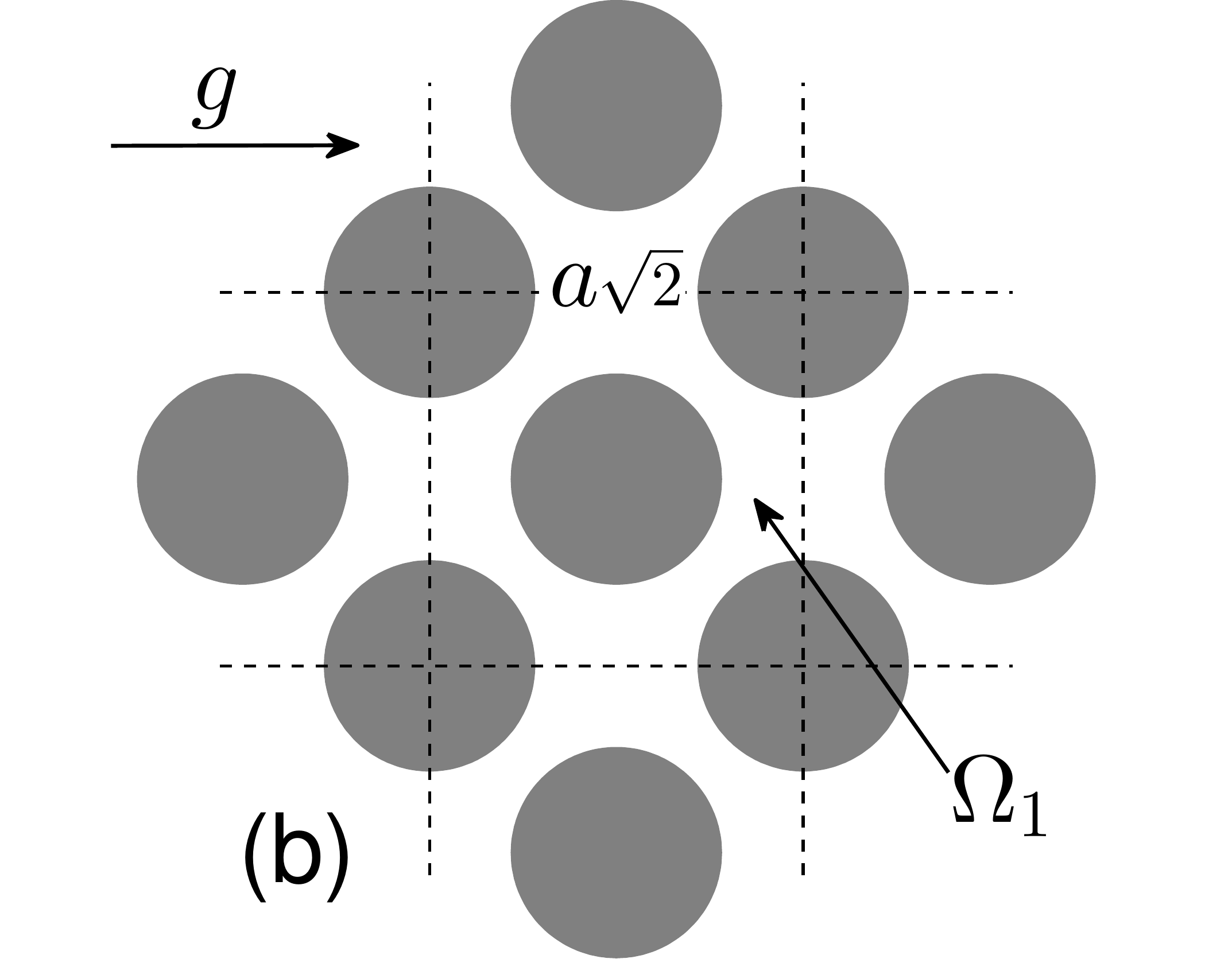}
\includegraphics[width=0.35\linewidth]{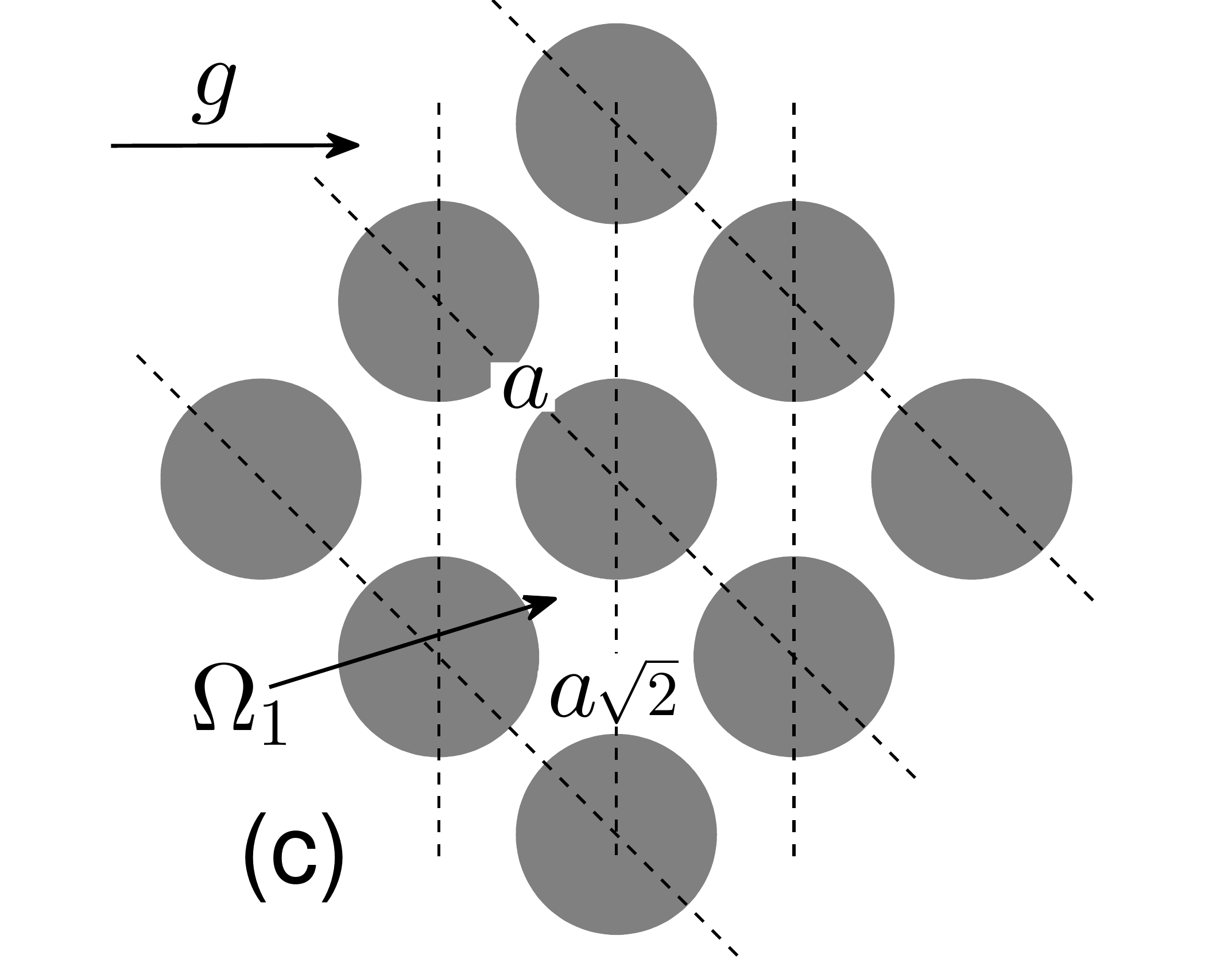}
\caption{This figure shows that the choice of the unit cell (delimited by dashed lines) is arbitrary and can be tailored to the particular orientation of the gradient. The \cc{situations} (a), (b) and (c) are geometrically equivalent but in (b), (c) the gradient is orthogonal to one of the lattice axes, that simplifies the computations. The unit cell in (b) has a convenient square shape but can be reduced further to (c), with a length along the gradient direction equal to $a/\sqrt{2}$.}
\label{fig:periodic_2d}
\end{figure}

Particular orientations of the gradient may simplify the computations. The simplest example is when one of the component of the gradient is zero, in that case no sampling needs to be done and the magnetization is at all times periodic along that direction. Another example is the gradient which is perpendicular to a lattice vector. In that case, it might be interesting to re-define the unit cell to cancel all components of the gradient except one (see Fig. \ref{fig:periodic_2d}). 

\section{Numerical Implementation and Results}
\label{section:numerical_results}
\subsection{Numerical implementation}
\label{section:numerical}

As discussed above, the computation of the magnetization and the normalized signal is reduced to matrix multiplications. However, all the matrices $G^x_{p\to p+q_0}$, $G^y_p$, $G^z_p$, $\Lambda_p$, depend on the Laplacian eigenfunctions with (pseudo-)periodic boundary conditions. Except for some trivial cases, these eigenfunctions are not known and need to be computed numerically. This computational step is usually the most time-consuming. However, once the eigenfunctions and the consequent matrices are computed for a given unit cell, one can apply them to various gradient sequences and strengths. 

We implemented the algorithm in Matlab by using the finite element PDE toolbox, but other numerical solvers could be used to compute the Laplacian eigenfunctions. The practical difficulty was that periodic/pseudo-periodic boundary conditions are not available in the Matlab PDE solver. Thus we generated the mass and stiffness matrices by imposing Neumann boundary conditions on the outer boundaries of the unit cell and then modified those matrices in order to account for the periodic or pseudo-periodic boundary conditions.
The search for eigenmodes and eigenvalues has to be truncated at \cc{some} threshold. Typically, any eigenvalue much larger than $1/\tau$, where $\tau$ is the duration between two Dirac peaks, can be omitted because its contribution to the final result will be negligible. Indeed, the diffusion step between two peaks corresponds to the multiplication by the matrix $\exp(-\tau \Lambda_p)$.
In practice, one can control the truncation error by increasing the truncation threshold and checking whether the \cc{difference between} computed quantities is small. We also employed this check to control the number of mesh points in the domain.

In the following, we will present numerical results for the particular example of a 2D square lattice of circular impermeable obstacles with no surface relaxivity (i.e., $\kappa=0$). 
For simplicity, we apply a PGSE sequence with rectangular gradient pulses of duration $\delta$ and no diffusion time between two pulses (i.e., $\Delta=\delta$), so that the problem is fully determined by three length ratios: $R/a$, $\ell_g/a$ and $\ell_\delta/a$, where $a_x=a_y=a$ is the lattice step, $R$ is the radius of obstacles and
\begin{equation*}
 \ell_g=(g/D_0)^{-1/3} \quad \mathrm{and} \quad \ell_\delta=(D_0\delta)^{1/2}
 \end{equation*}
 are respectively the gradient and diffusion lengths. The gradient length controls the competition between the Laplacian and gradient terms of the BT equation and can be interpreted as the typical length over which diffusing spins get uncorrelated phases.
Results for very short gradient pulses and non-zero diffusion time between pulses are presented in \ref{section:NPA}  and we show important and interesting qualitative differences in the transverse magnetization profile and the resulting signal.
The initial transverse magnetization is uniform and equal to $1$. We recall that the free-diffusion case (i.e., a periodic medium without any obstacle) would yield a uniform magnetization
\begin{equation}
m= \exp(-bD_0)=\exp\left(-\frac{2}{3}\left(\frac{\ell_\delta}{\ell_g}\right)^6\right)\;.
\label{eq:free_decay}
\end{equation}

In order to help relating the numerical results to practical experiments, let us compute typical values of $\ell_g$ and $\ell_\delta$ in the case of water diffusion (probed by the hydrogen resonance). We set $D_0=2~\mathrm{\mu m^2/ms}$, and we write $g=\gamma G$ with $\gamma=2.7\cdot 10^8~\mathrm{T^{-1}s^{-1}}$, where $G$ is the magnetic field gradient. Thus we have
\begin{eqnarray*}
2~\mathrm{ms} \leq \delta \leq 50~\mathrm{ms} & \quad \rightarrow \quad 2~\mathrm{\mu m} \leq \ell_\delta \leq 10~\mathrm{\mu m}\;,\\
1~\mathrm{mT/m} \leq G \leq 1~\mathrm{T/m} & \quad \rightarrow \quad 20~\mathrm{\mu m} \geq \ell_g \geq 2~\mathrm{\mu m}\;,
\end{eqnarray*}
where the inequalities were reversed for $\ell_g$ since it decreases with increasing gradient strength.
For experiments with xenon gas, such as in \cite{Moutal2019b}, $D_0=0.04~\mathrm{mm^2/ms}$ and $\gamma=7.4\cdot 10^7~\mathrm{T^{-1}s^{-1}}$, so that one gets:
\begin{eqnarray*}
2~\mathrm{ms} \leq \delta \leq 50~\mathrm{ms} & \quad \rightarrow \quad 0.3~\mathrm{mm} \leq \ell_\delta \leq 1.5~\mathrm{mm}\;,\\
1~\mathrm{mT/m} \leq G \leq 1~\mathrm{T/m} & \quad \rightarrow \quad 0.8~\mathrm{mm} \geq \ell_g \geq 0.08~\mathrm{mm}\;.
\end{eqnarray*}
Note the considerable upscaling (by a factor of about $100$) of gas experiments compared to water, due to the much larger diffusion coefficient.

We chose to sample $q$-values at multiples of $2\pi/(P a)$ with a rounding sampling scheme (see Sec. \ref{section:sampling_scheme}). Thus we computed  $P$ families of eigenmodes for a given geometry.
All computations were performed with $P=120$, about $6000$ mesh points in a single unit cell and $240$ Laplacian eigenmodes for each pseudo-periodic boundary condition. The computation of all eigenmodes and eigenvalues took about $5$ minutes on a standard desktop computer. Once this preliminary step has been performed, all computations of the magnetization took less than one second. For better visibility, we plot the magnetization inside one unit cell surrounded by its neighbors. We stress, however, that the computations were performed solely inside one unit cell and then the results were ``copy-pasted'' to other cells.

\subsection{Results}

\begin{figure}[t]
\centering
\includegraphics[width=0.24\linewidth]{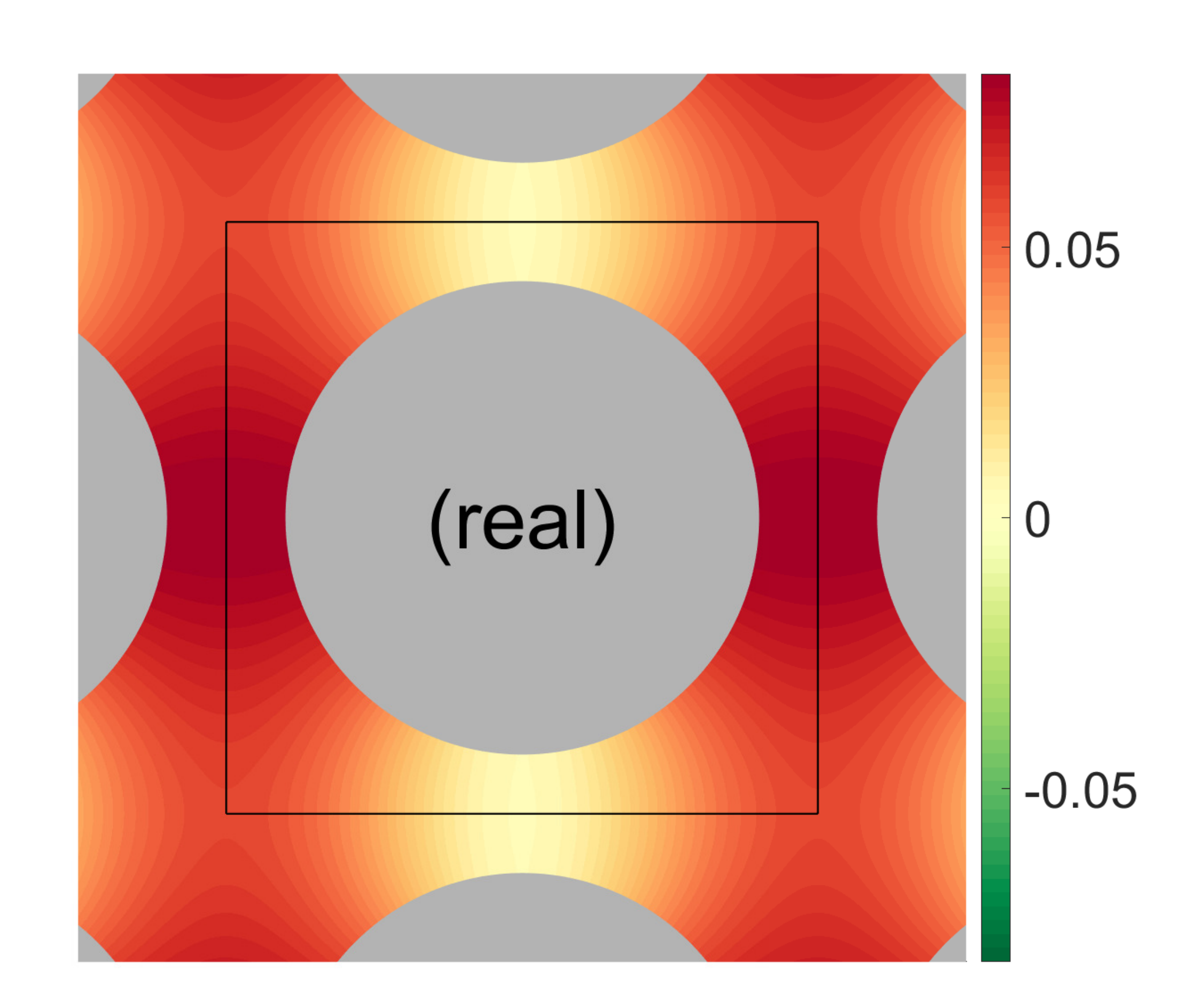}
\includegraphics[width=0.24\linewidth]{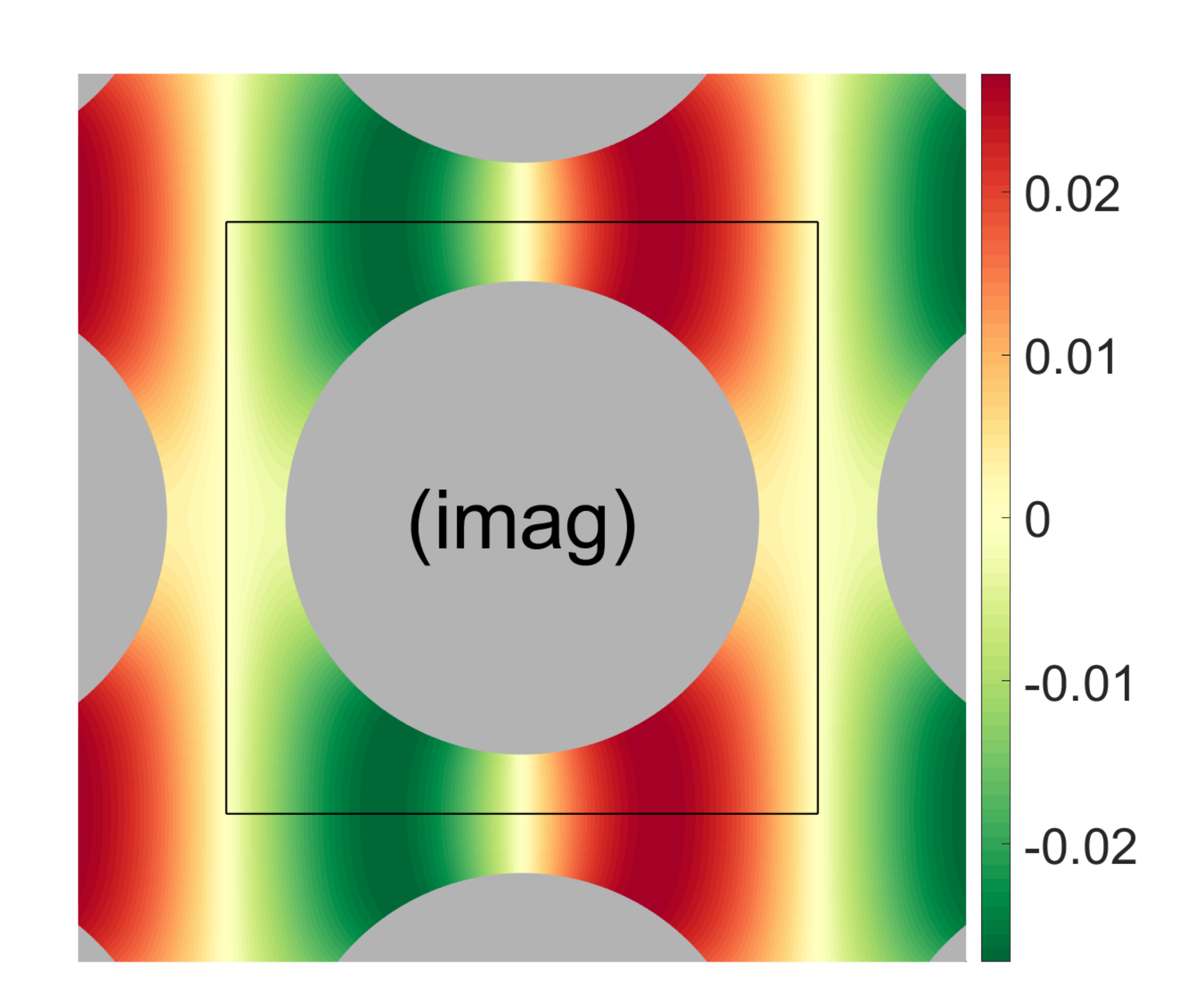}
\includegraphics[width=0.24\linewidth]{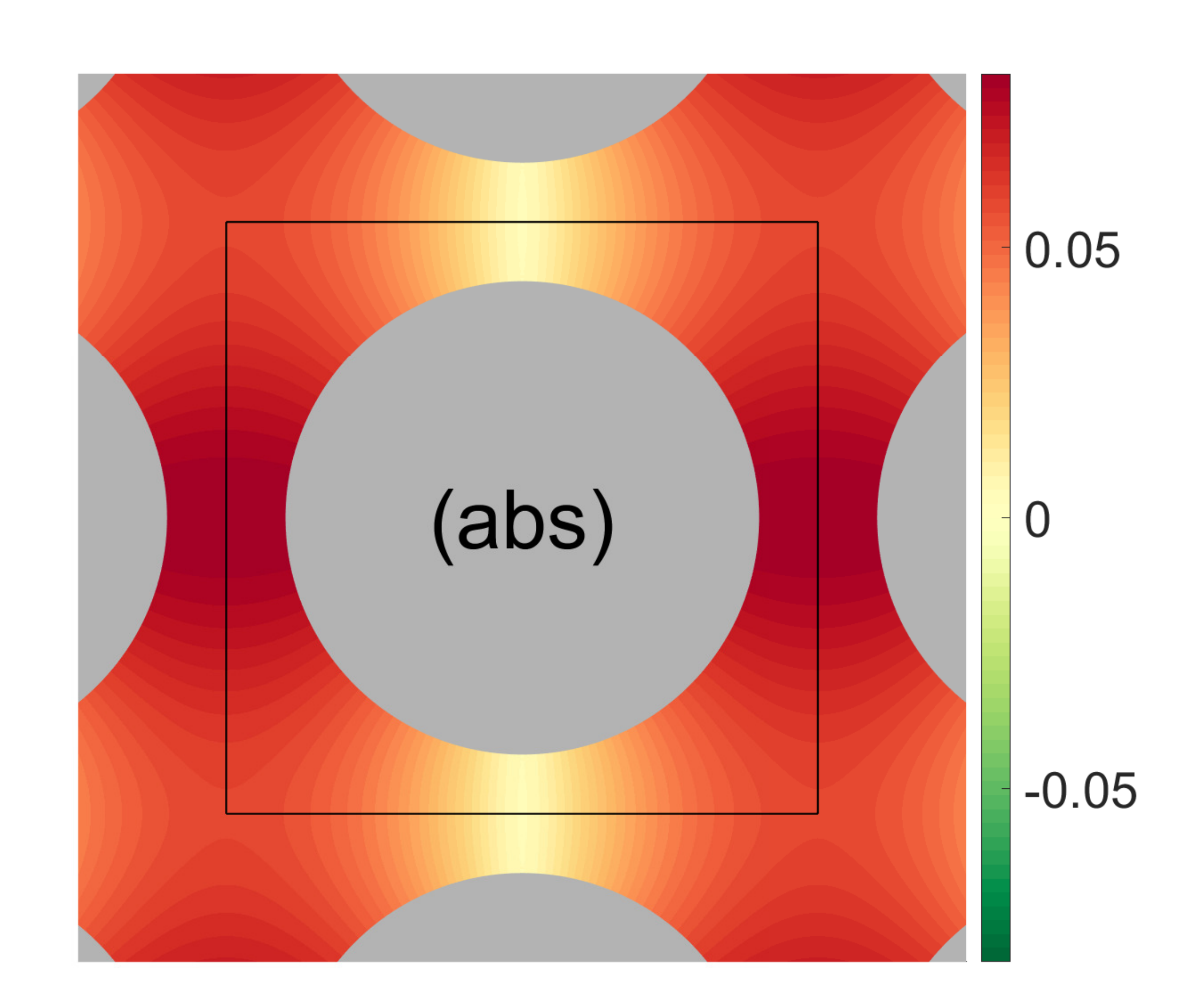}
\includegraphics[width=0.24\linewidth]{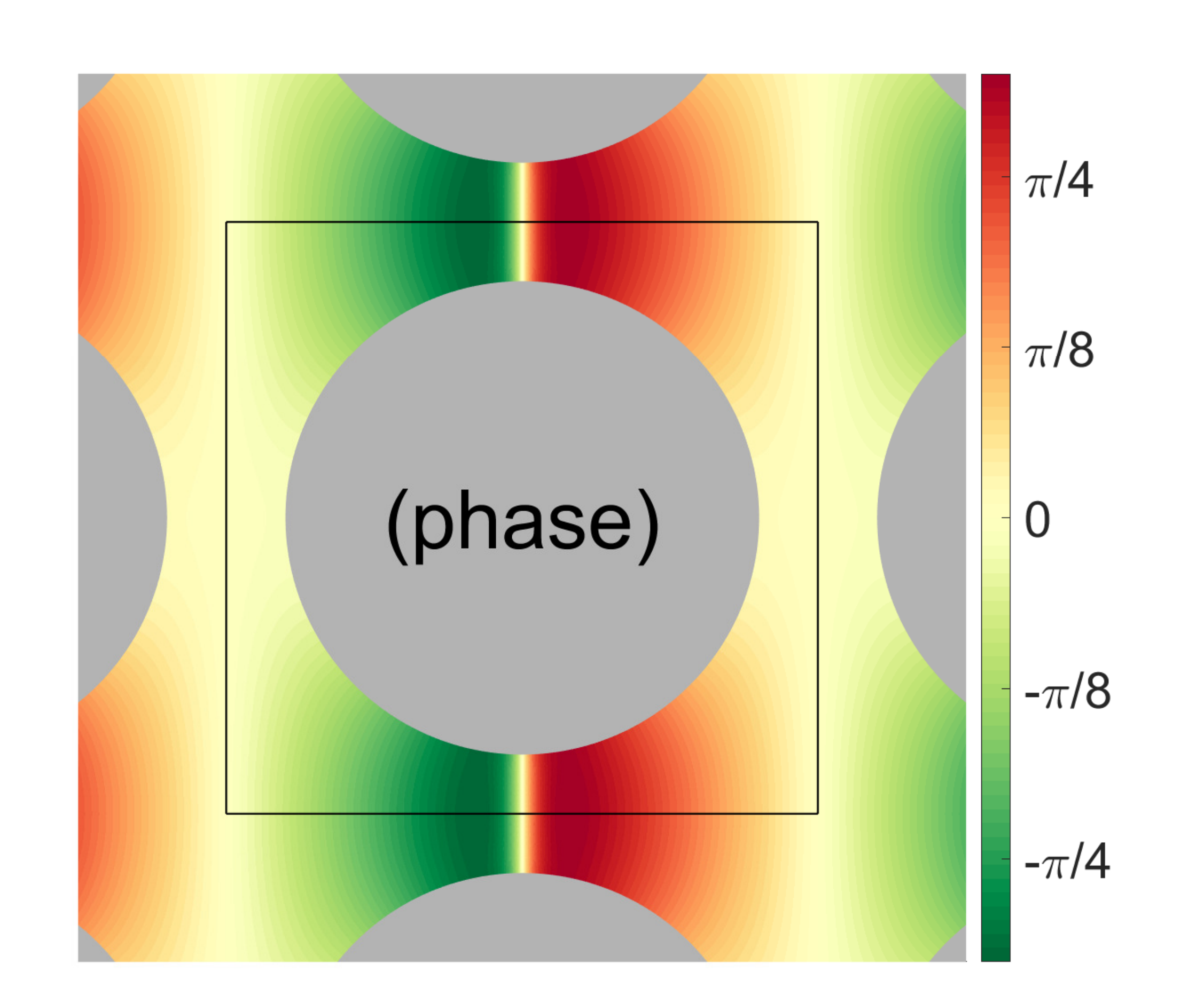}
\noindent\rule[5pt]{\linewidth}{0.4pt}
\includegraphics[width=0.24\linewidth]{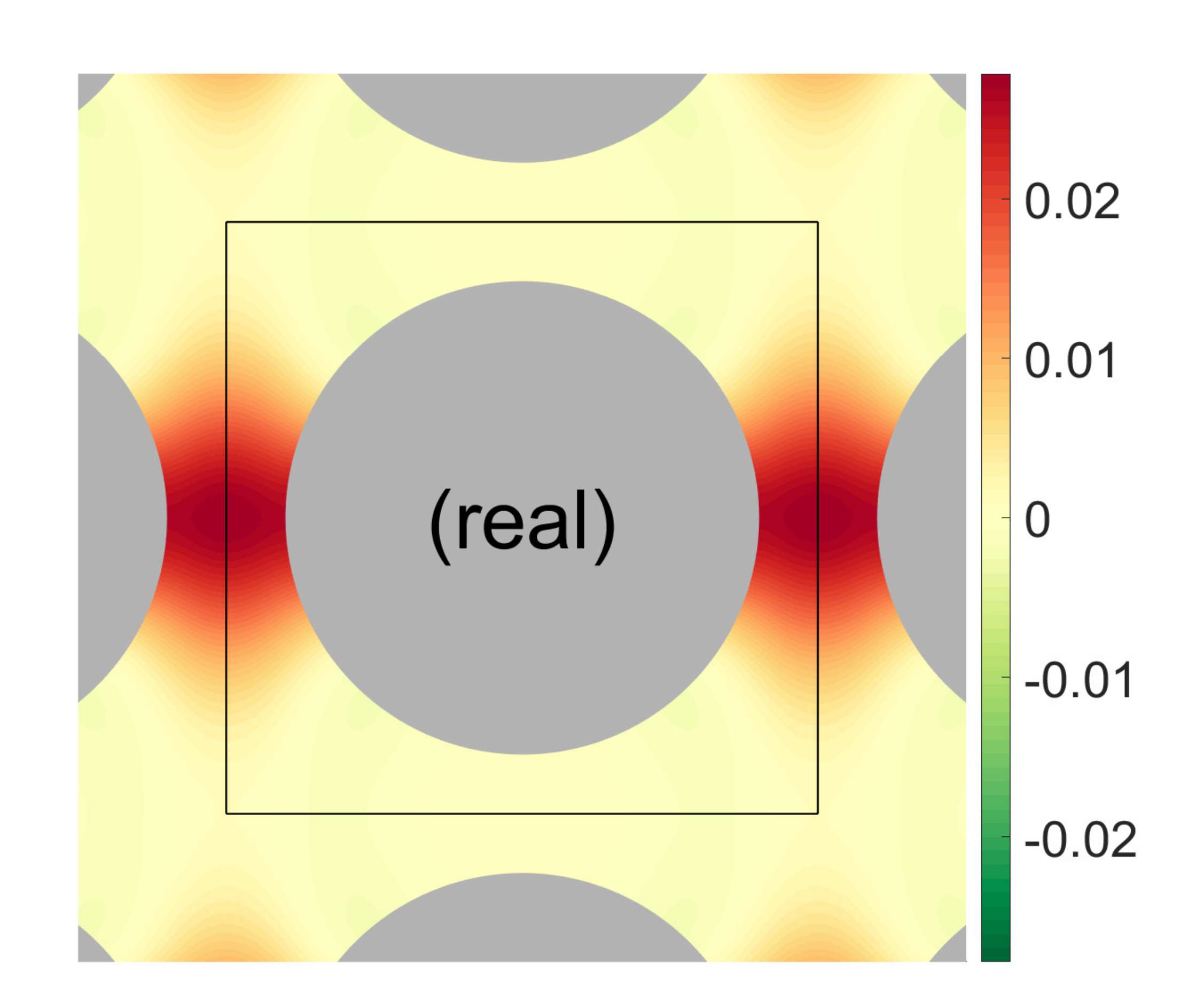}
\includegraphics[width=0.24\linewidth]{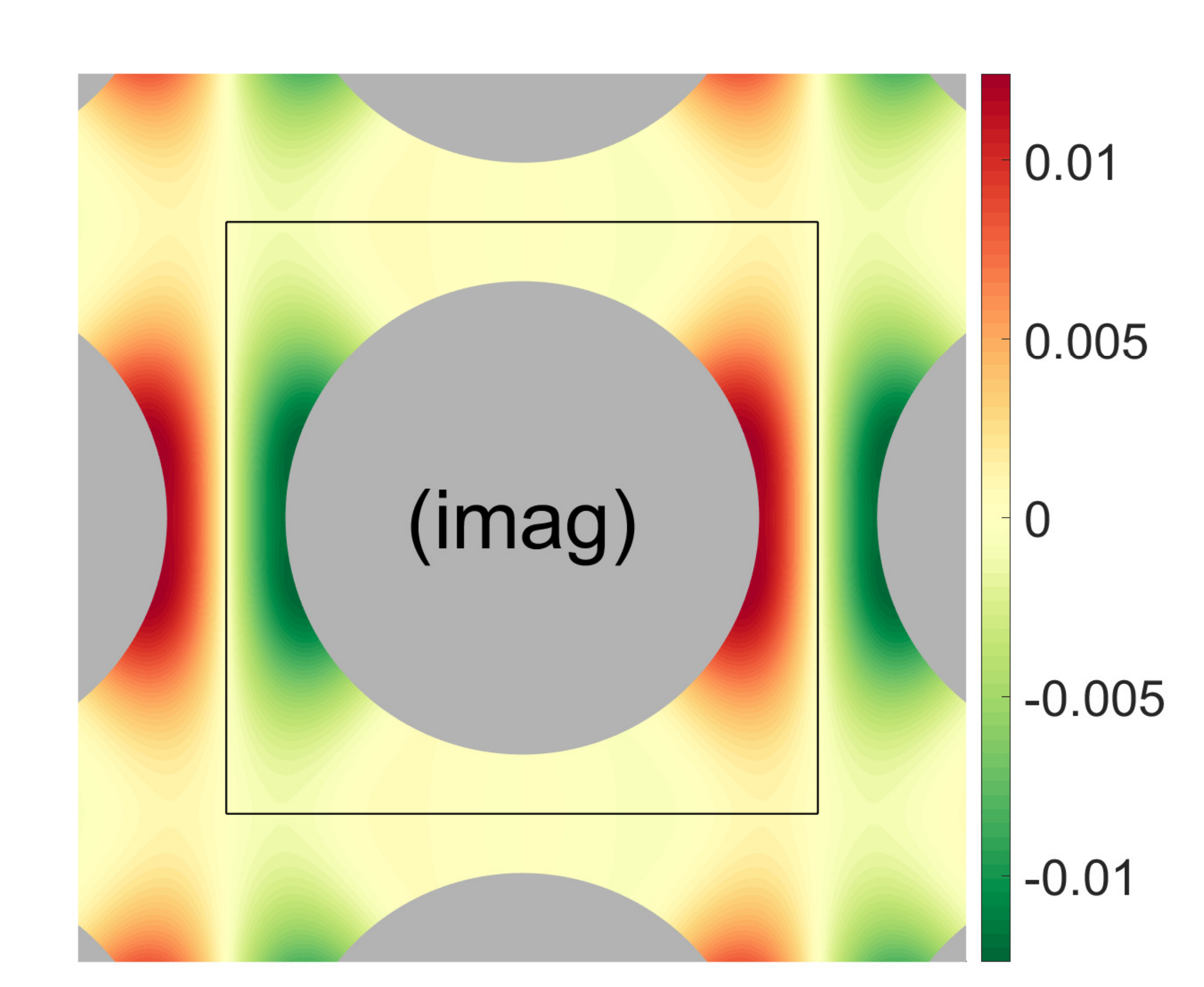}
\includegraphics[width=0.24\linewidth]{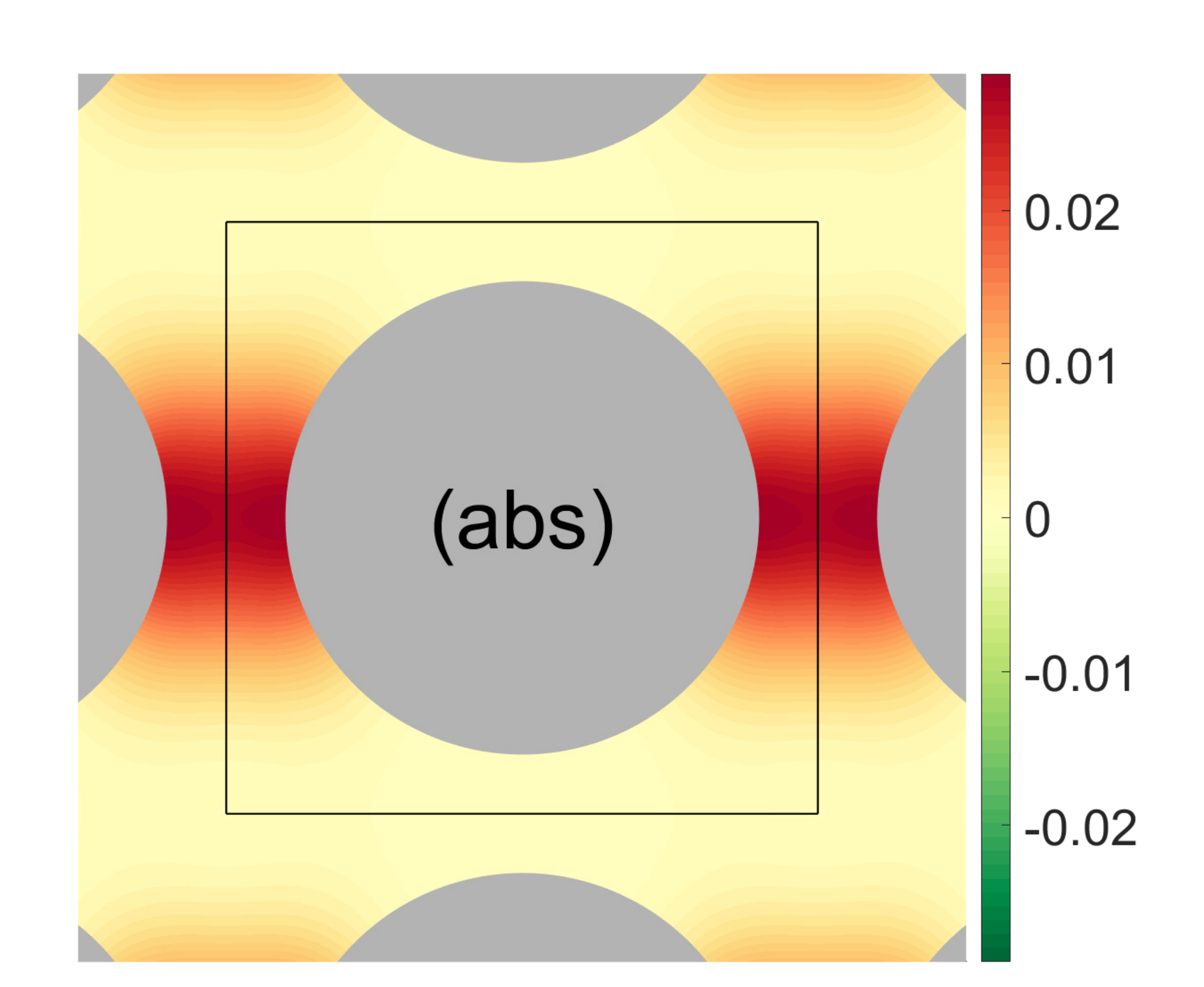}
\includegraphics[width=0.24\linewidth]{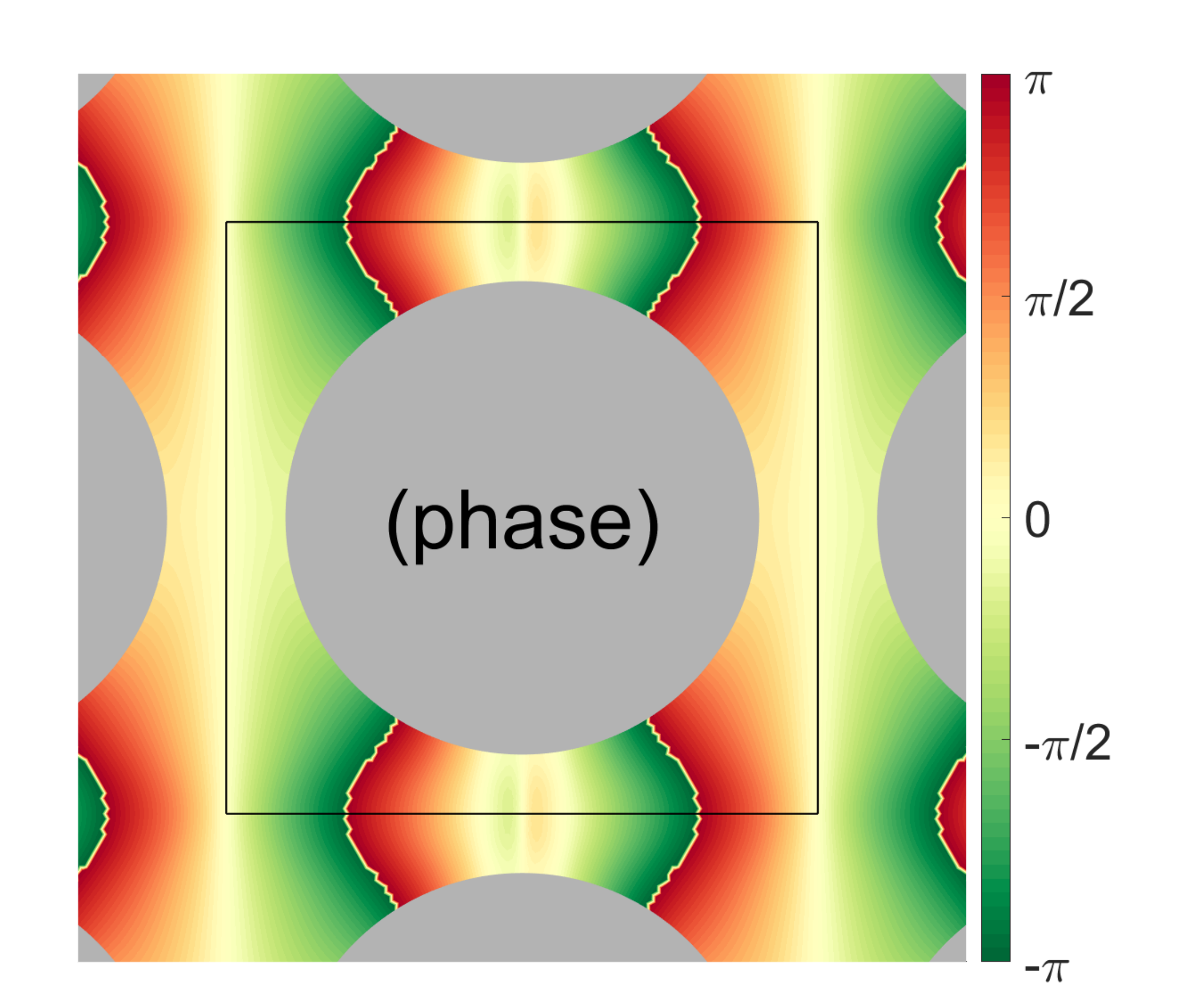}
\noindent\rule[5pt]{\linewidth}{0.4pt}
\includegraphics[width=0.24\linewidth]{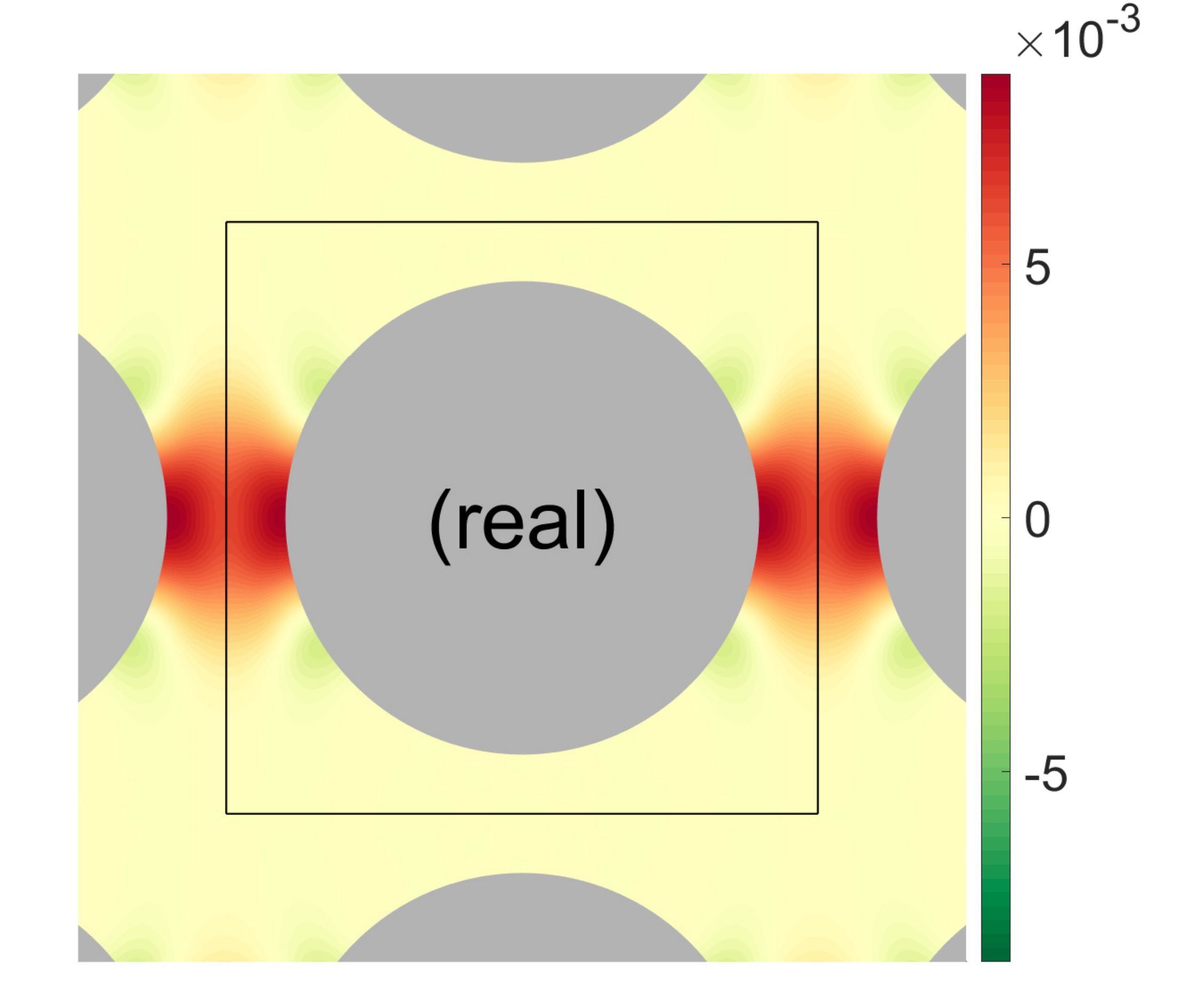}
\includegraphics[width=0.24\linewidth]{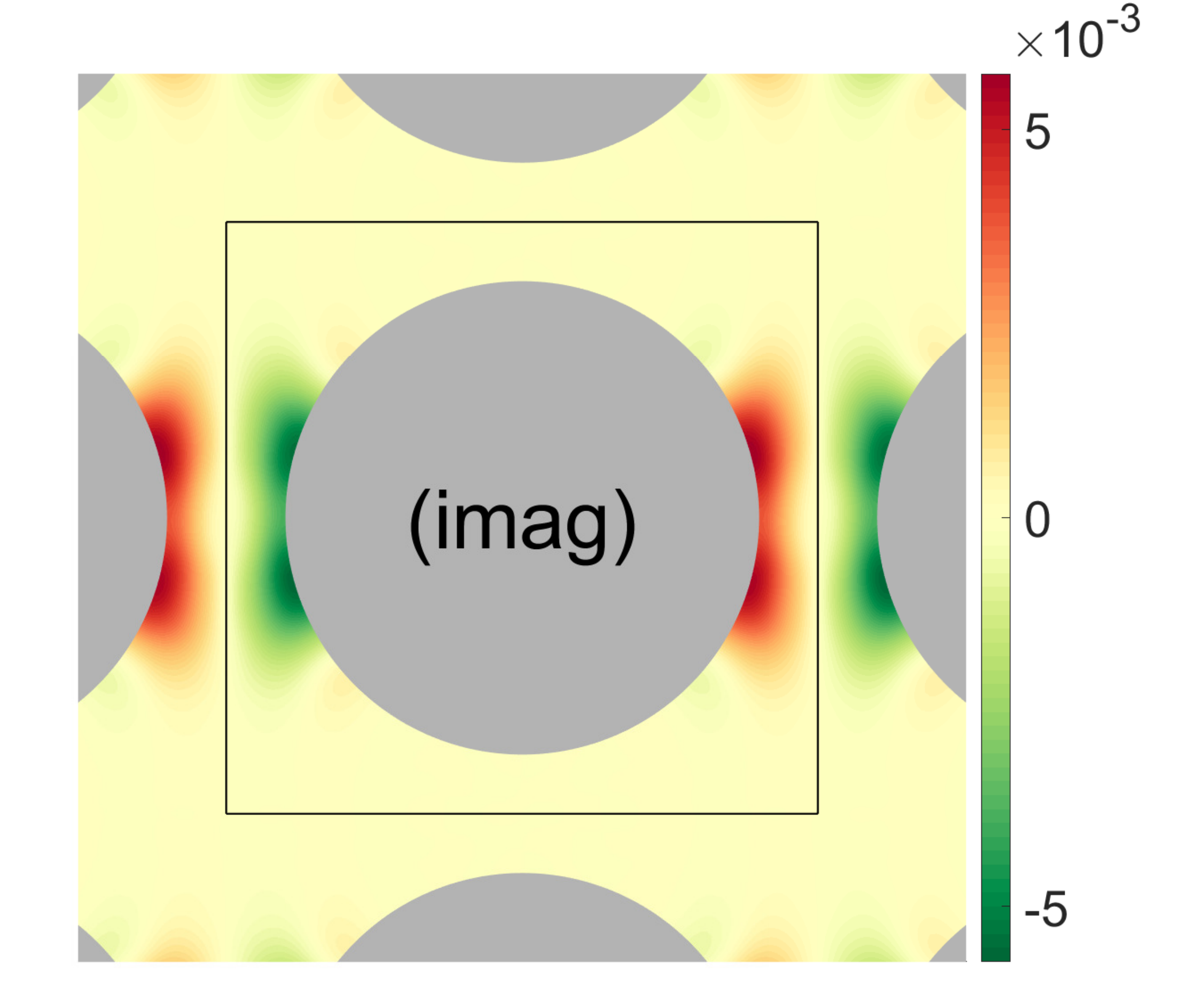}
\includegraphics[width=0.24\linewidth]{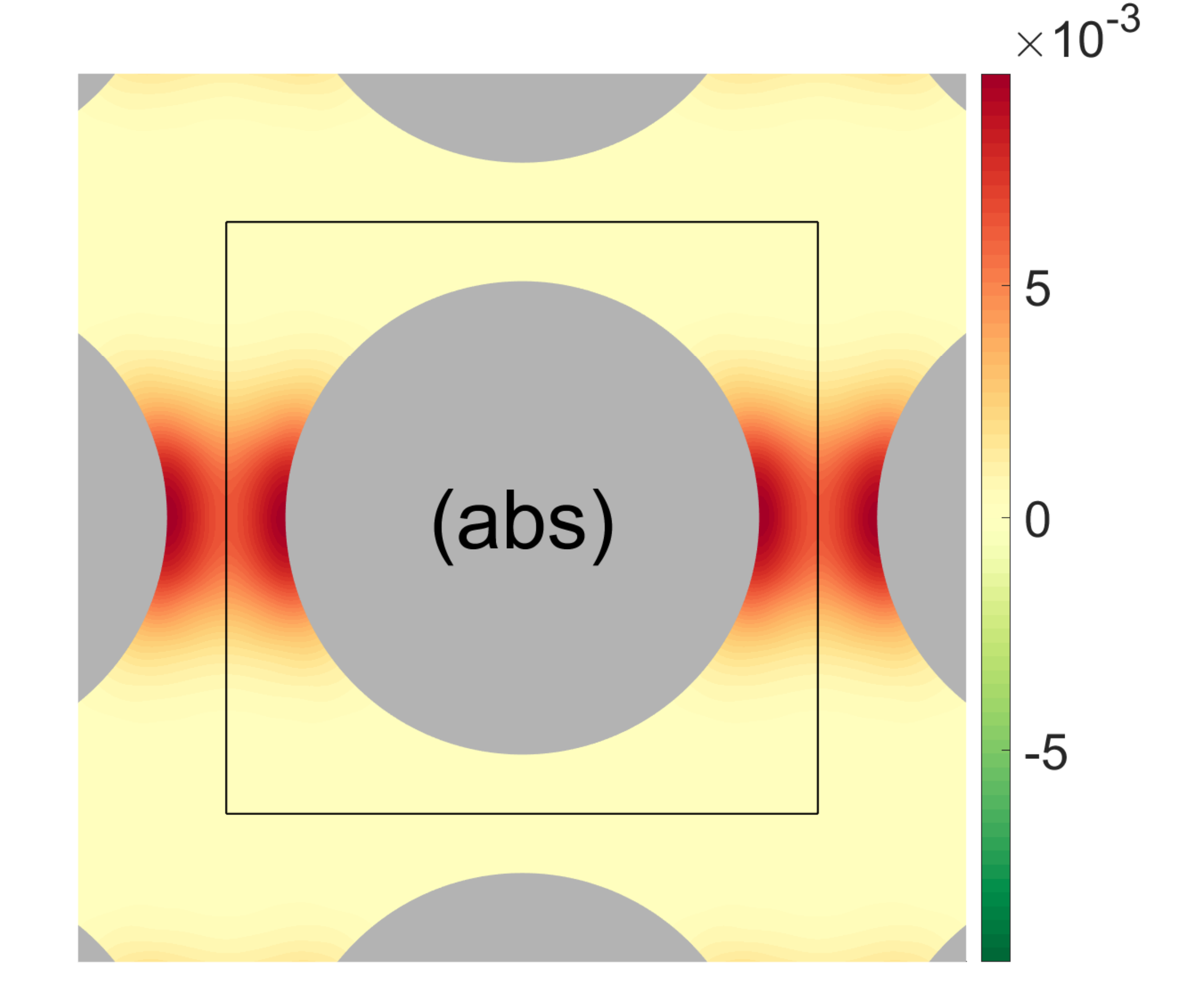}
\includegraphics[width=0.24\linewidth]{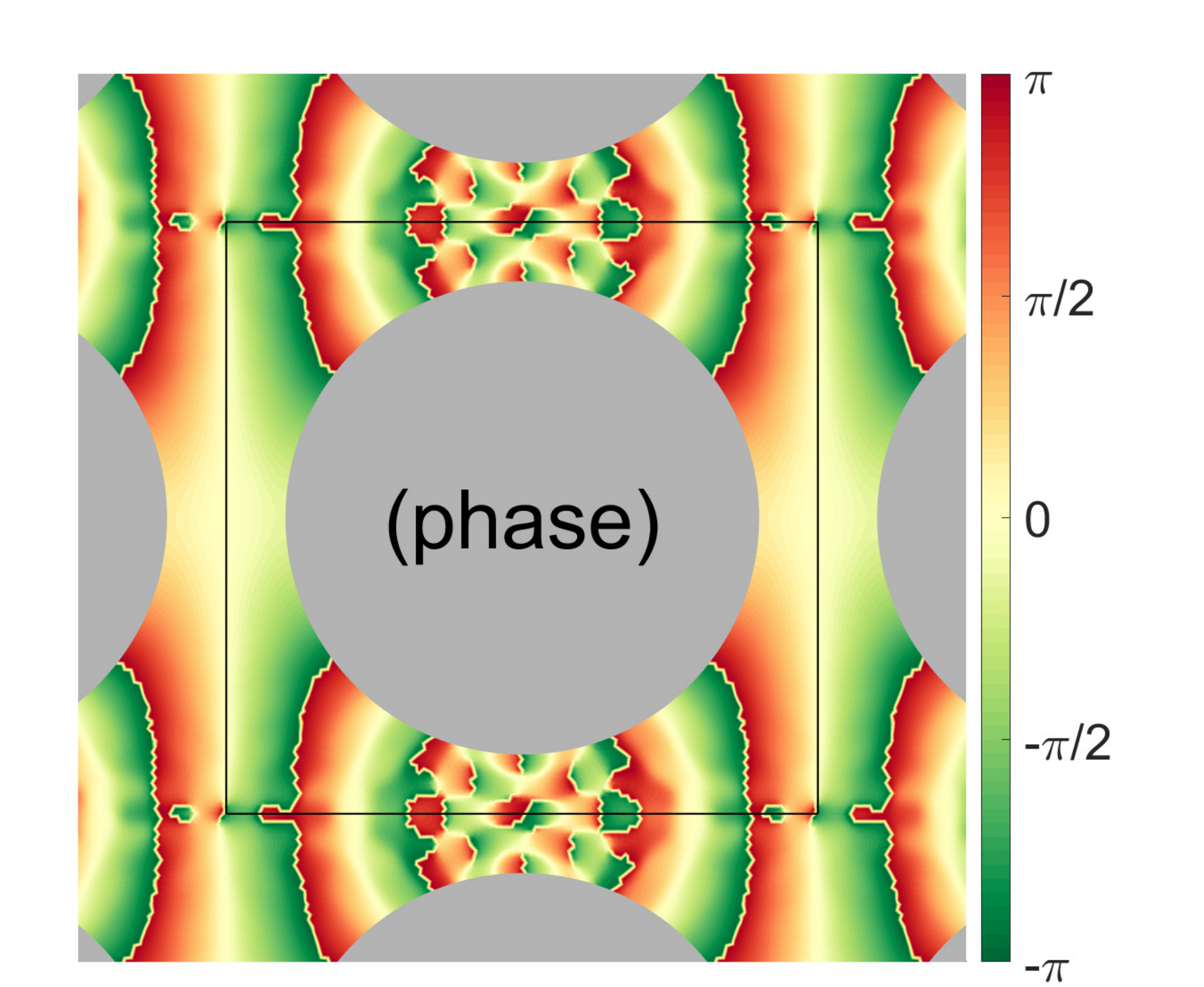}
\caption{Plot of the magnetization (real and imaginary part, absolute value and phase) after a PGSE sequence. The gradient is in the left to right horizontal direction. The black square indicates the unit cell in which the computation was performed. For all figures, $R/a=0.4$, and we kept a fixed ratio $\ell_\delta/\ell_g=2$. The corresponding normalized signal is shown on the left panel of Fig. \ref{fig:signal_b}.
(top) $\ell_g/a=0.25$; 
(middle) $\ell_g/a=0.1$; 
(bottom) $\ell_g/a=0.05$. 
\cc{Very fast phase variations result from rapidly varying imaginary part along with low real part of the magnetization (e.g., top row).}
}
\label{fig:horiz_magn}
\end{figure}

Figure \ref{fig:horiz_magn} shows the magnetization $m(T,x,y)$ after a PGSE sequence for a gradient in the left to right horizontal direction. 
This direction is expected to create the most important restriction to diffusion because of the proximity of neighboring obstacles along the gradient direction. Let us discuss first the top panel ($R/a=0.4$, $\ell_\delta/a=0.5$, $\ell_g/a=0.25$). One can see that the magnetization has been strongly attenuated in regions where there is almost no geometrical restriction by the obstacles. In contrast, one can interpret the areas with large magnetization (typically the red parts in the ``abs'' plot) as areas where the influence of the obstacles is strong. Thanks \cc{to} the large diffusion length, this red area is very broad. When one decreases both the gradient length and diffusion length (middle then bottom panel), the effect of the obstacle is less spread by diffusion and the localization of the magnetization between the neighboring obstacles becomes sharper. In the bottom panel, the magnetization is actually localized on each obstacle, with a small overlap between two neighboring localization pockets. \cc{The localized magnetization is a landmark of the localization regime which emerges at strong and extended gradient pulses.}

\begin{figure*}[t]
\centering
\includegraphics[width=0.24\linewidth]{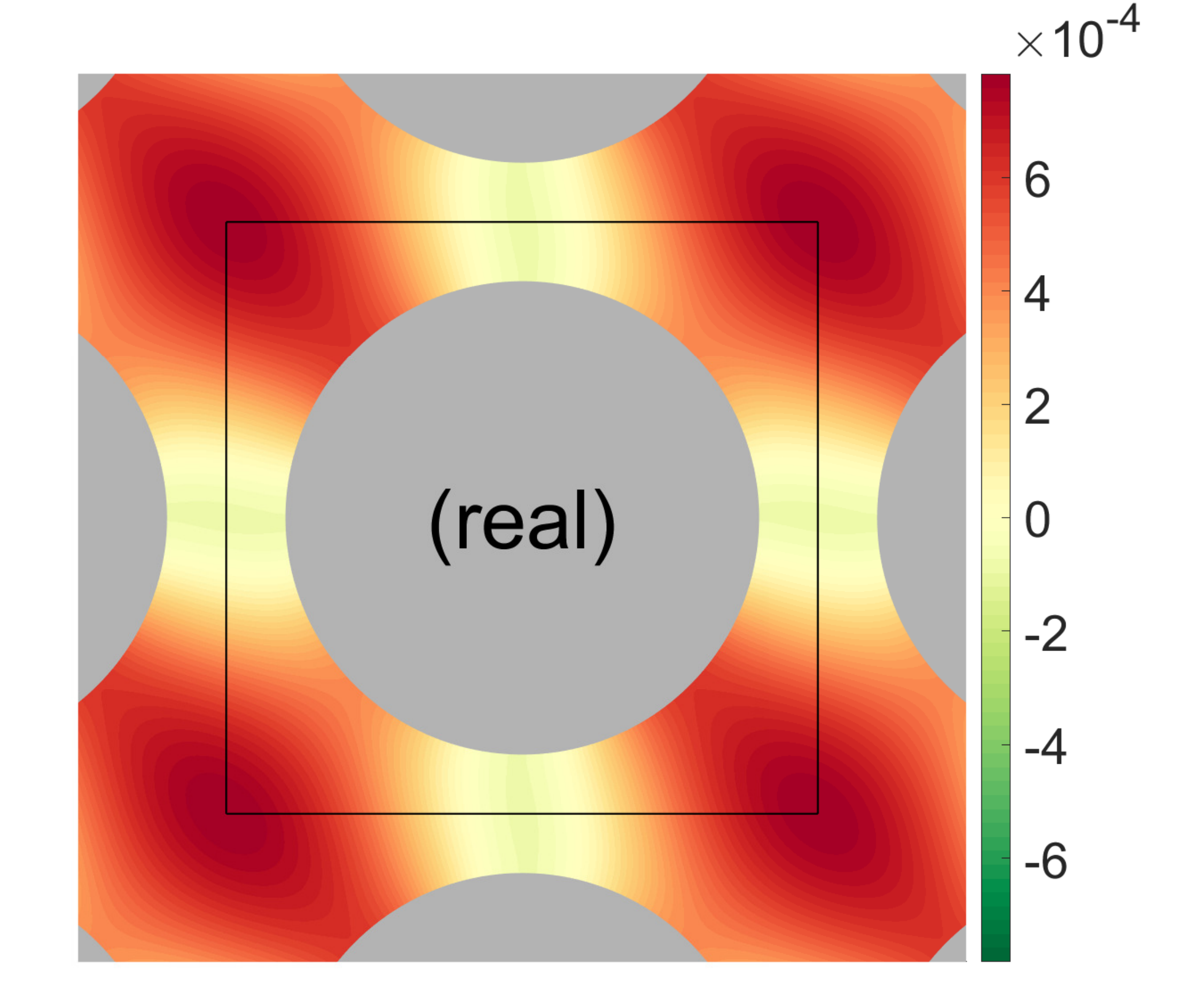}
\includegraphics[width=0.24\linewidth]{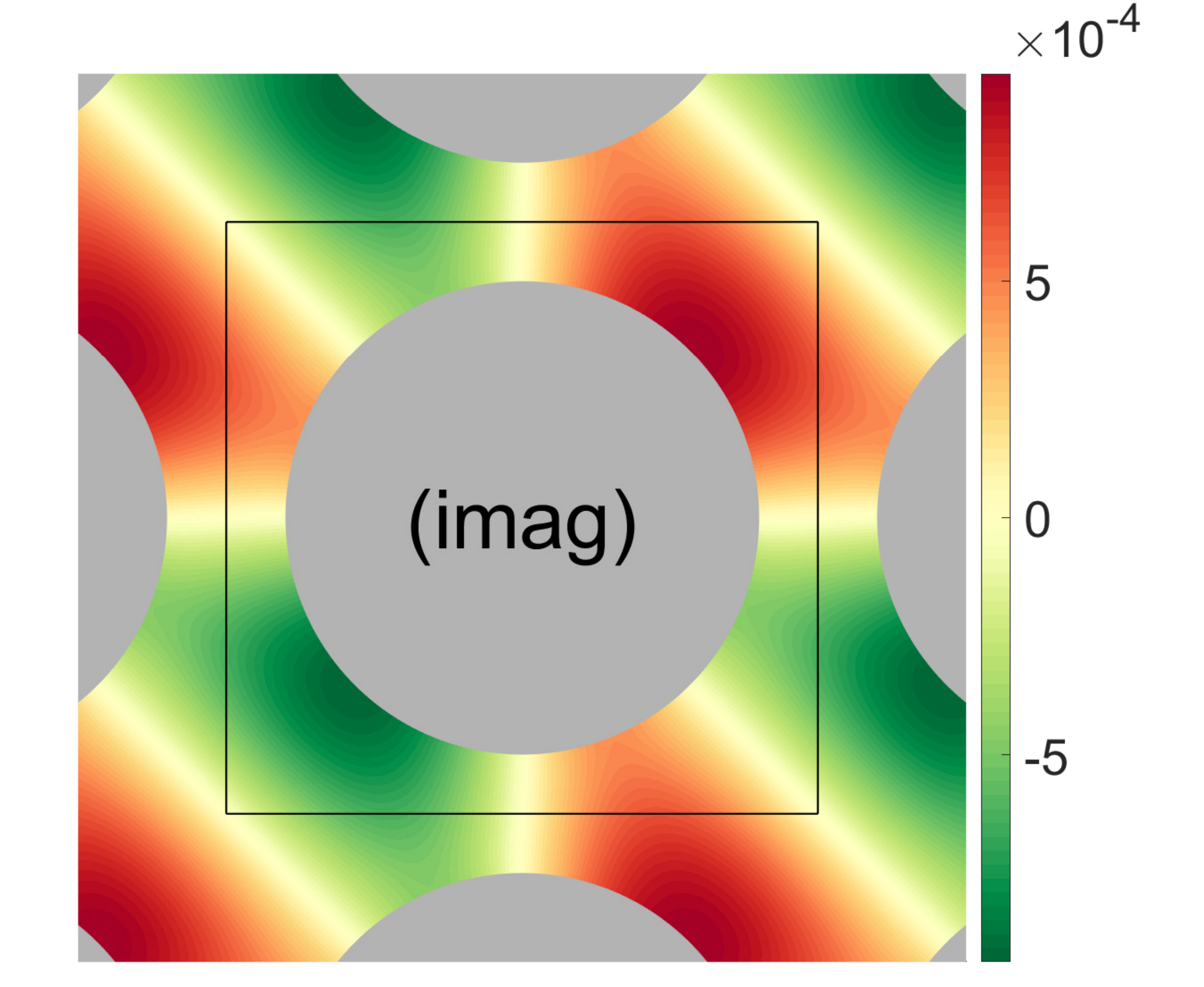}
\includegraphics[width=0.24\linewidth]{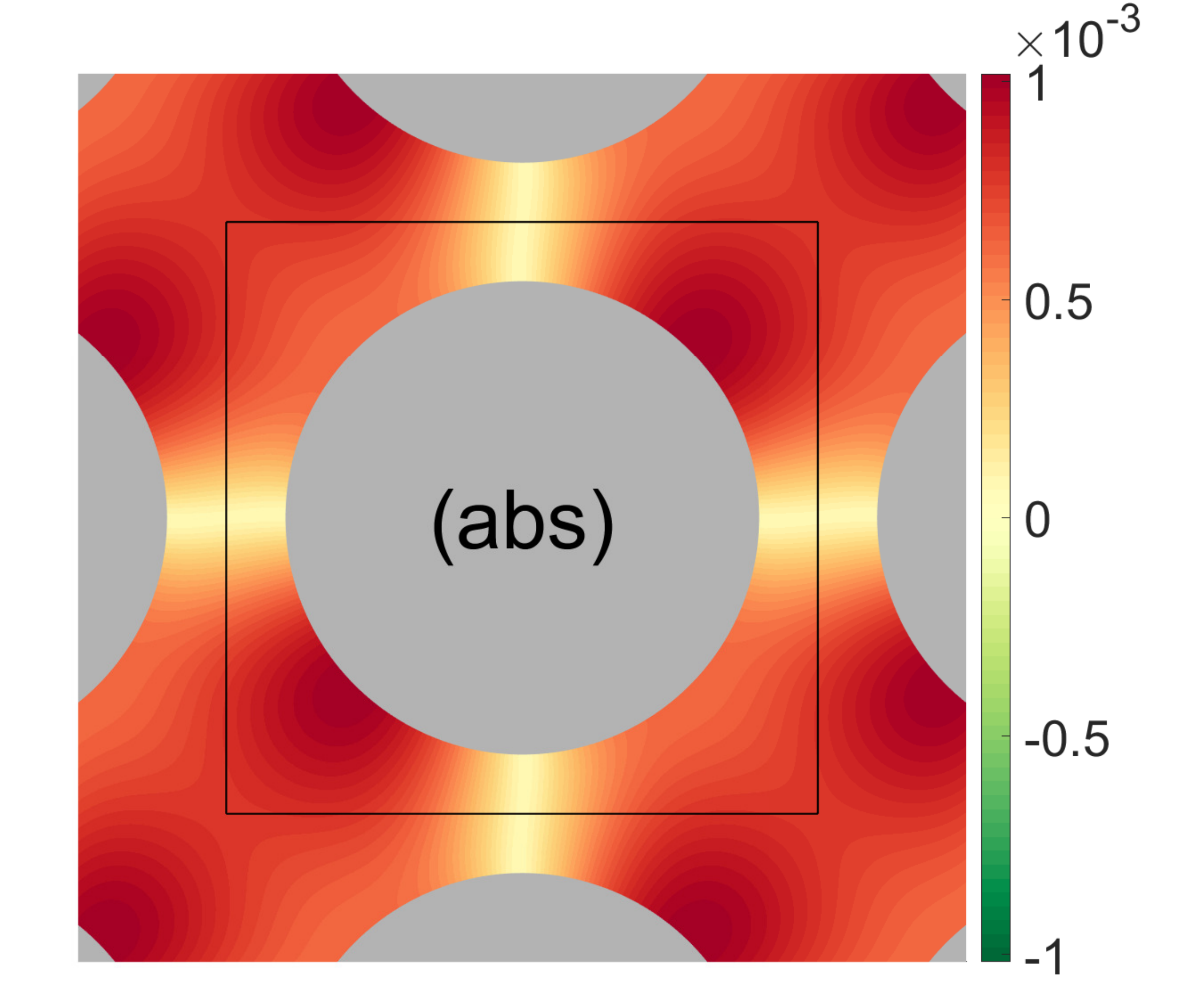}
\includegraphics[width=0.24\linewidth]{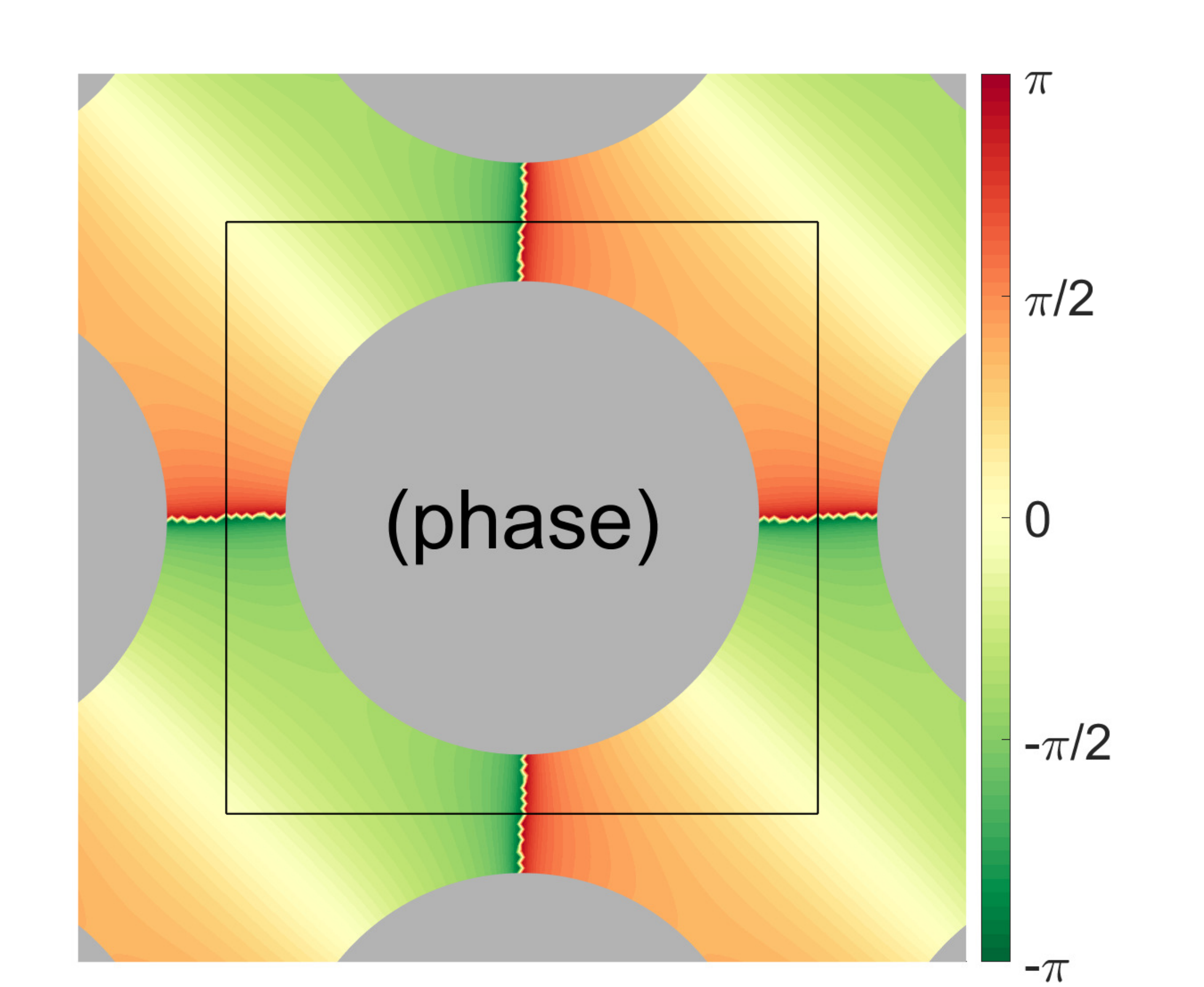}
\noindent\rule[5pt]{\linewidth}{0.4pt}
\includegraphics[width=0.24\linewidth]{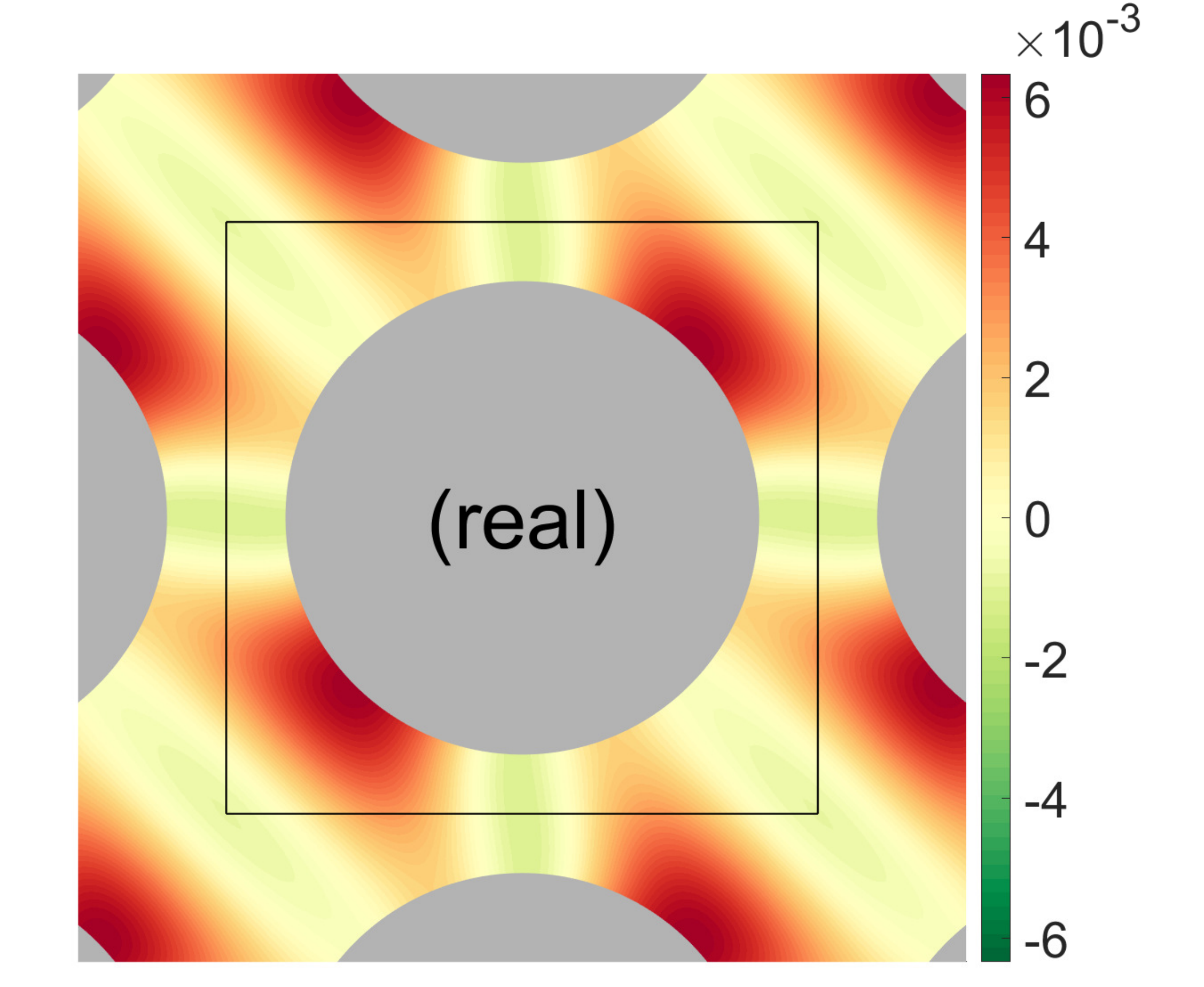}
\includegraphics[width=0.24\linewidth]{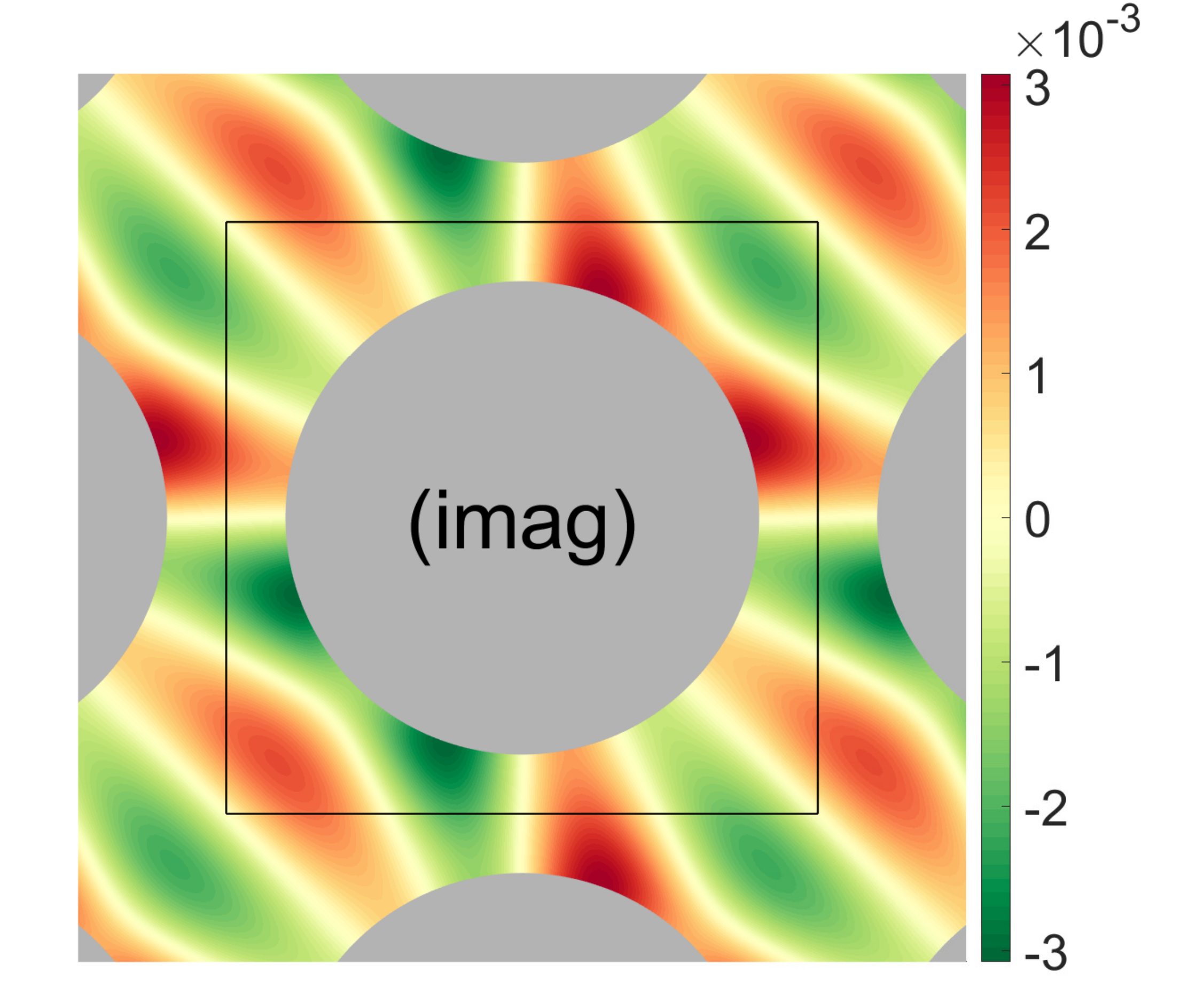}
\includegraphics[width=0.24\linewidth]{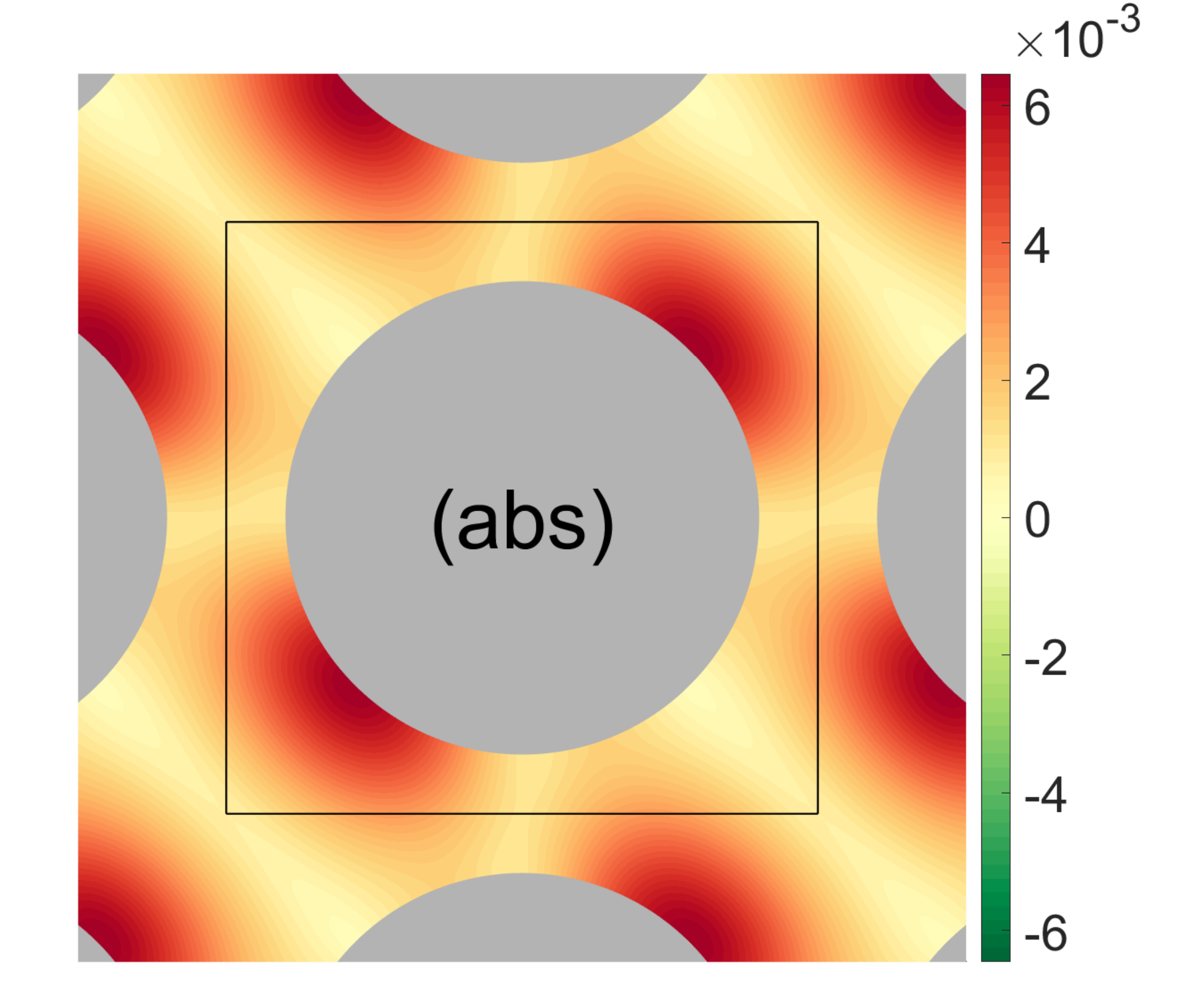}
\includegraphics[width=0.24\linewidth]{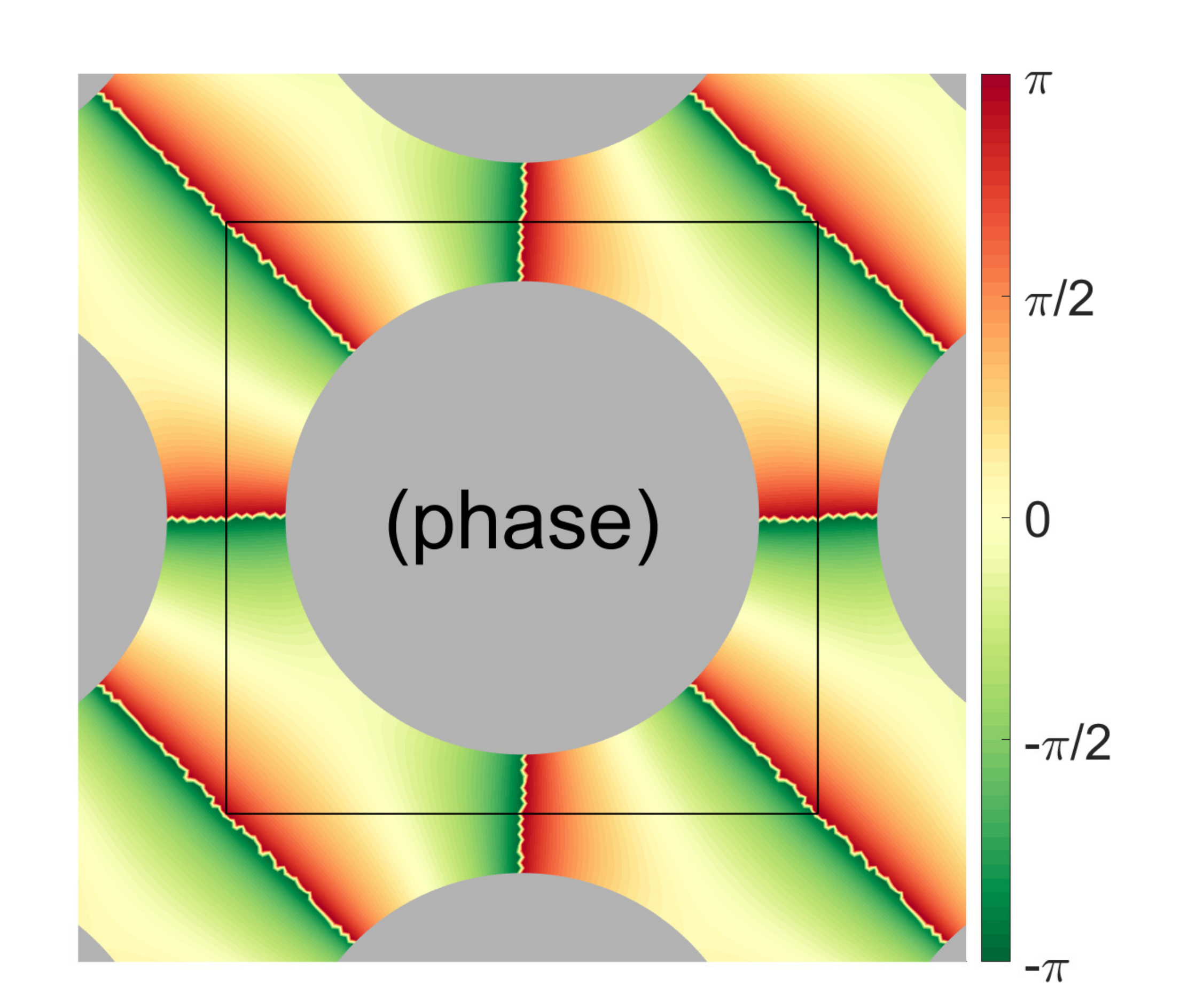}
\noindent\rule[5pt]{\linewidth}{0.4pt}
\includegraphics[width=0.24\linewidth]{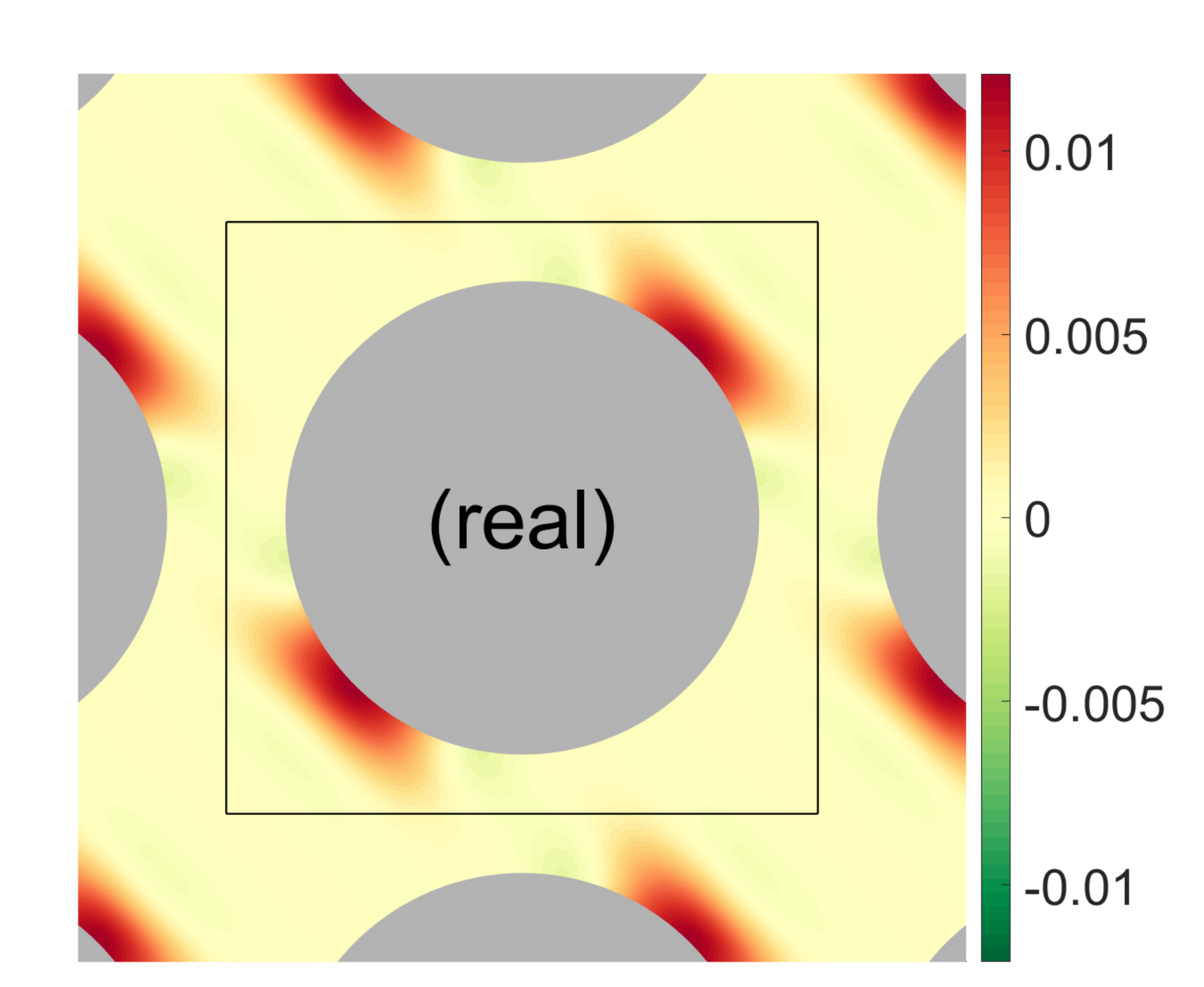}
\includegraphics[width=0.24\linewidth]{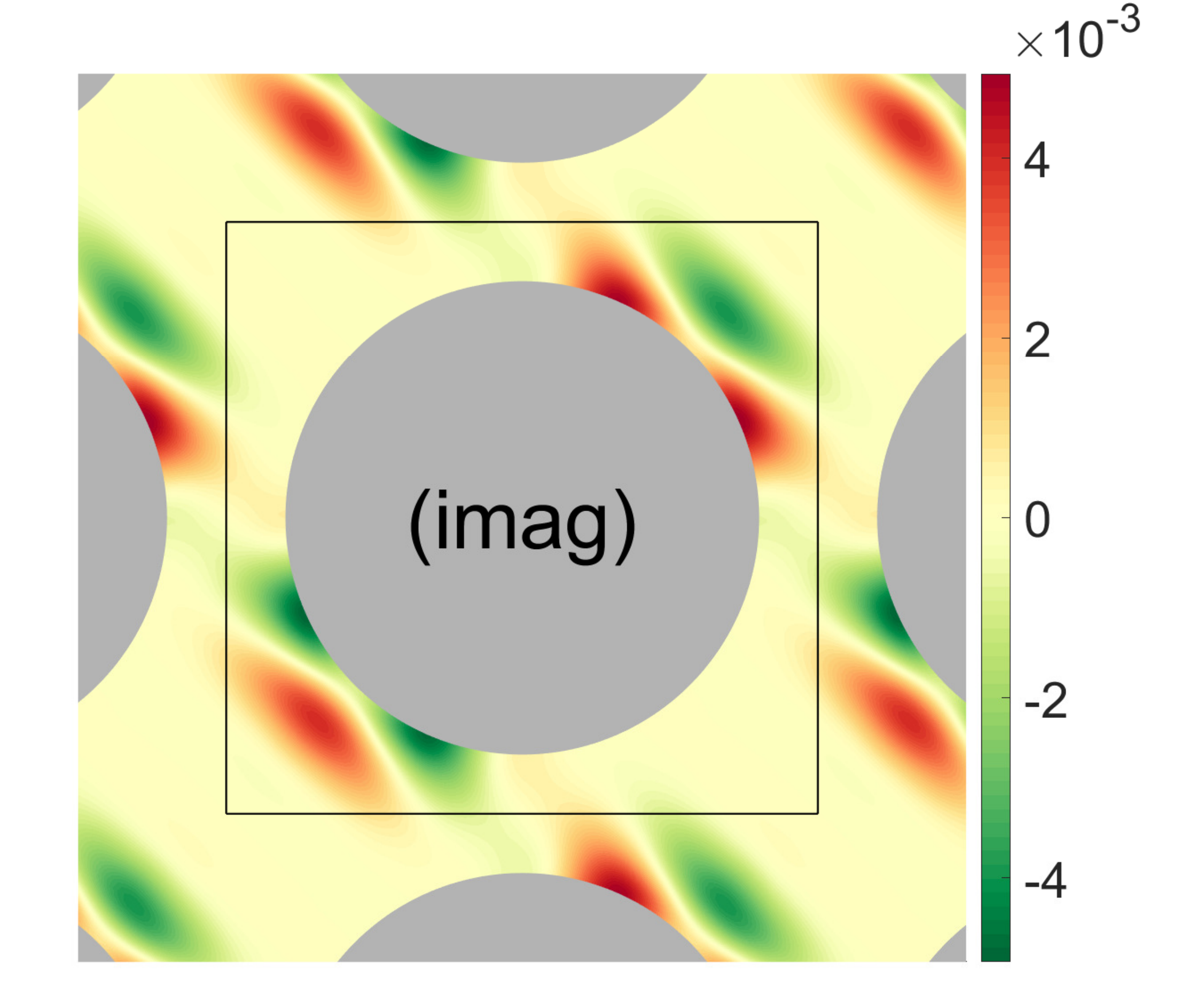}
\includegraphics[width=0.24\linewidth]{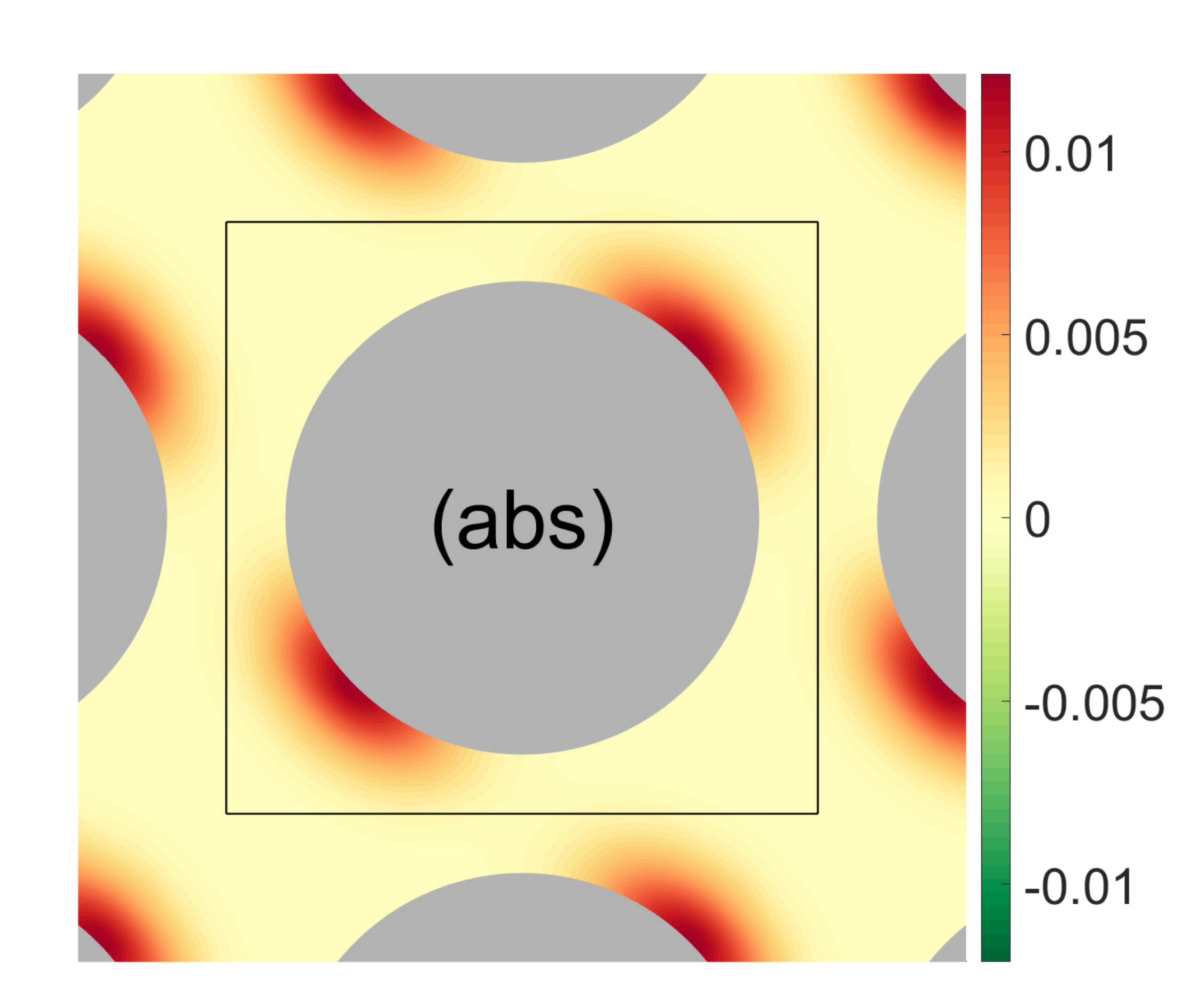}
\includegraphics[width=0.24\linewidth]{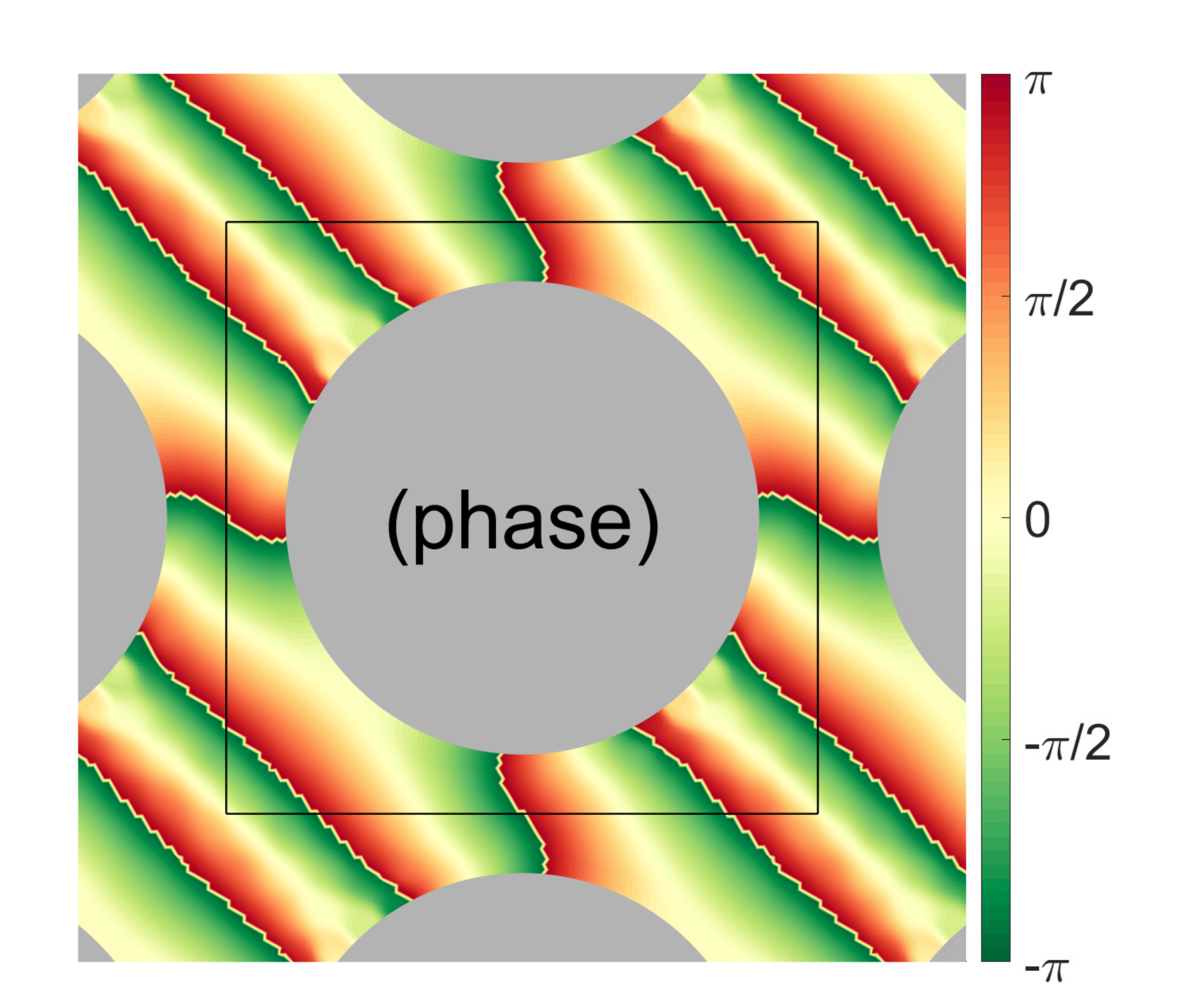}
\caption{Plot of the magnetization (real and imaginary part, absolute value and phase) after a PGSE sequence. The gradient is in the bottom-left to top-right diagonal direction. The black square indicates the unit cell in which the computation was performed. For all figures, $R/a=0.4$, and we kept a fixed ratio $\ell_\delta/\ell_g=2$. The corresponding normalized signal is shown on the right panel of Fig. \ref{fig:signal_b}.
(top) $\ell_g/a=0.25$; 
(middle) $\ell_g/a=0.1$; 
(bottom) $\ell_g/a=0.05$. 
}
\label{fig:diag_magn}
\end{figure*}

Figure \ref{fig:diag_magn} shows the magnetization for the same set of parameters but with the gradient in the bottom-left to top-right diagonal direction. In that case, the geometrical restriction by the disks is much weaker. This is especially visible on the top panel, where the magnetization is two orders of magnitude lower than that in the horizontal gradient case presented above. One still observes the same pattern as above on the top panel, with almost zero magnetization where there is no geometrical restriction along the gradient direction, and the largest magnetization inbetween two neighboring obstacles. As the gradient length and diffusion length decrease, the magnetization localizes more sharply near the obstacles. On the middle panel, one can already see magnetization pockets on each obstacle, with almost no overlap between neighboring obstacles. \cc{Thus, the localization regime emerges at larger gradient length $\ell_g$ (i.e., lower gradient strength) in this setting than for the horizontal gradient.}

\cc{In both cases of horizontal and diagonal gradient direction, with high gradient strength,} the localization along the gradient direction is much sharper than in the orthogonal direction (parallel to the \cc{boundary of the obstacles}). A computation of the magnetization around a curved boundary shows that the magnetization localizes on the scale $\ell_g$ along the gradient direction and on the scale
\begin{equation}
\ell_{g,\parallel}=(2\ell_g^{3}R)^{1/4}
\end{equation}
parallel to the \cc{boundary of the obstacles} \cite{Moutal2019b}. In particular, the bottom panels of Figs. \ref{fig:horiz_magn} and \ref{fig:diag_magn} correspond to a ratio $\ell_{g,\parallel}/\ell_g=2$, which is visually consistent with the figures.

%

\begin{figure}[h]
\centering
\includegraphics[width=0.49\linewidth]{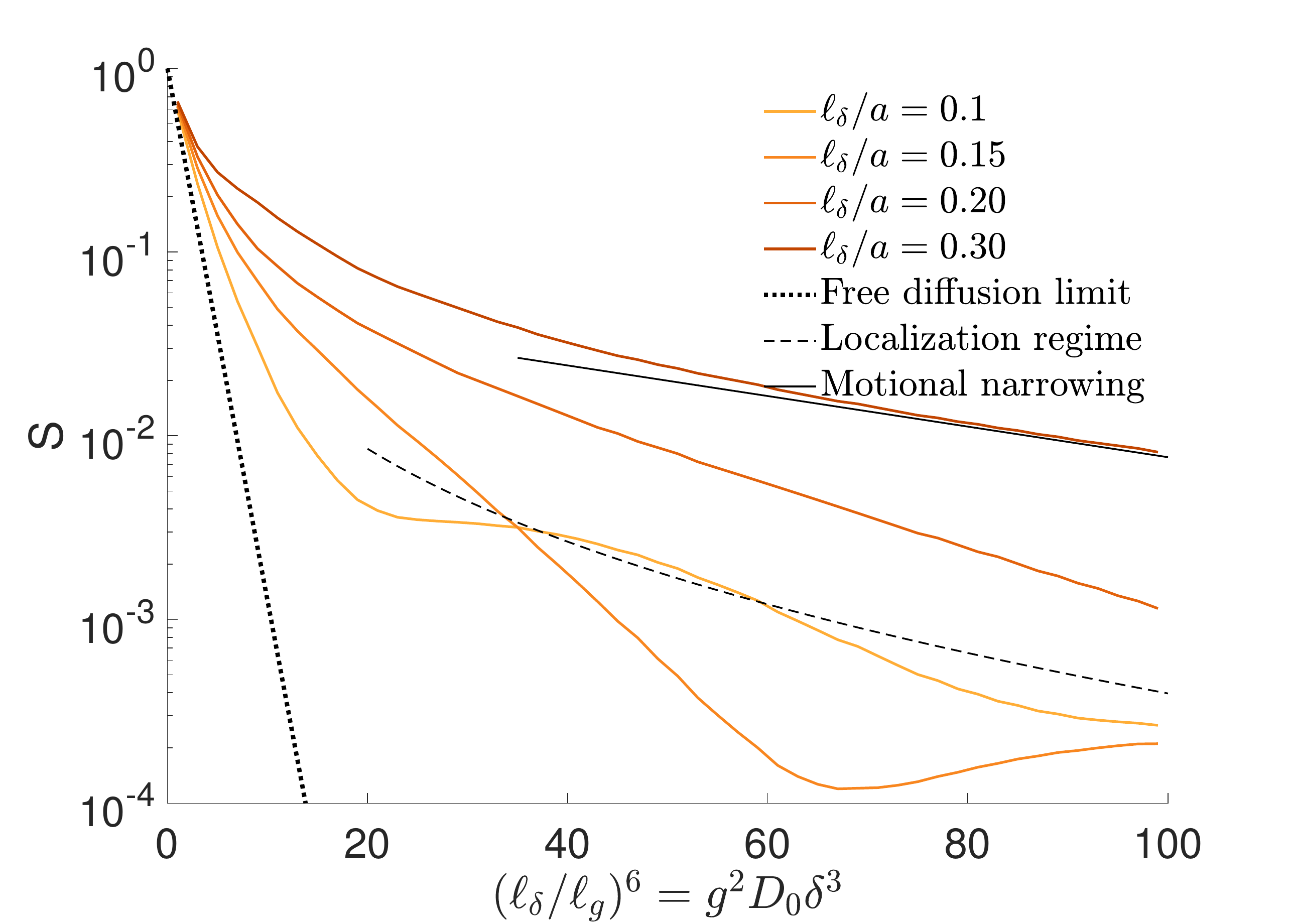}
\includegraphics[width=0.49\linewidth]{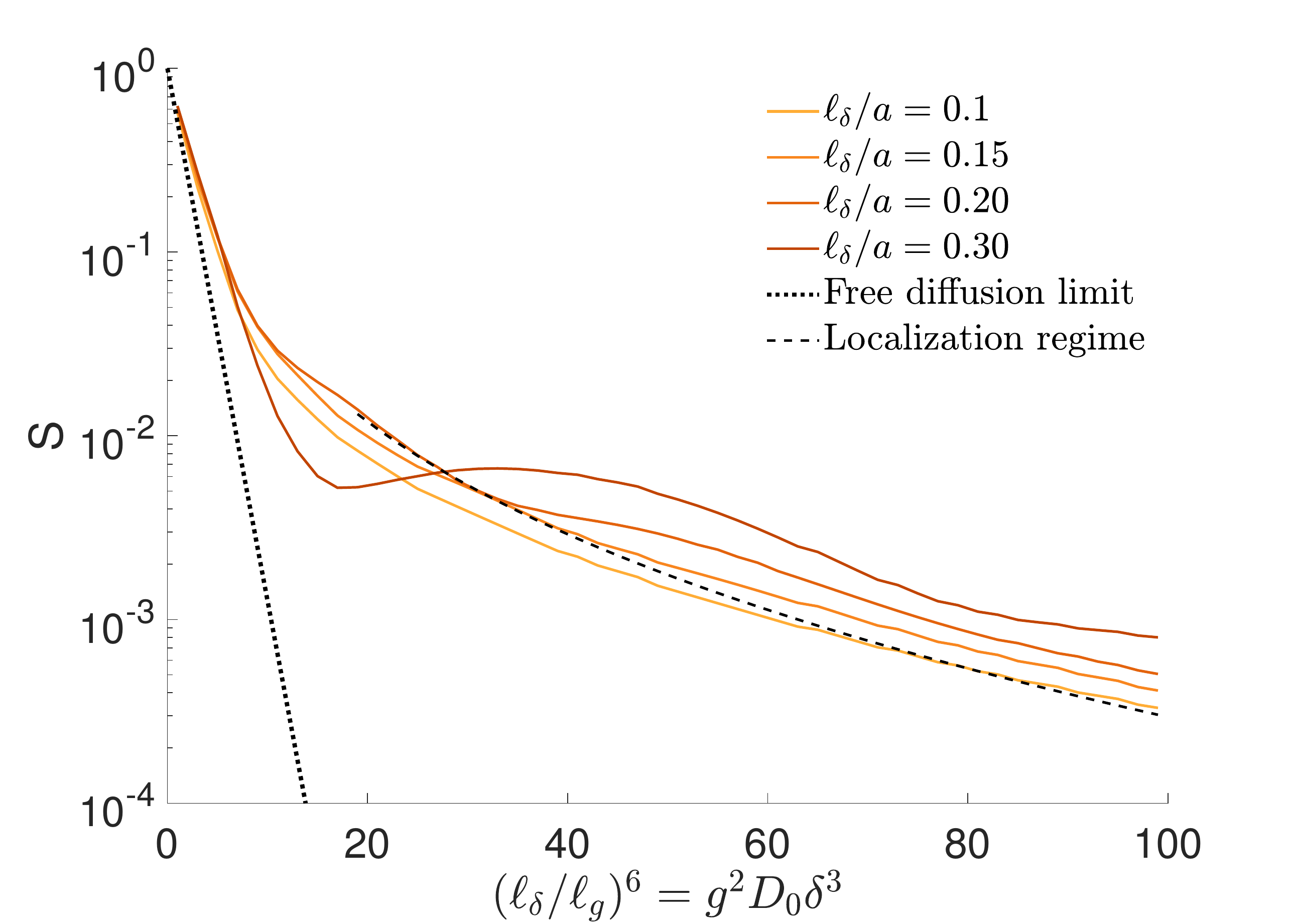}
\caption{Signal as a function of $(\ell_\delta/\ell_g)^6\propto bD_0$ for different values of $\ell_\delta$ as well as asymptotic regimes \eqref{eq:free_decay_bis} (in the limit $\ell_\delta/a\to0$), \eqref{eq:motional_narrowing} (for $\ell_\delta/a=0.3$) and \eqref{eq:asymptot_localization} (for $\ell_\delta/a=0.1$). (left) The gradient is in the horizontal direction. (right) The gradient is in the diagonal direction.
Refer to the text for discussion of the figure.}
\label{fig:signal_b}
\vskip 3mm
\end{figure}

The (normalized) signal is presented in Fig. \ref{fig:signal_b} as a function of $(\ell_\delta/\ell_g)^6$ for different fixed values of $\ell_\delta/a$ \cc{with horizontal and diagonal gradient directions}. At low values of $\ell_\delta/\ell_g$ the gradient encoding is weak so that the signal is well represented by an expression similar to the free diffusion decay \eqref{eq:free_decay}:
\begin{eqnarray}
S&\approx\exp(-\alpha(\ell_\delta/a) bD_0)\nonumber\\
&\approx\exp\left(- \frac{2}{3}\,\alpha(\ell_\delta/a) \left(\frac{\ell_\delta}{\ell_g}\right)^6\right)\;,
\label{eq:free_decay_bis}
\end{eqnarray}
where $0<\alpha(\ell_\delta/a)<1$ is the ratio of effective diffusion coefficient to intrinsic diffusion coefficient $D_0$ that accounts for the restriction by obstacles in the domain. At infinitely short diffusion time, i.e. $\ell_\delta/a\to 0$, the effect of the obstacles becomes negligible so that $\alpha(\ell_\delta/a) \to 1$. This limit is plotted as a dotted line on Fig. \ref{fig:signal_b}. The short-time behavior of $\alpha(\ell_\delta/a)$ was shown to be linear in $\ell_\delta \sigma$, where $\sigma$ is the surface-to-volume ratio of the domain \cite{Mitra1992a,Mitra1993a,Moutal2019c}. For unbounded domains such as the one considered here, $\alpha(\ell_\delta/a)$ has a positive limit at infinitely long times that can be interpreted as the tortuosity of the domain \cite{Haus1987a,Latour1995a}. Furthermore, the long-time asymptotic behavior of $\alpha(\ell_\delta/a)$ is related to the structural disorder of the medium \cite{Novikov2014a}. In that regard, periodic media present a special case of perfectly ordered media, however this is of little importance as long as the diffusion length is at most of the order of the lattice step (i.e. $\ell_\delta \lesssim a$). As we argued in the introduction and in Sec. \ref{section:fine_sampling}, this is a natural assumption in the context of this paper as otherwise the effect of microstructure is averaged out by the diffusion, as it is discussed in \cite{Novikov2014a}.

At large values of $\ell_\delta/\ell_g$, the decay of the signal is much slower than the free diffusion decay \eqref{eq:free_decay}. 
As we plot the signal in terms of $(\ell_\delta/\ell_g)^6=D_0 g^2 \delta^3$ for different fixed values of $\ell_\delta$, smaller values of $\delta$ (i.e., smaller $\ell_\delta$) correspond to a larger range of values of $g$ (i.e., smaller values of $\ell_g$ are attained). Therefore, in this representation, a sharp localization phenomenon is obtained at large values of $\ell_\delta/\ell_g$ and small $\ell_\delta$. Bearing that in mind, we observe two distinct behaviors depending on the gradient direction.

(i) For the gradient in the horizontal direction (left panel of Fig. \ref{fig:signal_b}), the decay of the signal as a function of $\ell_\delta/\ell_g$ changes significantly when $\ell_\delta$ decreases, and the signal displays oscillations at the lowest considered value of $\ell_\delta$. This behavior can be related to the previous observations about Fig. \ref{fig:horiz_magn}, that corresponds to $(\ell_\delta/\ell_g)^6=64$. At $\ell_\delta/a \geq 0.2$, the gradient length is too large compared to the inter-obstacle spacing so that the magnetization is not localized on each obstacle's boundary but rather inside the small slab-like space between two neighboring obstacles.
As the diffusion length $\ell_\delta$ is larger than the inter-obstacle spacing, one can interpret this regime as a motional narrowing regime in an effective slab of width $L$:
\begin{eqnarray}
S&\approx C_{\rm mn}\exp\left(-\frac{1}{60} \frac{\ell_\delta^2 L^4}{\ell_g^6}\right)\nonumber\\
&\approx C_{\rm mn}\exp\left(-\frac{1}{60} \left(\frac{L}{\ell_\delta}\right)^4\left(\frac{\ell_\delta}{\ell_g}\right)^6\right)\;,
\label{eq:motional_narrowing}
\end{eqnarray}
where the above formula is valid in the regime $L/\ell_\delta\lesssim 1$ \cite{Wayne1966a,Robertson1966a,Neuman1974a} and $C_{\rm mn}$ represents here the fraction of spins inside the small inter-obstacle space. A rough fitting of the signal at the longest diffusion length, i.e. $\ell_\delta/a=0.3$, yields $L/a \approx 0.3$, that is larger than the inter-obstacle spacing $1-2R/a=0.2$ as expected from the curvature of obstacles. This asymptotic regime is plotted as solid line on the left panel of Fig. \ref{fig:signal_b} for $\ell_\delta/a=0.3$.
In contrast, at smaller gradient length the localization regime emerges and the signal from localized magnetization pockets decays as 
\begin{eqnarray}
S&\approx C_{\rm loc}\exp\left(-|a'_1|\frac{\ell_\delta^2}{\ell_g^2} - \frac{\ell_\delta^2}{R^{1/2}\ell_g^{3/2}} - \frac{\sqrt{3}\ell_\delta^2}{2|a'_1|R\ell_g}\right)
 \nonumber\\
&\approx C_{\rm loc}\exp\left(-\frac{\ell_\delta^2}{\ell_g^2}\left(|a'_1|+\frac{\ell_g^{1/2}}{R^{1/2}}+\frac{\sqrt{3}}{2|a'_1|}\frac{\ell_g}{R} \right)\right)
\label{eq:asymptot_localization}
\end{eqnarray}
$a'_1\approx -1.02$ is the first zero of the derivative of the Airy function \cite{Almog2018a,Moutal2019b}.
The prefactor $C_{\rm loc}$ represents the fraction of spins in the localization pockets and one has $C_{\rm loc} \sim \ell_g \ell_{g,\parallel} \sim \ell_g^{7/4}R^{1/4}$. This asymptotic regime is plotted on Fig. \ref{fig:signal_b} for $\ell_\delta/a=0.1$.
Truncating this expansion to the first term yields the formula for a flat boundary that was discovered by Stoller \textit{et al.} in \cite{Stoller1991a} and naturally corresponds to the limit $R/\ell_g\to\infty$. In turn, the curvature of the boundary enters in correction terms. Note that the sign of the third term would be reversed if the obstacle's boundary was concave (i.e. inside the disk).
Moreover, the signal exhibits some oscillations that are related to overlapping of magnetization pockets, as we showed in \cite{Moutal2019b}. This is consistent with Fig. \ref{fig:horiz_magn} where the magnetization pockets on two neighboring obstacles have some significant overlapping even at the highest gradient strength. Note that these oscillations may lead to a significant signal attenuation for some particular values of $\ell_\delta$ and $\ell_g$.

(ii) For the gradient in the diagonal direction (right panel of Fig. \ref{fig:signal_b}), the spacing between neighboring obstacles is much larger so that the localization regime emerges at larger gradient length. Correspondingly, all curves follow the asymptotic decay \eqref{eq:asymptot_localization} and one observes some oscillations only for the largest diffusion length $\ell_\delta/a=0.3$. This is consistent with Fig. \ref{fig:diag_magn} where the magnetization pockets on neighboring obstacles have almost no overlap even at the lowest gradient strength (top panel).

%
%

\begin{figure}[t]
\centering
\includegraphics[width=0.40\linewidth]{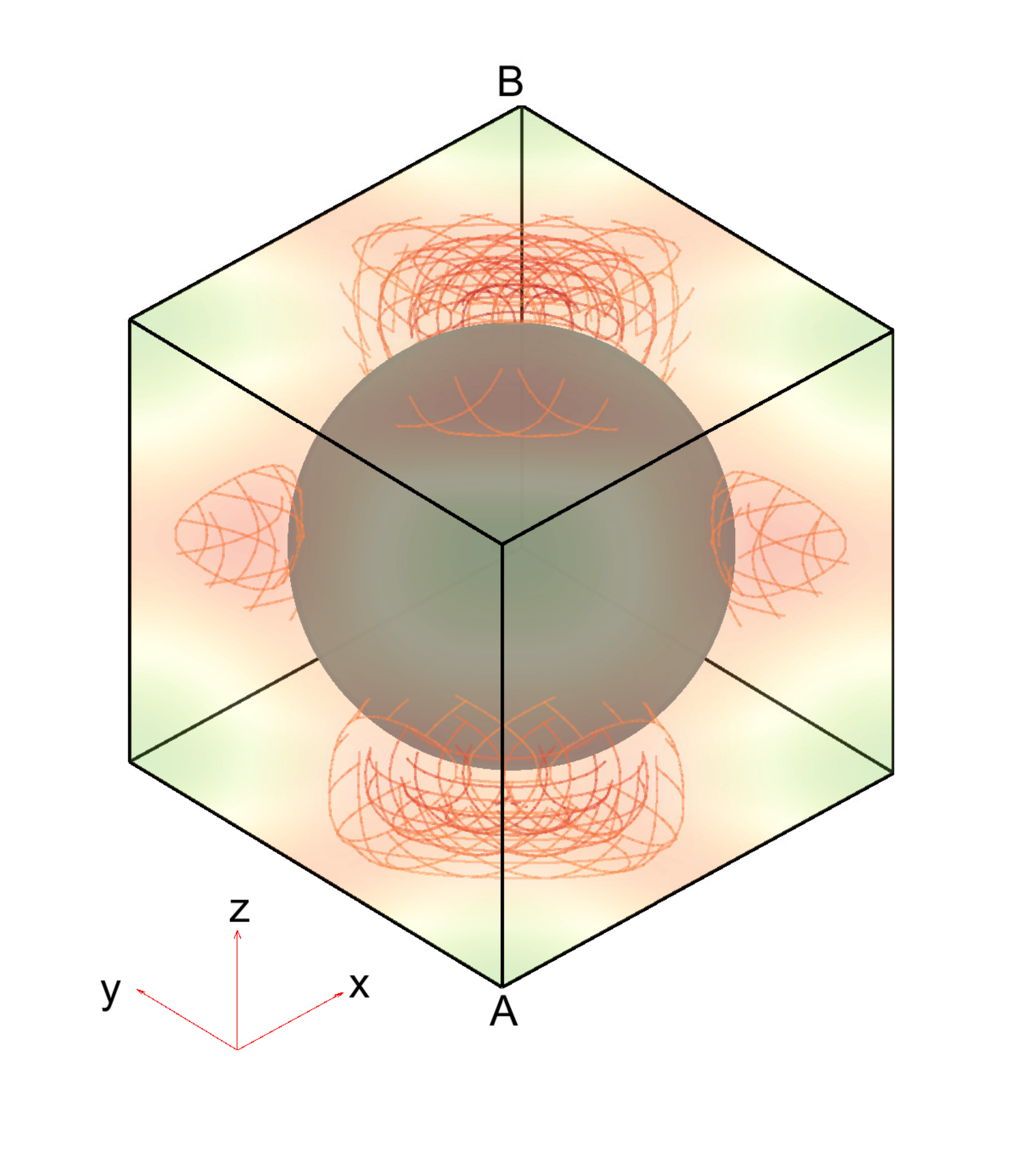}
\includegraphics[width=0.59\linewidth]{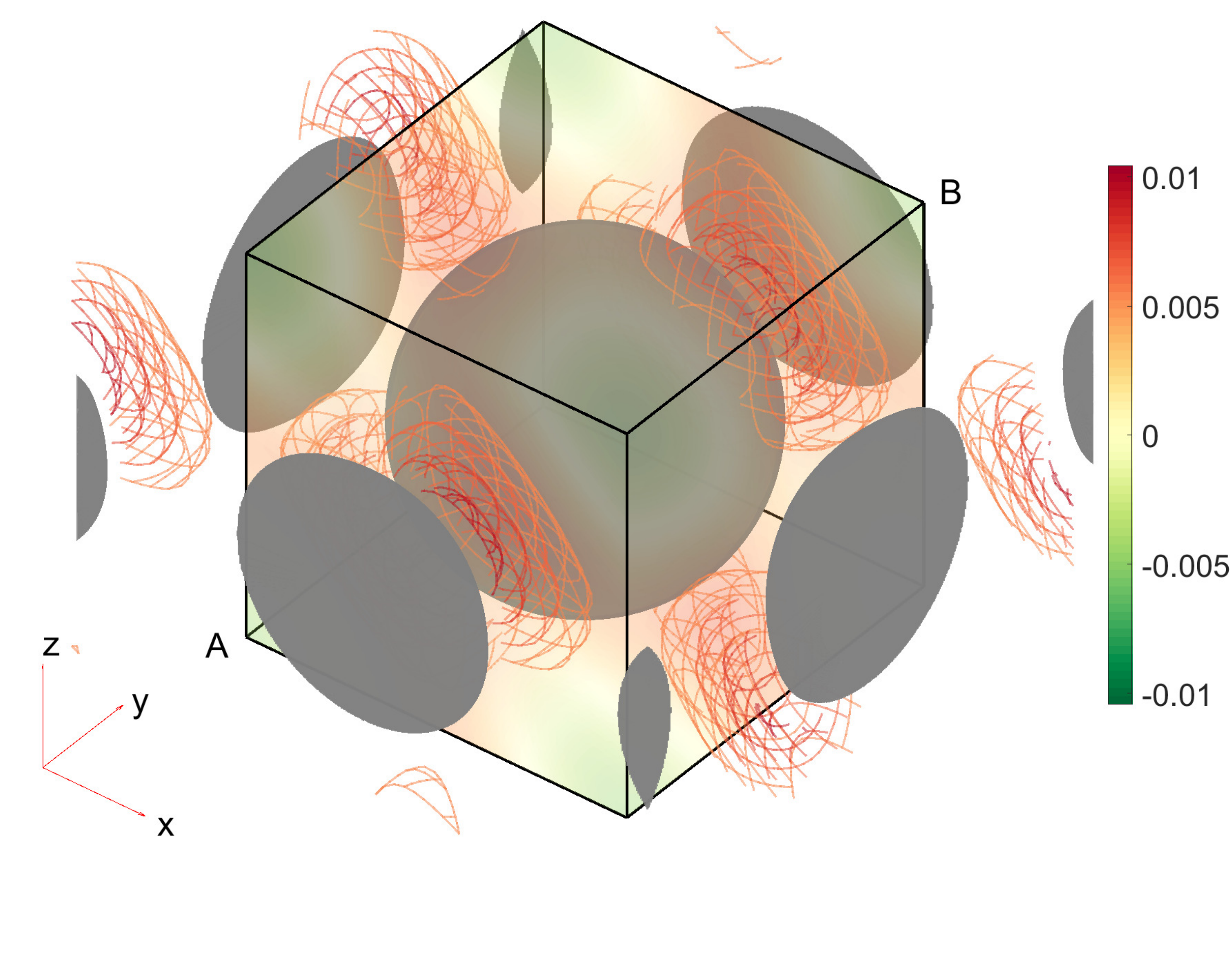}
\caption{Real part of the magnetization $m(T,x,y,z)$ in a 3D cubic lattice of spherical obstacles after a PGSE sequence, plotted as a set of colored wired isosurfaces as well as volume colors (dark colors represent intense magnetization). The left plot represents a single unit cell and the right plot represents a different view with neighboring cells (the black cube helps to visualize a unit cell). 
The gradient is along $\mathbf{e}_x+\mathbf{e}_y+\mathbf{e}_z$ (the diagonal of the cube from A to B). 
The parameters are $R/a=0.4$, $\ell_g/a=0.15$, $\ell_\delta/a=0.225$ and the normalized signal is $2.75\cdot 10^{-3}$.
One can see that the magnetization forms two localization pockets near the ``poles'' of the spheres, where the gradient is orthogonal to the boundary of obstacles. The magnetization is also high near the ``equatorial plane'' of the spheres; the right plot reveals that this is caused by the overlapping of neighboring localization pockets on the central unit cell.
}
\label{fig:magn_3D}
\end{figure}

For completeness we have also performed some numerical simulations in a 3-dimensional cubic lattice with spherical impermeable obstacles (see Fig. \ref{fig:magn_3D}). We used about $29000$ mesh points, $P=12$ and $350$ Laplacian eigenmodes for each pseudo-periodicity condition. The physical parameters used were $R/a=0.4$, $\ell_g/a=0.15$, $\ell_\delta/a=0.225$.
 The magnetization displays similar features compared to the 2D case. In particular it takes maximum values around ``poles'' of the spherical obstacles (i.e., points where the gradient is perpendicular to the boundary of obstacles). These magnetization pockets are well localized along the gradient direction on the scale $\ell_g$ but they are rather delocalized in the orthogonal plane (one can compute $\ell_{g,\parallel}/a=0.23$). Thus they overlap on neighboring cells, that creates a pattern similar to the top panel of Fig. \ref{fig:diag_magn}, with a rather intense magnetization in the ``equatorial plane'' of the obstacle, where one would expect a very weak magnetization if the obstacles were much further apart from each other. The right plot of Fig. \ref{fig:magn_3D} reveals this overlapping effect from neighboring cells.



\section{Eigenmodes of the Bloch-Torrey operator in a periodic medium}
\label{section:eigenmodes}

In this section we study eigenmodes and eigenvalues of the BT operator $\mathcal{B}$ defined in Eq. \eqref{eq:BT_operator}:
\begin{eqnarray}
&\mathcal{B}v_n(x,y)=\mu_n v_n(x,y)\label{eq:BT_eigen}\\
&\restr{\mathbf{n}\cdot D_0\nabla v_n + \kappa v_n}{\partial \Omega} = 0\;.
\end{eqnarray}
\cc{The BT operator is defined with $-\nabla^2$ in order to ensure eigenvalues $\mu_n$} with a positive real part (as shown in Sec. \ref{section:basic}).
By convention, the eigenvalues are sorted by increasing real part.
The existence and the properties of eigenmodes of the BT operator in a bounded domain have been studied in \cite{Huerlimann1995a,Swiet1994a,Grebenkov2017a,Grebenkov2018b,Stoller1991a,Grebenkov2014b,Herberthson2017a},
whereas a class of unbounded domains (exterior of an obstacle) has been investigated in \cite{Grebenkov2017a,Grebenkov2018b,Almog2018a,Almog2019a}.
 To summarize, when the gradient length $\ell_g=(D_0/g)^{1/3}$ is sufficiently small compared to the relevant geometrical scales in the domain, the eigenmodes are localized near the boundary regions that are perpendicular to the gradient direction. The length $\ell_g$ is the typical scale over which these modes localize along the gradient direction. This localization effect can be qualitatively understood as a competition between the delocalized eigenmodes of the Laplace operator and the Dirac peaks eigenmodes of the gradient term.
\cc{To our knowledge}, no theoretical or numerical studies were devoted to the spectral properties of the BT equation in the periodic case.
\cc{We recall that a discrete spectrum allows one to express the solution of the BT equation as a spectral decomposition. In turn, the behavior of the magnetization for long pulse duration is governed by few eigenmodes with the lowest eigenvalues (in real part). Therefore the study of the spectrum of the BT operator is of significant interest to understand the signal formation, particularly for extended-gradient pulses.}

For the sake of clarity, we restrict our discussion in this section to a 2D medium, periodic along $x$ and $y$ with periods $a_x$ and $a_y$, and the gradient is aligned along $x$ (i.e., $g_y=g_z=0$). This particular case allows us to describe the effect of periodicity along the gradient and perpendicular to the gradient. Note that all of our discussions and results are actually valid for any 2D periodic medium with one periodicity axis orthogonal to the gradient and can be extended to any 3D medium with two periodicity axes orthogonal to the gradient.
The general case of an arbitrary gradient direction is briefly discussed in Sec. \ref{section:periodicity_general}.
We emphasize that the results of this section require further mathematical analysis on the existence of the eigenmodes of the BT operator in periodic media. Throughout this section, we conjecture that these eigenmodes exist, and we shall provide strong numerical support to this conjecture.

\subsection{Basic properties}
\label{section:basic}

We shall recall some basic properties of the eigenmodes and eigenvalues of the BT operator. First, as $\mathcal{B}$ is not Hermitian but symmetric, two eigenmodes $v_n$ and $v_{n'}$ with distinct eigenvalues $\mu_n\neq \mu_{n'}$ are ``orthogonal'' for the real scalar product:
\begin{equation}
(v_n | v_{n'}) = \int_\Omega v_n v_{n'} \,\mathrm{d}x\,\mathrm{d}y\,\mathrm{d}z = 0\;.
\end{equation}
As we shall see later, there are exceptional values of the gradient at which two eigenvalues and eigenmodes can collapse (``bifurcation points'') but this leads to a Jordan block of dimension $2$ instead of two distinct eigenmodes. This behavior is illustrated on a simple matrix model in \ref{section:bifurcation_simple}.
Multiplying Eq. \eqref{eq:BT_eigen} by $v_n^*$ and integrating yields
\begin{eqnarray}
\mathrm{Re}(\mu_n) = \frac{D_0\int_\Omega |\nabla v_n|^2  +\kappa \int_{\partial \Omega} |v_n|^2
}{\int_\Omega |v_n|^2 
} \geq 0\;,\\
\cc{\mathrm{Im}(\mu_n) = -g_x \frac{\int_\Omega x |v_n|^2 
}{\int_\Omega |v_n|^2 
}\;,}
\end{eqnarray}
where we used the Robin boundary condition \eqref{eq:robin_boundary} and Green's formula to write the first relation.
If the integrals in the above formulas are well-defined, then the BT eigenmode $v_n$ is localized and its mean position is given by \cc{$-\mathrm{Im}(\mu_n)/g_x$}. Moreover, if the surface relaxivity $\kappa$ is zero, then the approximate width of the mode is given by $\sqrt{D_0/\mathrm{Re}(\mu_n)}$.


\subsection{Periodicity versus gradient direction}

\subsubsection{Periodicity perpendicular to the gradient}
\label{section:periodicity_perp}

Since the gradient is along $x$, the BT operator is invariant under any translation $y\to y+a_y$. From the theory of Bloch bands in condensed matter physics \cite{Bloch1929a,Kittel2004a}, we deduce that any eigenmode of $\mathcal{B}$ can be written in the form
\begin{equation}
v_{p_y,n}(x,y)=e^{ip_y y}w_{p_y,n}(x,y)\;,
\end{equation}
where $w_{p_y,n}$ is periodic along $y$, $p_y \in [0,2\pi/a_y)$ is the wavenumber associated to the eigenmode, and the index $n$ is integer. As a consequence, \cc{the spectrum of $\mathcal{B}$ is made of continuous bands}, each band being indexed by the integer $n$.

Note that if one considers a uniform initial magnetization, then only eigenmodes with $p_y=0$ (i.e., periodic along $y$) will be populated. As such, eigenmodes with $p_y\neq 0$ do not play any role in the signal formation. In the following, we discard the index $p_y$ from notations for brevity and all our numerical results \cc{are obtained for} $p_y=0$.

\subsubsection{Periodicity along the gradient}
\label{secion:periodicity_gradient}

The translation $x\to x+a_x$ modifies the BT operator as \cc{$\mathcal{B}\to \mathcal{B} - ig_x a_x$}. Hence, we can conclude that any eigenmode $v_n(x,y)$ and eigenvalue $\mu_n$ of $\mathcal{B}$ \cc{belongs} to a family of eigenmodes $v_n(x-ka_x,y)$ and eigenvalues \cc{$\mu_n-ikg_x a_x$}, where $k\in\mathbb{Z}$. This is consistent with the idea that the BT eigenmode $v_n(x,y)$ is localized near an obstacle of the medium and that $v_n(x-ka_x,y)$ is localized on the same obstacle but in a different unit cell. Indeed, as we showed in Sec. \ref{section:basic}, the imaginary part of $\mu_n$ can be interpreted as \cc{$-g_x$} times the position along $x$ of the localized mode $v_n$. \cc{For brevity, we discarded} the index $k$ from notations.

Moreover, if the unit cell $\Omega_1$ is not irreducible along the gradient direction, i.e. if there exists a lattice vector $\mathbf{e}$ such that $0<e_x<a_x$, then all eigenmodes $v_n$ can be translated by multiples of $\mathbf{e}$ that lead to $a_x/e_x$ families of eigenvalues 
\begin{equation*}\cc{
\mu=\mu_n- ikg_x a_x - ik'g_x e_x\;, \quad k \in \mathbb{Z}\;, \quad k'=0,\ldots,a_x/e_x-1\;,}
\end{equation*}
where $a_x/e_x$ is necessarily integer because of the hypothesis $g_y=0$ and the properties of additive groups. To avoid this artificial splitting of one family of eigenvalues into $a_x/e_x$ different families, we assume in the following that $\Omega_1$ is irreducible along the gradient direction. An example of this situation is illustrated on Fig. \ref{fig:periodic_2d} where the case (b) is reducible to (c).

\subsubsection{General gradient direction}
\label{section:periodicity_general}

As we explained in Sec. \ref{section:multi-periodic}, if the gradient is perpendicular to a generating vector of the lattice, one can redraw the unit cell and the previous discussion will be valid. Here we discuss the case of an arbitrary gradient direction and we assume that no lattice vector is perpendicular to the gradient. In that case, one cannot find any translation that leaves the BT operator invariant. However, the set $\{\mathbf{e}\cdot \mathbf{g}\}$, where the vector $\mathbf{e}$ spans all possible vectors of the lattice, is known to be a dense set in $\mathbb{R}$. Therefore to any eigenvalue $\mu_n$ is associated an infinite band $\mu_n+ i\nu$, where $\nu$ spans a dense set in $\mathbb{R}$. 
Although this case is formally the most general one (in the sense that a randomly chosen gradient direction always falls into that situation), we discard it in our analysis for two reasons: (i) as this paper represents the first step in the study of the spectrum of the BT operator in periodic media, we focus on a simpler but physically relevant situation and postpone the general case for future research; (ii) slightly changing the gradient direction allows returning to the case discussed in this section where the gradient is orthogonal to all periodicity axes but one.

\subsection{Numerical computation}
\label{section:numerical_eigenmodes}

\cc{We} use our numerical technique to investigate the properties of eigenmodes of the BT operator on a periodic medium. Let us stress again that since $\mathcal{B}$ does not respect the periodicity of the medium, it is impossible to study its eigenmodes and eigenvalues directly on a unit cell. However, the eigenmodes of the BT operator $\mathcal{B}$ are also the eigenmodes of its semi-group operator $\exp(-\tau \mathcal{B})$, whereas the eigenvalues $\mu_n$ are transformed into $\exp(-\tau \mu_n)$. 
Note that the minus sign comes from the definition \eqref{eq:BT_operator} of $\mathcal{B}$ so that $\exp(-\tau \mathcal{B})$ represents the effect of a $g_x$ gradient pulse of duration $\tau$.
If $\tau$ and $g_x$ satisfy the condition 
\begin{equation}
g_x \tau a = 2\pi\;,
\label{eq:periodicity_condition}
\end{equation}
then the semi-group operator respects the periodicity of the medium and one can study its eigenmodes and eigenvalues on a unit cell.
Note that one can impose any $p$-pseudo-periodic boundary conditions on the unit cell, not only periodic ones. In other words, one can study the eigenmodes and eigenvalues of the semi-group operator $\exp(-\tau\mathcal{B})$ on the space of $p$-pseudo-periodic functions for any value of $p$.
The application of a gradient pulse of a given duration is represented by the multiplication by a matrix (see Eq. \eqref{eq:matrix_magn_bis}), hence the study of the eigenmodes and eigenvalues of the BT operator is reduced to the study of the eigenvectors and eigenvalues of a matrix. Performing this study for a given $p$-pseudo-periodic boundary condition, one obtains a family $(v'_{p,n}(x,y),\,\mu'_{p,n})$ of the $p$-pseudo-periodic eigenmodes and associated eigenvalues of the semi-group $\exp(-\tau \mathcal{B})$ on $\Omega_1$:
\begin{eqnarray}
\exp(-\tau\mathcal{B})v'_{p,n}=\mu'_{p,n}v'_{p,n}\;,\\
\restr{\mathbf{n}\cdot D_0\nabla v'_{p,n} + \kappa v'_{p,n}}{\partial \Omega_1} = 0\;,\\
v'_{p,n}(a_x/2,y)=e^{ipa_x}v'_{p,n}(-a_x/2,y)\;.
\end{eqnarray}
In the following, we call them ``numerical'' eigenmodes and eigenvalues, to distinguish them from ``true'' eigenmodes and eigenvalues of the BT operator.
It is quite easy to see that $\mu'_{p,n}$ does not depend on $p$, hence we will denote it by $\mu'_n$ in the following.

The accuracy of the numerical computation can be assessed using Eq. \eqref{eq:error_estimation} combined with Eq. \eqref{eq:periodicity_condition}, which yields a relative error:
\begin{equation}
\epsilon \approx \frac{1}{P^2} \frac{D_0}{g_x a^3} = \frac{1}{P^2}\left(\frac{\ell_g}{a}\right)^3\;.
\end{equation}
This formula implies that the numerical computation of eigenmodes and eigenvalues of the BT operator is more accurate at high gradients. In the following, we assume that the sampling of $q_x(t)$ is fine enough so that this error is negligible.
Moreover, because of the condition \eqref{eq:periodicity_condition}, low gradients $g_x$  require long pulse duration $\tau$ and increase the relative difference between the eigenvalues $\exp(-\tau \mu_n)$ of the semi-group operator $\exp(-\tau \mathcal{B})$. 
As eigenvalues are sorted by increasing real part, the accuracy in the numerical computation of $\mu_n$ is limited by the ratio $|\exp(-\mu_n\tau)/\exp(-\mu_0\tau)|$.
If one denotes by $\eta$ the relative precision of numerical computations \cc{(usually, $\eta =2^{-52}\approx 2\cdot 10^{-16}$)}, then any eigenvalues $\mu_n$ such that
\begin{equation}
\mathrm{Re}(\mu_n-\mu_0) > -\frac{\log(\eta)}{\tau}
\label{eq:condition_mu_tau}
\end{equation}
is ``lost'' because of the finite precision of numerical computations. The above equation, combined with Eq. \eqref{eq:periodicity_condition}, can be rewritten as
\begin{equation}
\mathrm{Re}(\mu_n-\mu_0) > -\frac{g_x a \log(\eta)}{2\pi}\;,
\label{eq:condition_mu_g}
\end{equation}
so that the limit between computable and non-computable eigenvalues is a line in a $\mu_n(g_x)$ plot (see Fig. \ref{fig:spectrum} below).

The numerical eigenmodes $v'_{p,n}$ are pseudo-periodic, hence delocalized, that means that they are not eigenmodes of the BT operator. However, they are formed by a superposition of translated BT operator eigenmodes. In fact, let us assert the following formula
\begin{equation}
v'_{p,n}(x,y)\overset{?}{=}K_{p,n}\sum_{k\in\mathbb{Z}} e^{ipka_x} v_n(x-ka_x,y)\;,
\label{eq:eigen_superposition}
\end{equation}
with $K_{p,n}$ a normalization constant. First, one can note that the right-hand side of Eq. \eqref{eq:eigen_superposition} is $p$-pseudo-periodic. Moreover, it is an eigenmode of $\exp(-\tau \mathcal{B})$ with the eigenvalue
\begin{equation}\cc{
\mu'_n=\exp\left(-\tau(\mu_n - ikg_x a_x)\right)\;, \quad k\in \mathbb{Z}\;.
\label{eq:eigenvalue_collapse}}
\end{equation}
Indeed, the right-hand side of Eq. \eqref{eq:eigenvalue_collapse} does not depend on $k$ according to Eq. \eqref{eq:periodicity_condition}. This proves that Eq. \eqref{eq:eigen_superposition} is correct.

From the ``numerical'' eigenvalues $\mu'_n$, one can deduce the eigenvalues of the BT operator according to
\begin{eqnarray}
\cc{\mu_n }&\cc{= -\log(\mu'_n)/\tau - 2ik\pi/\tau\;, \quad k\in\mathbb{Z} \nonumber}\\
&\cc{= -\frac{g_x a_x\log(\mu'_n)}{2\pi} - ikg_x a_x\;, \quad k\in\mathbb{Z}}\;.
\label{eq:true_eigenvalues}
\end{eqnarray}
As explained in Sec. \ref{secion:periodicity_gradient}, the above formula describes an infinite family of eigenvalues corresponding to eigenmodes localized on the same obstacle's boundary region but at different \cc{unit cells}. We applied the convention that the imaginary part of the complex logarithm belongs to $(-\pi, \pi]$ so that $k=0$ corresponds to the smallest imaginary part in absolute value and to a mode centered on the unit cell $\Omega_1$ ($-a_x/2 \leq x \leq a_x/2$).
An example of spectrum obtained numerically is shown on Figs. \ref{fig:spectrum} and \ref{fig:spectrum_long} (discussed below).

\begin{figure}[t]
\centering
\includegraphics[width=0.99\linewidth]{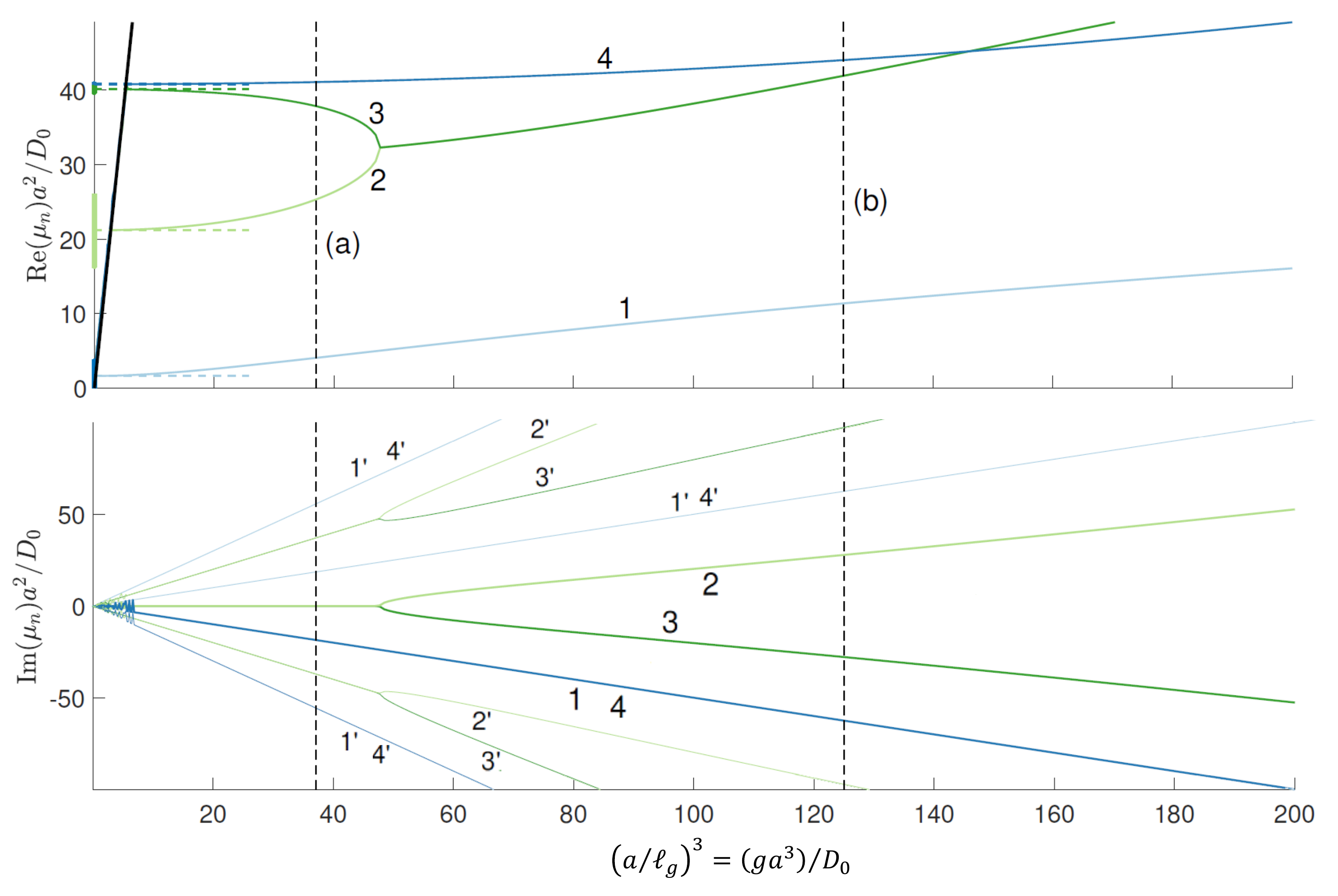}
\caption{Several eigenvalues of the BT operator on a square lattice of circular impermeable obstacles with $R/a=0.4$ and the gradient in the horizontal direction. 
The dimensionless eigenvalues $\mu_n a^2/D_0$ and dimensionless gradient $(a/\ell_g)^3$
ensure that the plot is independent of the actual value of $a$ used in the computation.
The numbers and colors help to associate the top plot to the bottom plot. (top) Real part of the spectrum. The numerical limit \eqref{eq:condition_mu_g} is represented by a thick black line above which the computation of eigenvalues is limited by numerical accuracy. Moreover, dashed horizontal lines show the low gradient limit \eqref{eq:asymptot_low_g} and the bands $\lambda_{p,n}$ of the Laplace operator are plotted as vertical segments at $g=0$. (bottom) Imaginary part of the spectrum. Equation \eqref{eq:true_eigenvalues} is plotted for $k=-1,0,1$  and branches of $\mu_n$ with $k\neq 0$ are denoted by ``$n'$''. Spurious fluctuations at small $g$ are caused by difficulties in ordering complex eigenvalues with identical real parts. Vertical dashed lines indicate the values of the gradient used in Fig. \ref{fig:eigenmodes}: (a) $\ell_g/a=0.3$; (b) $\ell_g/a=0.2$.}

\label{fig:spectrum}
\end{figure}

\begin{figure}[t]
\centering
\includegraphics[width=0.99\linewidth]{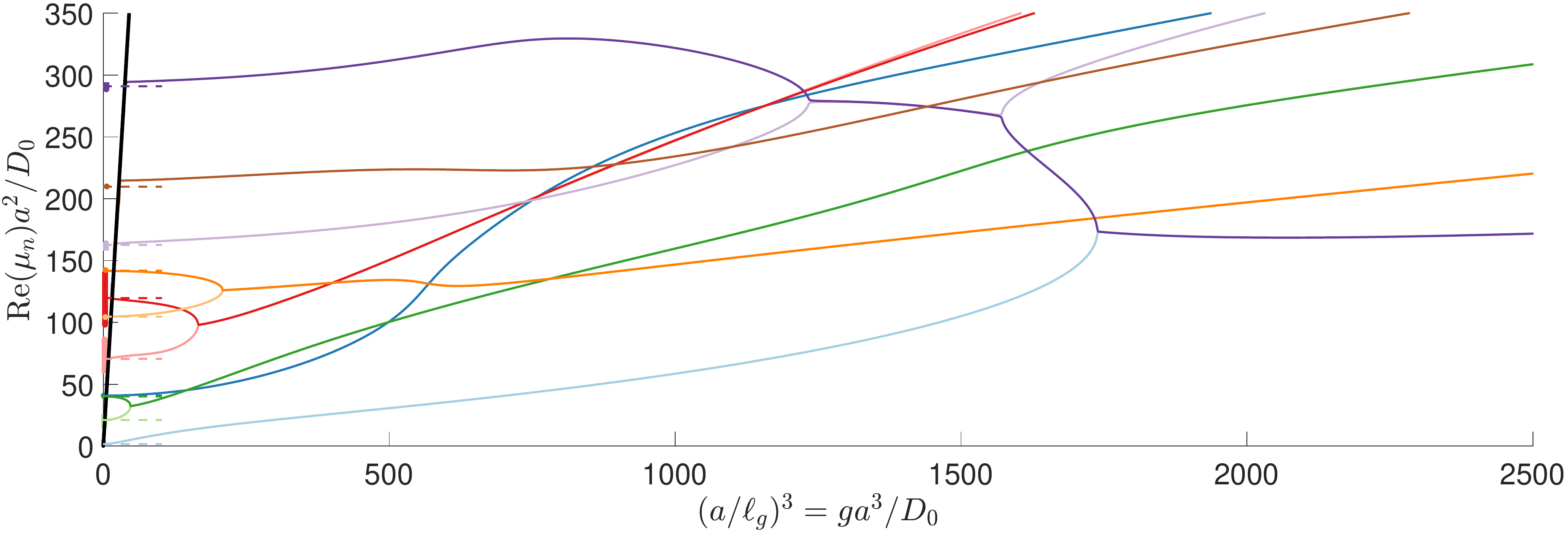}
\vskip 3mm
\includegraphics[width=0.99\linewidth]{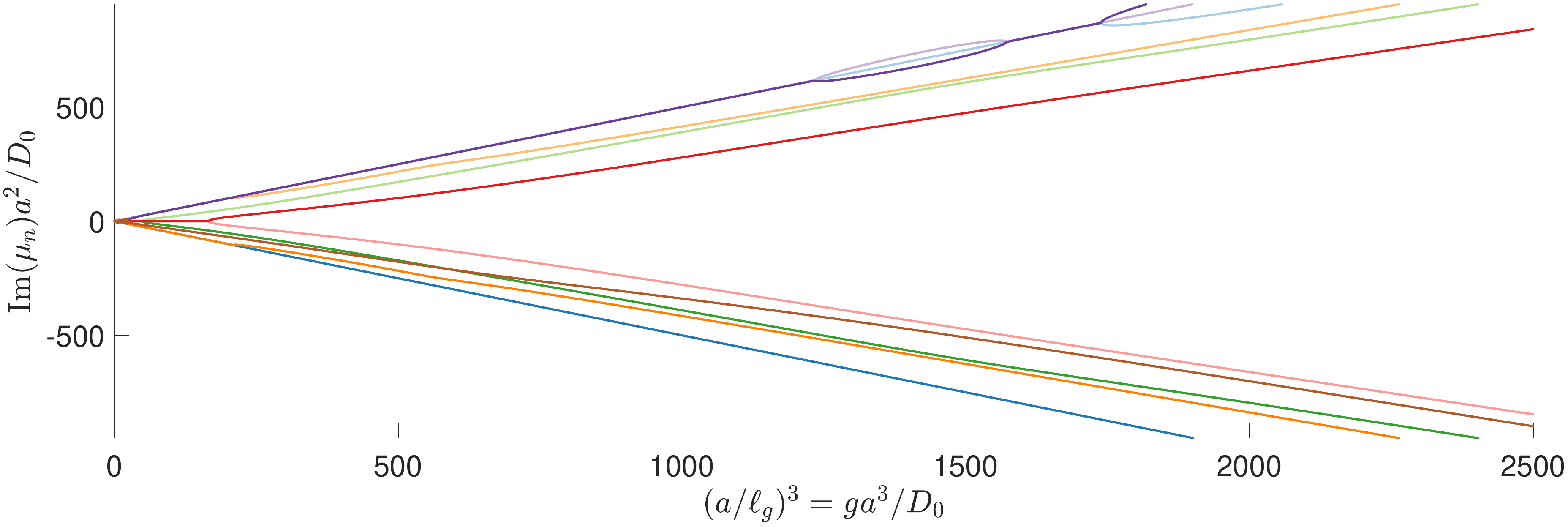}
\caption{Same plot as in Fig. \ref{fig:spectrum} but with a larger range of gradient values and additional branches of $\mu_n$ (some were omitted to improve visibility). The figure reveals a rich structure of bifurcation points.}
\label{fig:spectrum_long}
\end{figure}

\begin{figure}[th]
\centering
\includegraphics[width=0.24\linewidth]{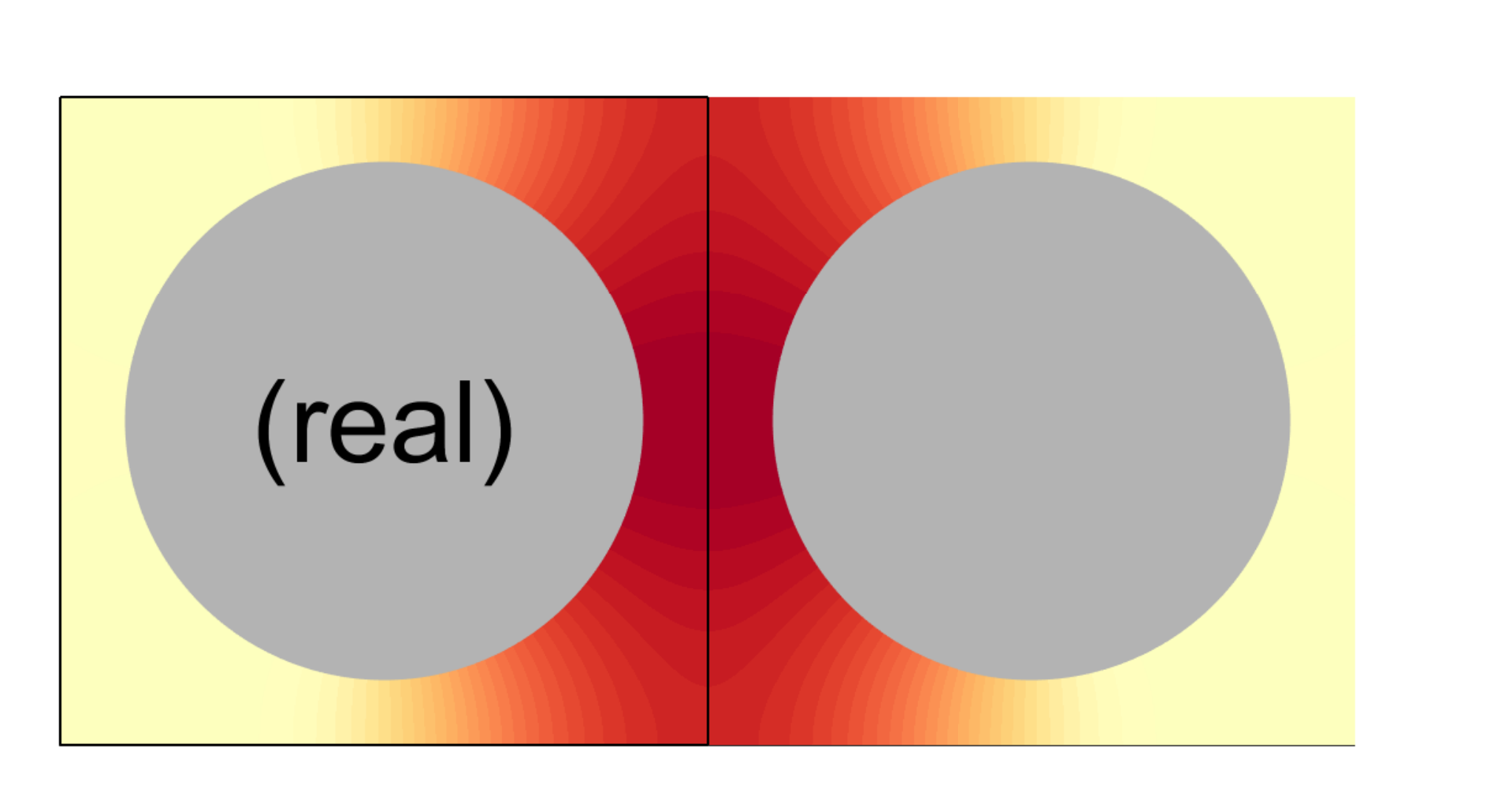}
\includegraphics[width=0.24\linewidth]{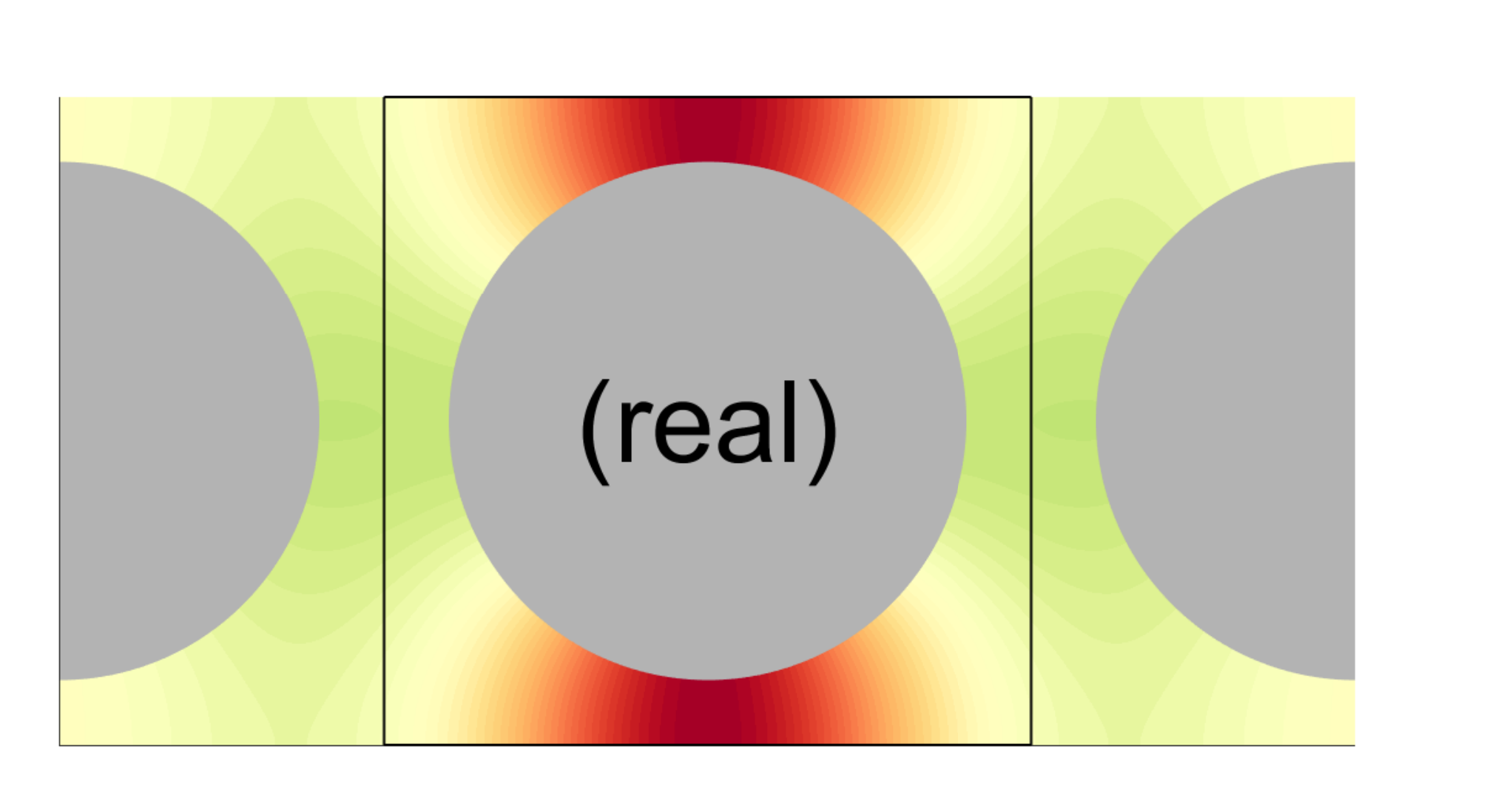}
\includegraphics[width=0.24\linewidth]{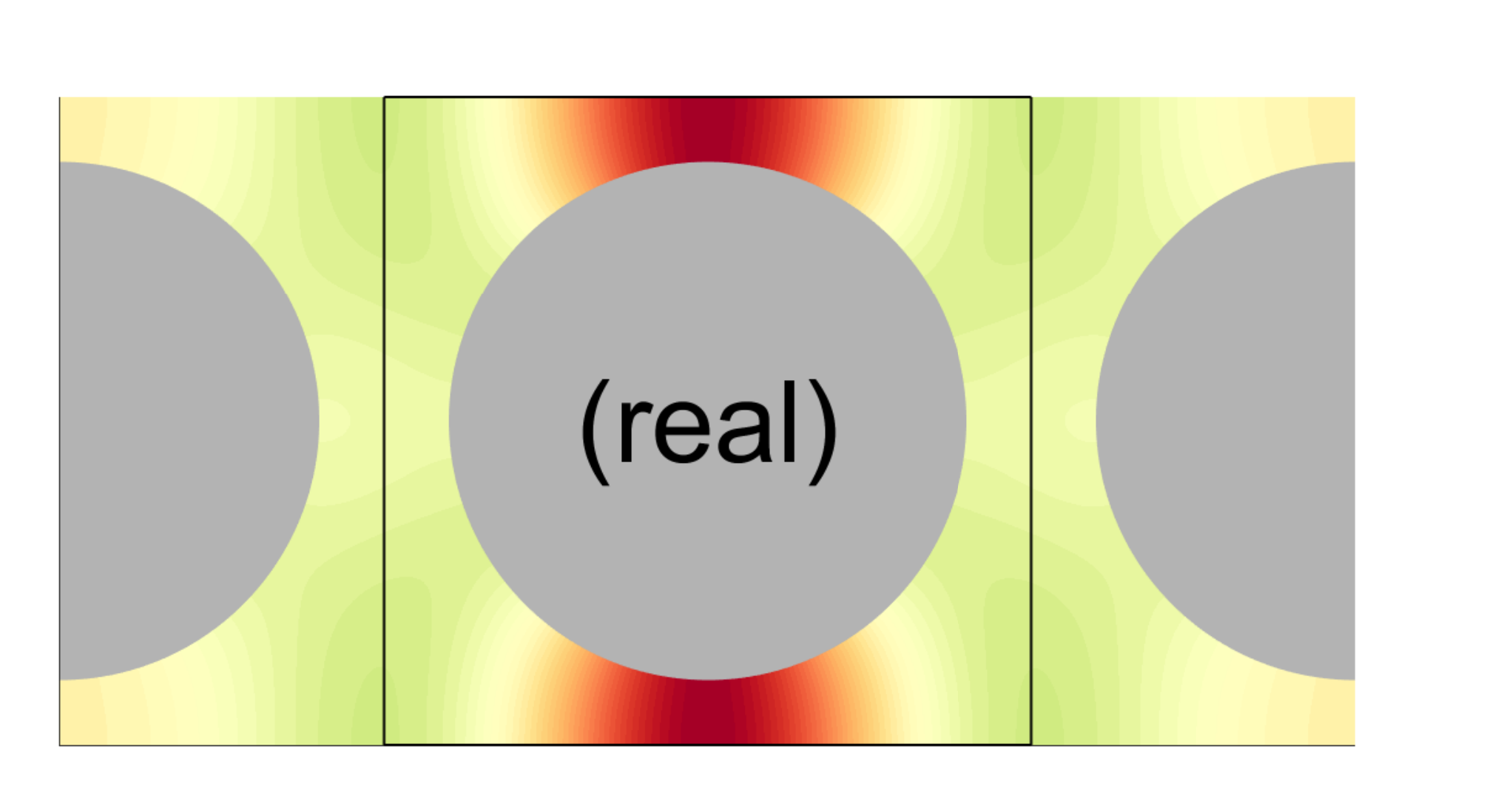}
\includegraphics[width=0.24\linewidth]{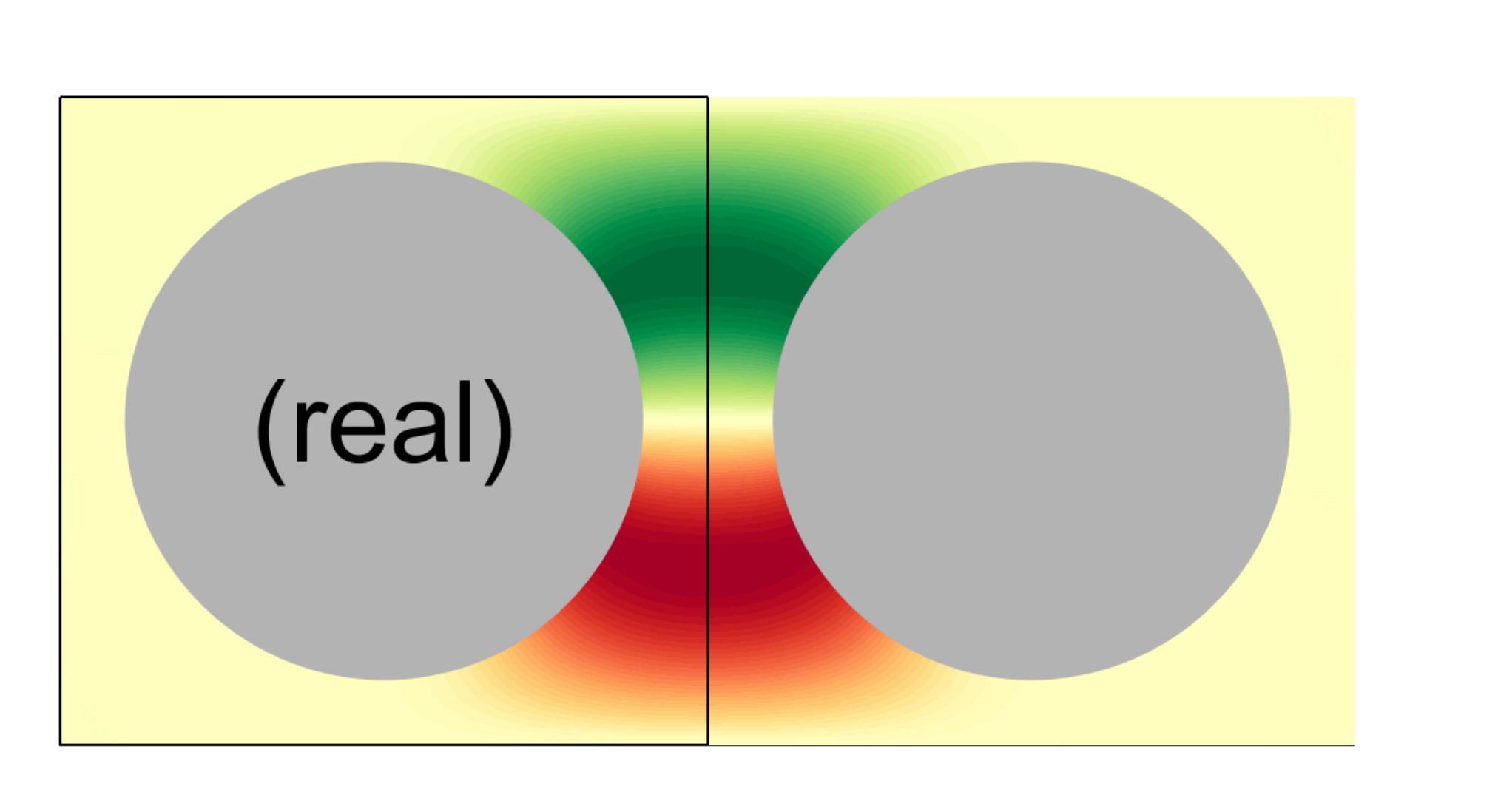}
(a1) \hspace{0.185\linewidth} (a2) \hspace{0.185\linewidth} (a3) \hspace{0.185\linewidth} (a4) \\
\includegraphics[width=0.24\linewidth]{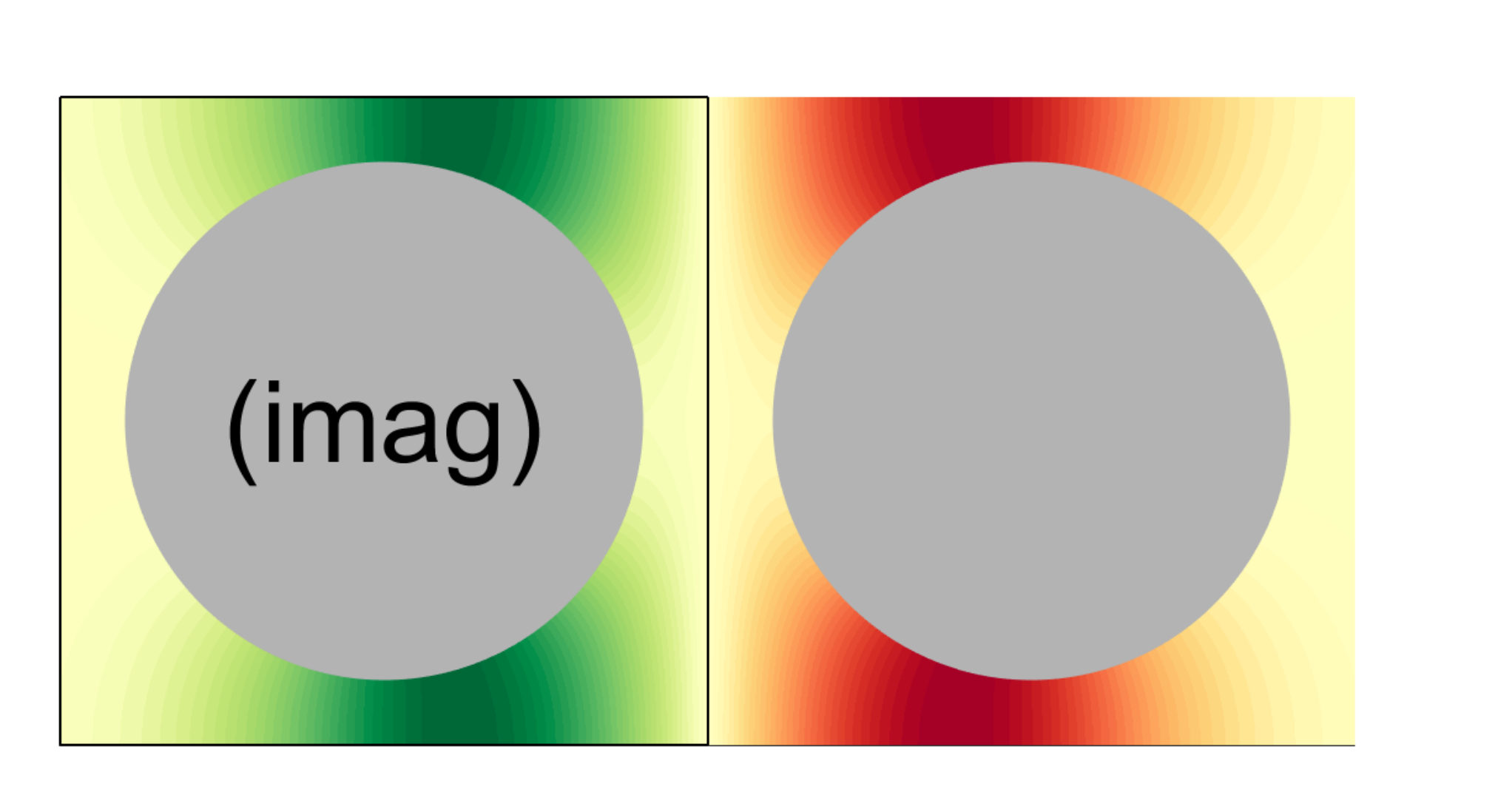}
\includegraphics[width=0.24\linewidth]{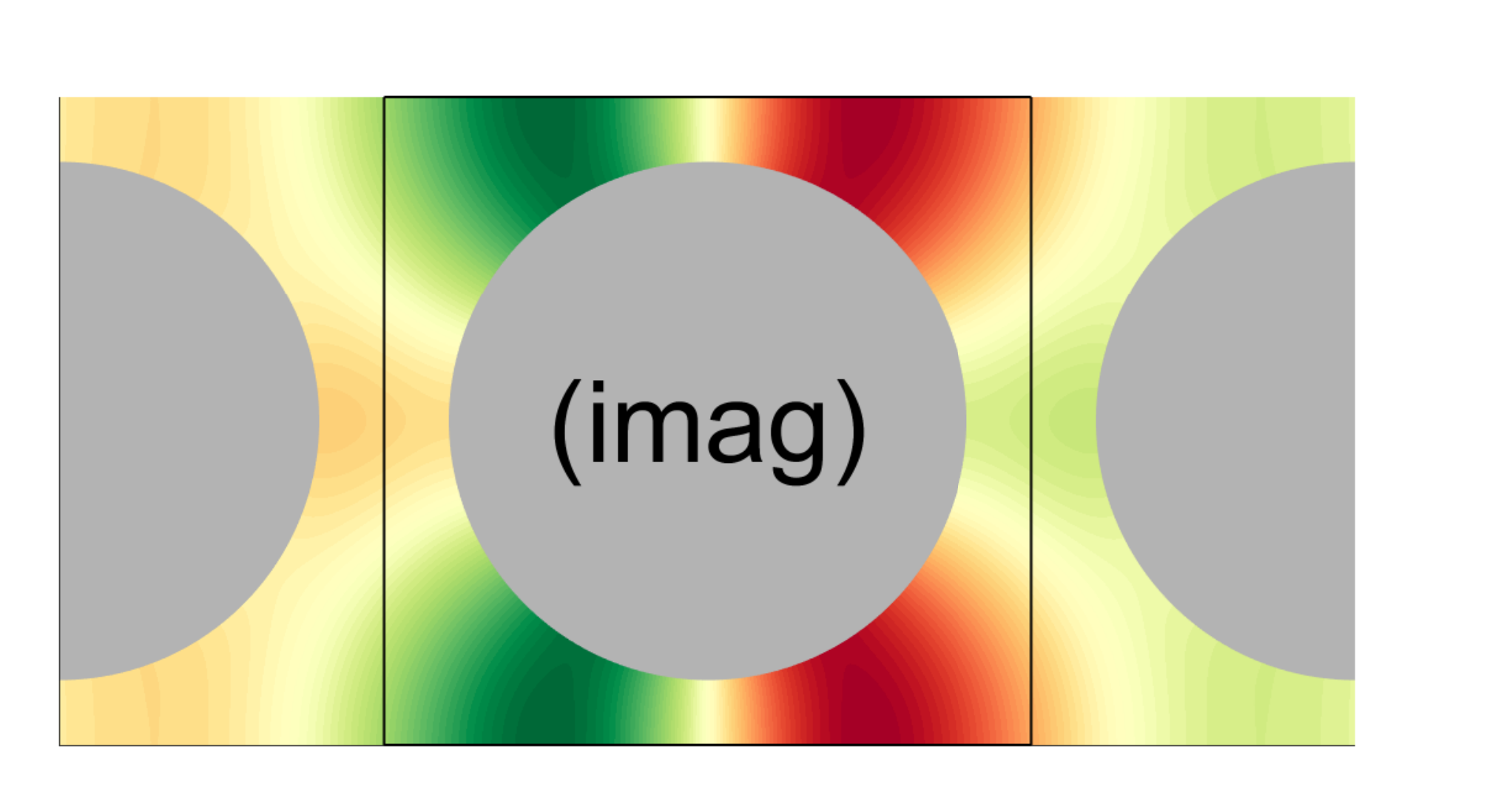}
\includegraphics[width=0.24\linewidth]{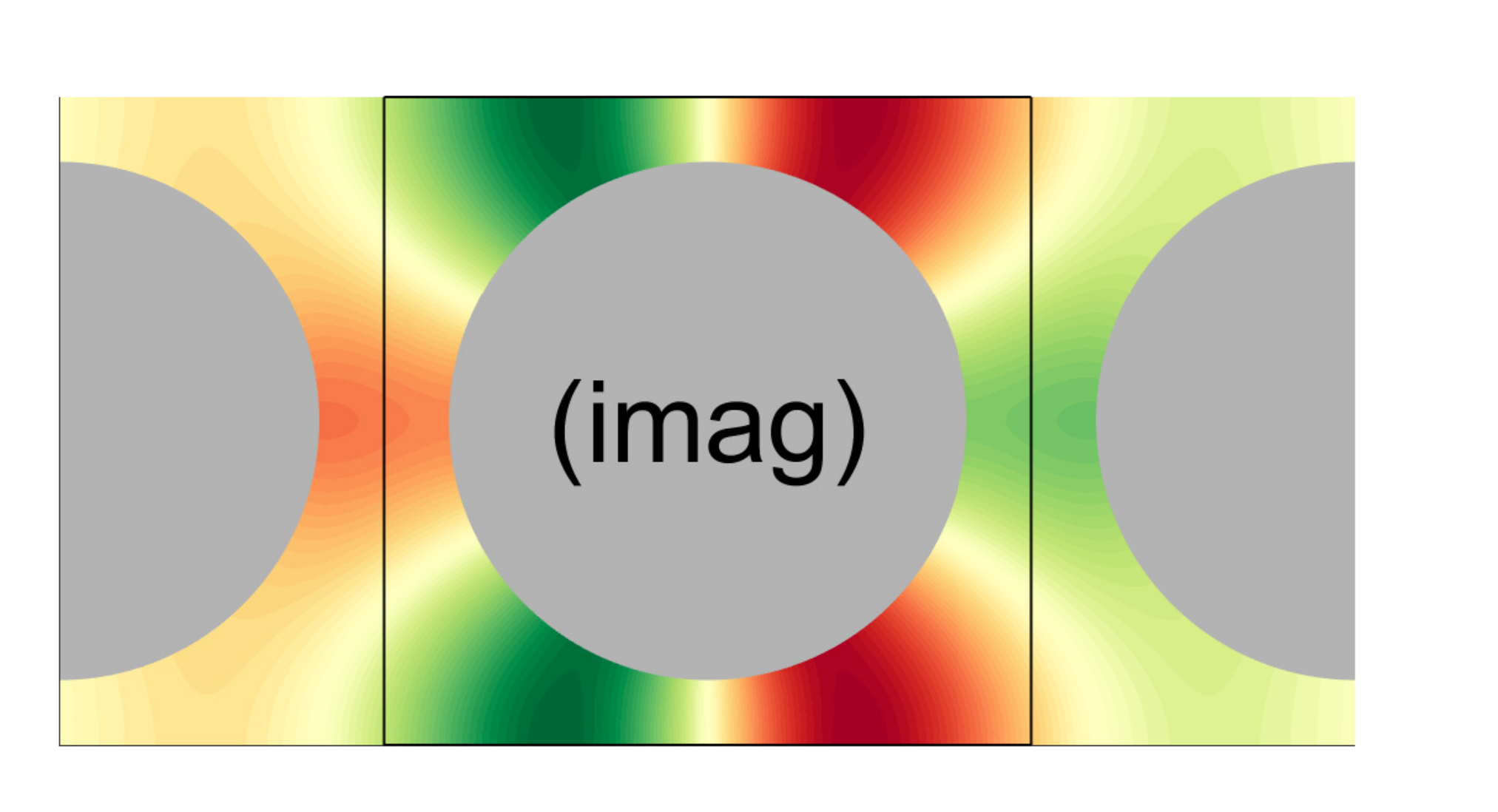}
\includegraphics[width=0.24\linewidth]{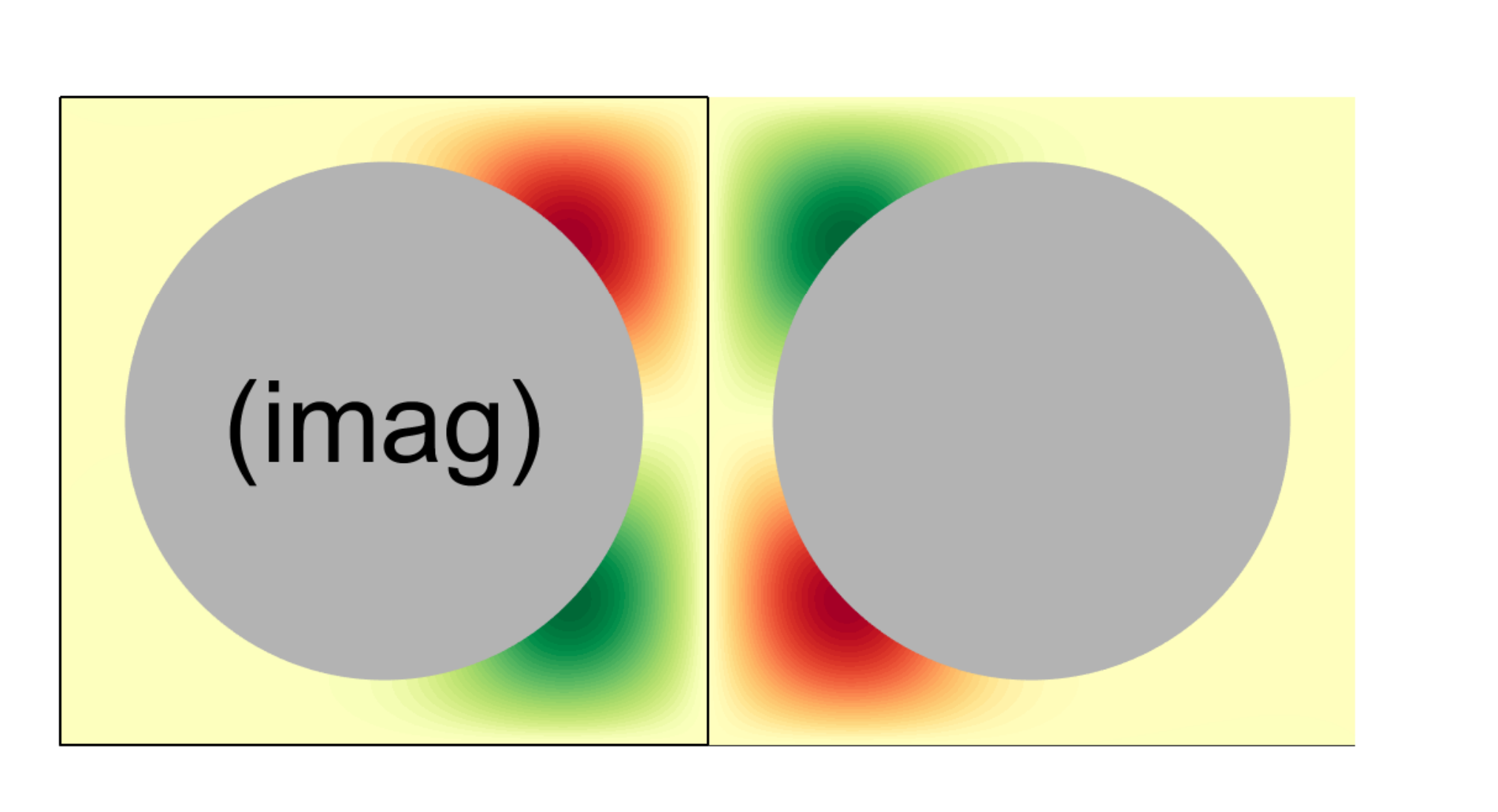}
\noindent\rule[5pt]{\linewidth}{0.4pt}
\includegraphics[width=0.24\linewidth]{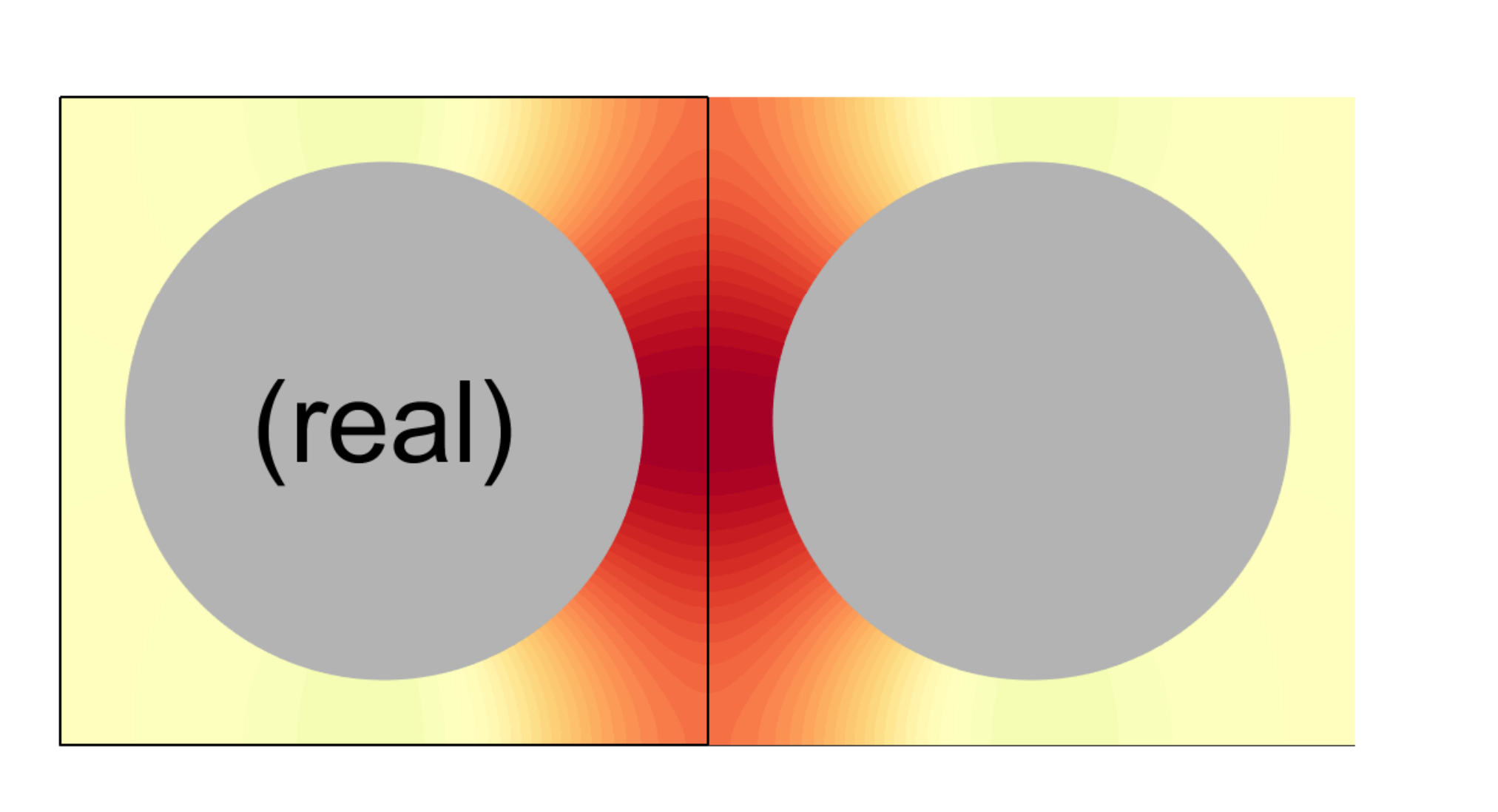}
\includegraphics[width=0.24\linewidth]{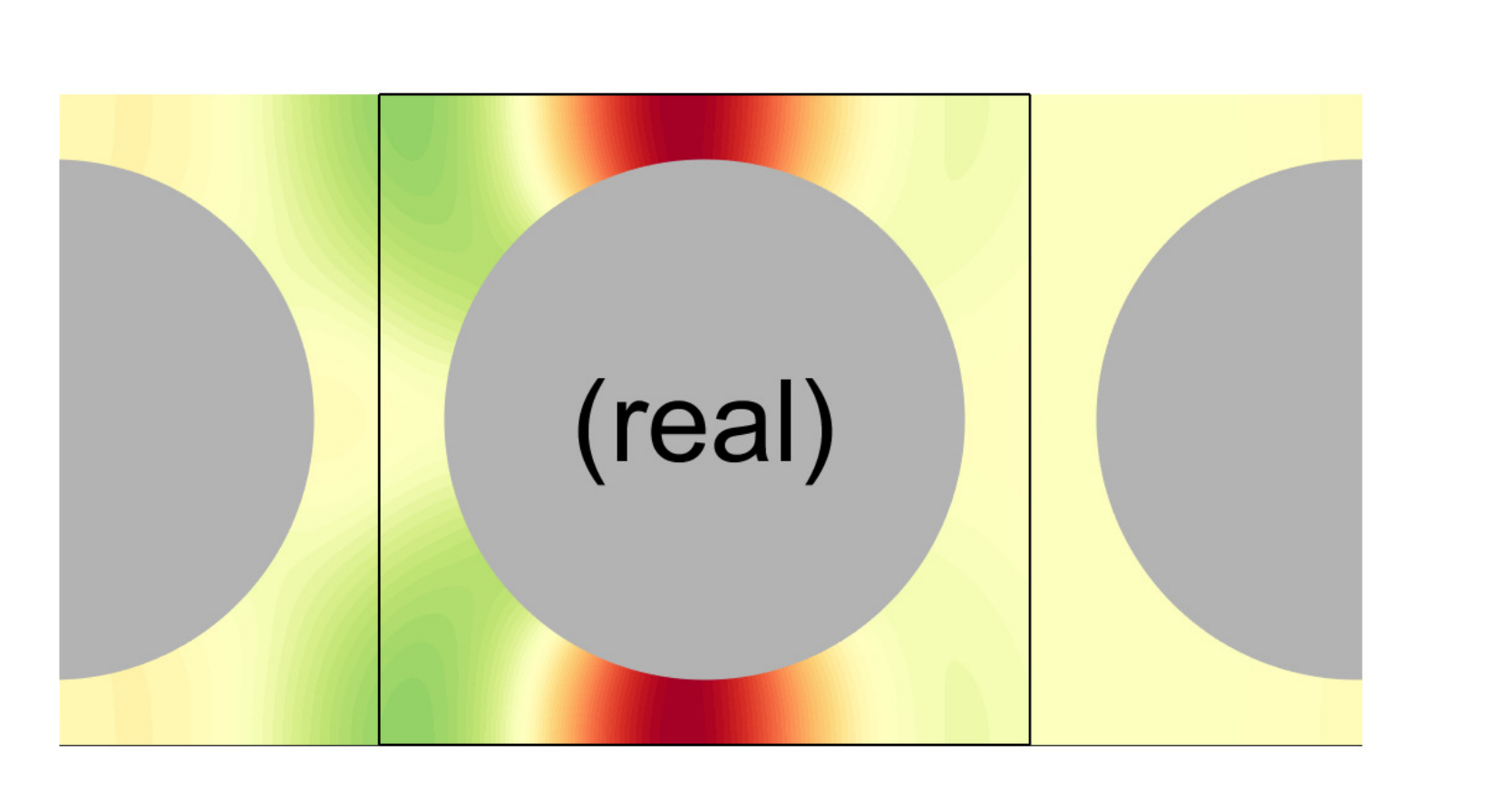}
\includegraphics[width=0.24\linewidth]{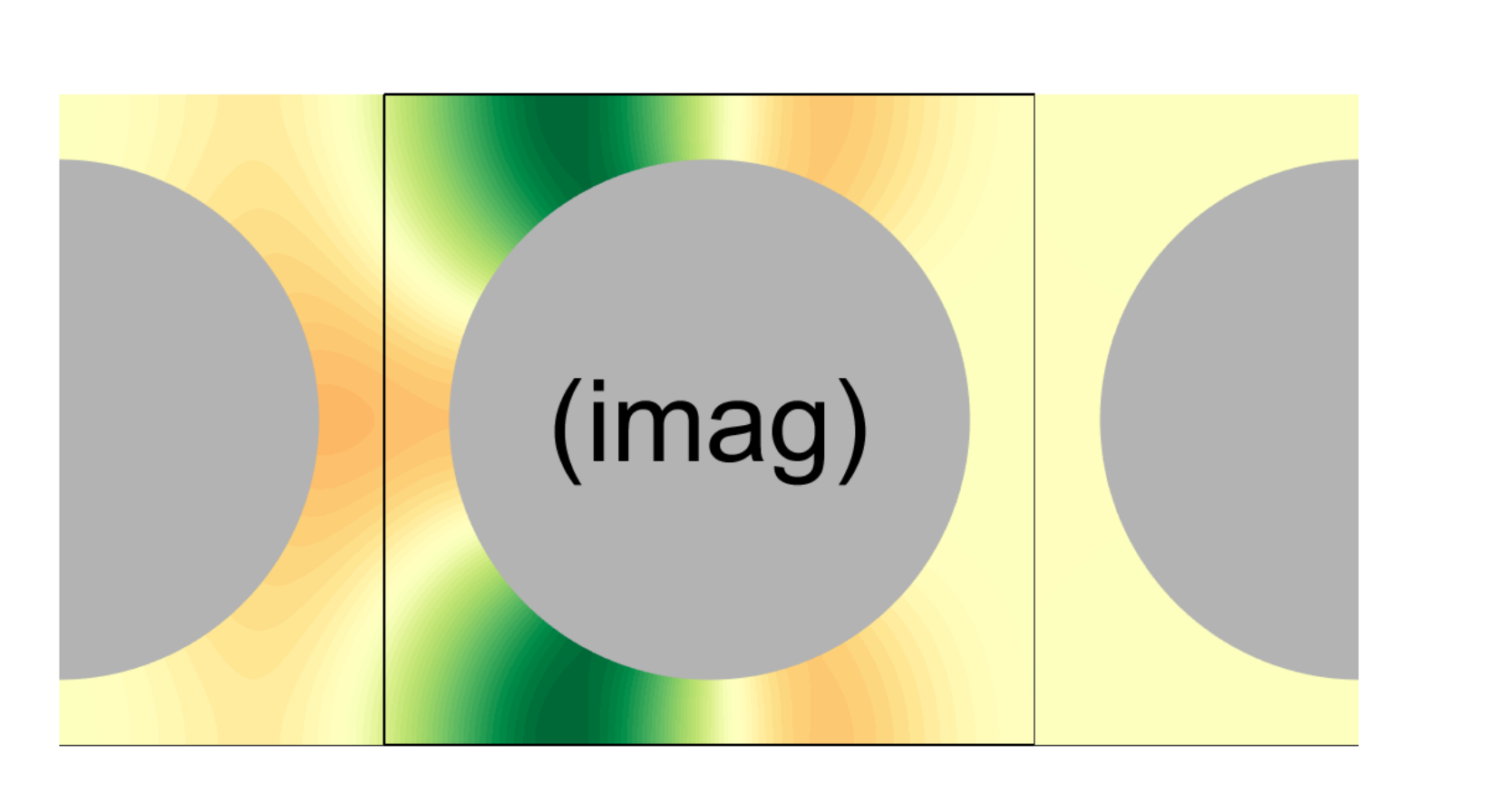}
\includegraphics[width=0.24\linewidth]{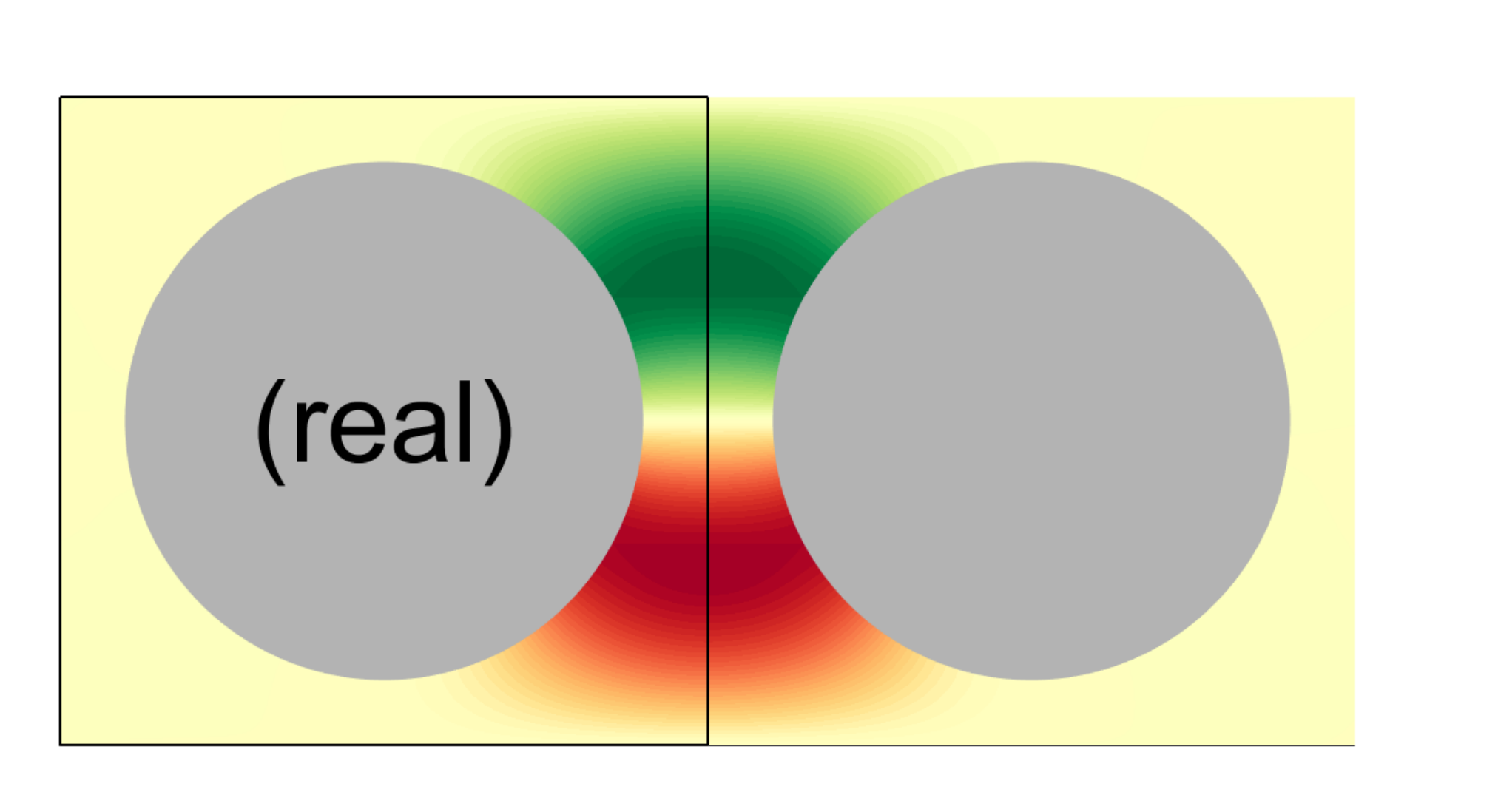}
(b1) \hspace{0.185\linewidth} (b2) \hspace{0.185\linewidth} (b3) \hspace{0.185\linewidth} (b4) \\
\includegraphics[width=0.24\linewidth]{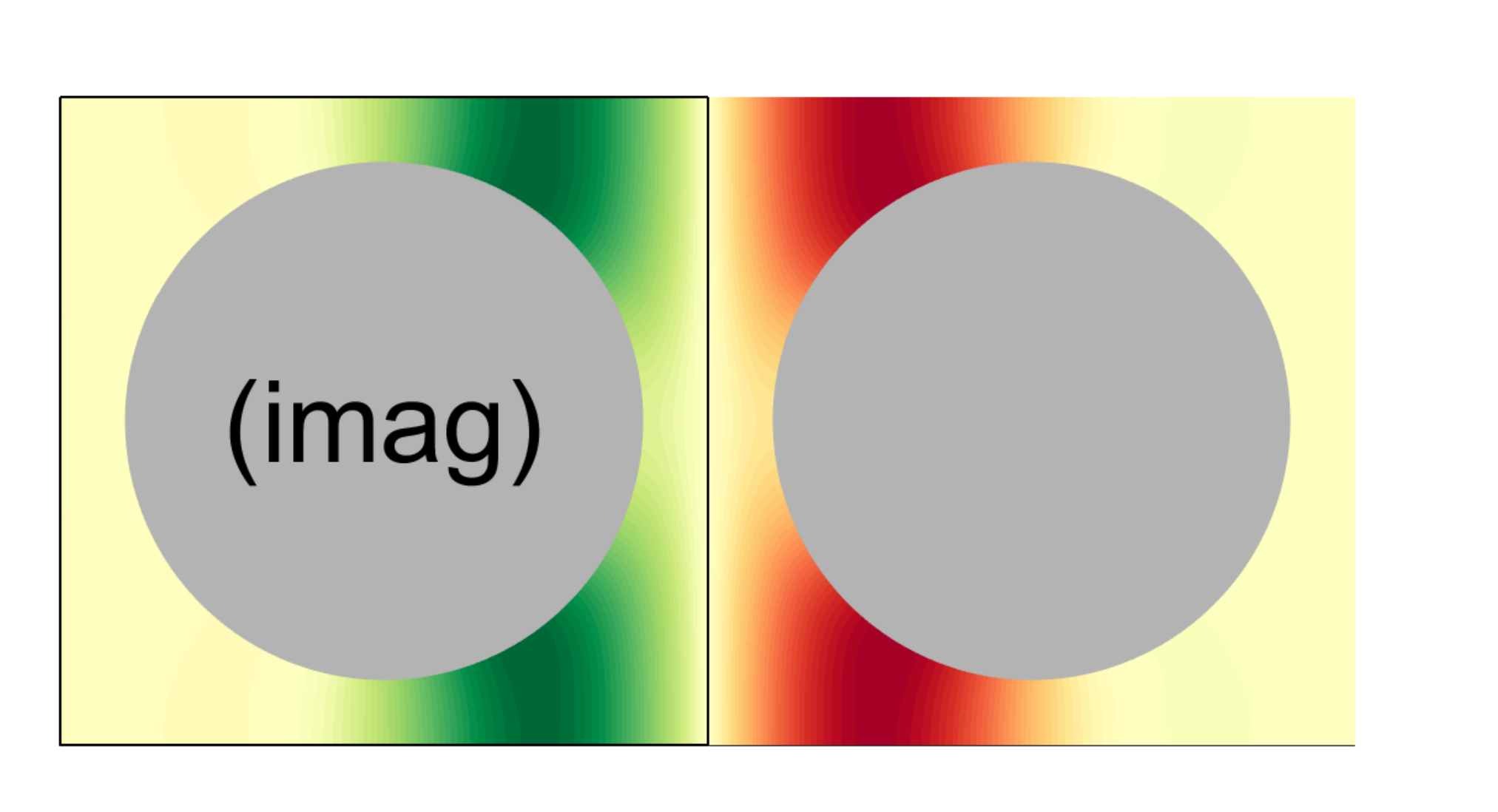}
\includegraphics[width=0.24\linewidth]{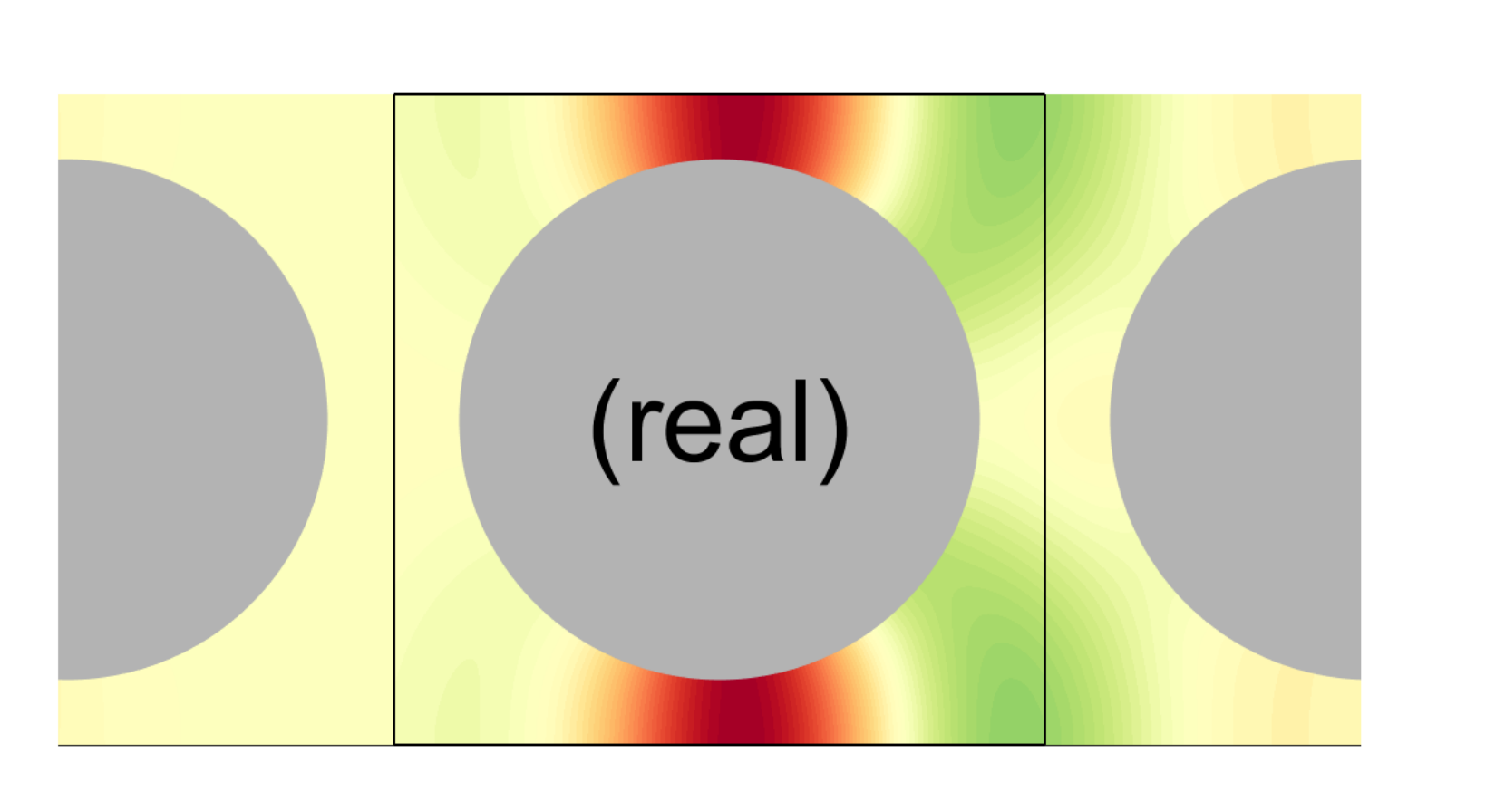}
\includegraphics[width=0.24\linewidth]{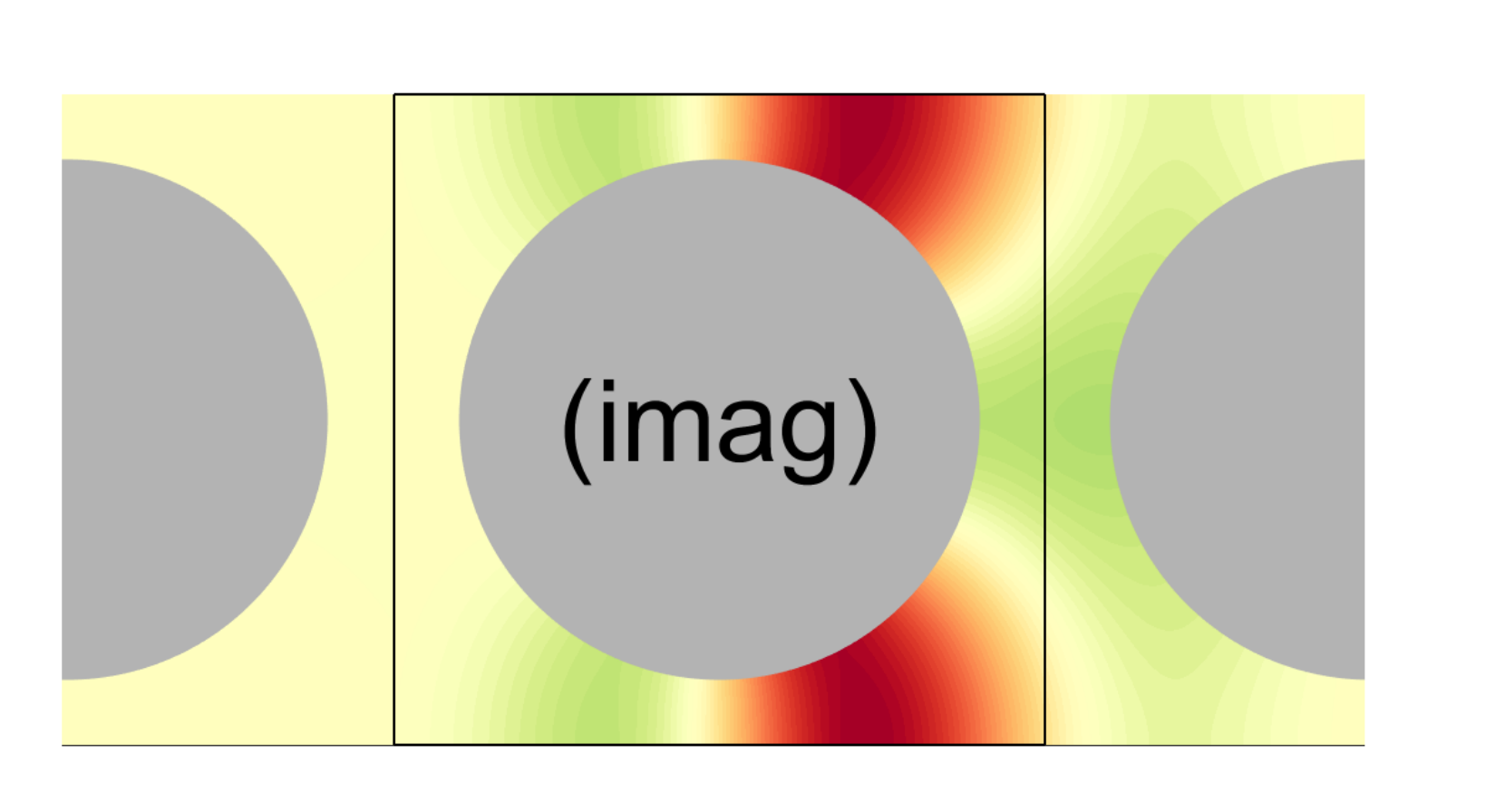}
\includegraphics[width=0.24\linewidth]{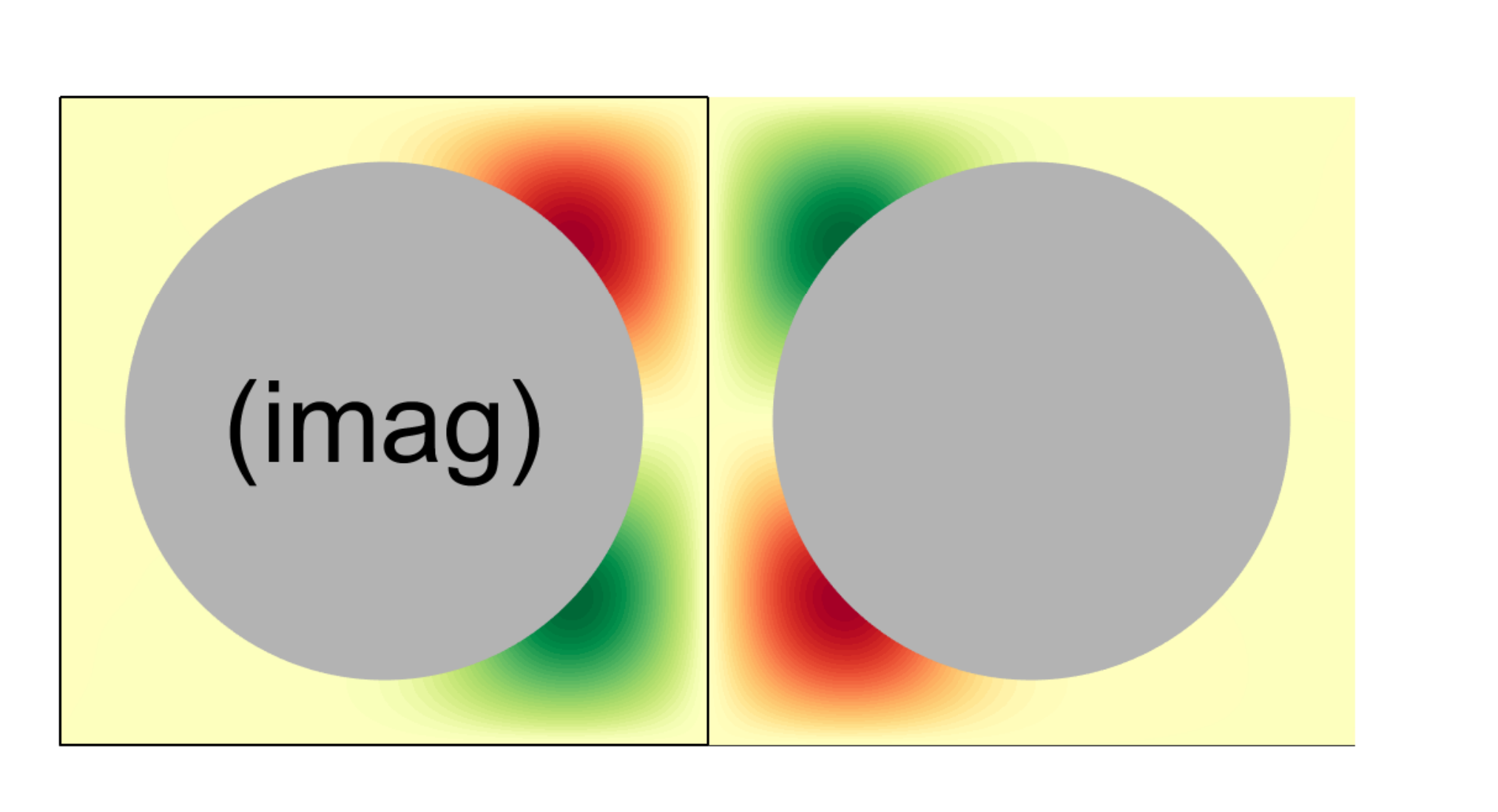}
\caption{Real and imaginary part of the first four eigenmodes of the BT operator on a square lattice of circular impermeable obstacles with the gradient in the horizontal direction, for two different gradient strengths: (a) $\ell_g/a=0.3$; (b) $\ell_g/a=0.2$. Refer to Fig. \ref{fig:spectrum} for the corresponding eigenvalues. \cc{While} modes $1$ and $4$ show little variation from (a) to (b), the pair $2,3$ undergoes a bifurcation that dramatically affects its symmetry properties. The black square helps to visualize the unit cell $\Omega_1$ and interpret the imaginary part of $\mu_n$ on Fig. \ref{fig:spectrum}. The color scale is the same as on Figs. \ref{fig:horiz_magn} and \ref{fig:diag_magn}: green for negative, red for positive, intense colors correspond to large absolute value.
}
\label{fig:eigenmodes}
\end{figure}

Now we will show how one can recover the true eigenmodes $v_n$ from the numerical ones $v'_{p,n}$. First, Eq. \eqref{eq:eigen_superposition} implies that $v_n$ can be computed as an infinite superposition of $v'_{p,n}$: 
\begin{equation}
v_n(x,y)=\frac{a_x}{2\pi}\int_0^{2\pi/a_x} \frac{v'_{p,n}}{K_{p,n}} \,\mathrm{d}p\;.
\label{eq:true_eigenmode}
\end{equation}
However, this is clearly impractical from a numerical point of view because one would have to compute $v'_{p,n}$ for infinitely many values of $p$. Thus, let us consider the discrete version of the above formula, where $p=2l\pi/(Pa_x)$, $l=0,1,\ldots,P-1$, and define
\begin{eqnarray}
v'_n(x,y)&=\frac{1}{P}\sum_{l=0}^{P-1} \frac{v'_{2l\pi/(Pa_x),n}(x,y)}{K_{2l\pi/(Pa_x),n}}
\label{eq:eigenbis_superposition}\\
&=\sum_{k\in\mathbb{Z}} v_n(x-kPa_x,y)\;.
\label{eq:eigenbis_superposition_2}
\end{eqnarray}
Note that compared to Eq. \eqref{eq:true_eigenmode} where integrating over all $p$ leads to a perfect cancellation for all $k\neq 0$, the discrete sum in Eq. \eqref{eq:eigenbis_superposition} generates a $Pa_x$ periodic pattern where the eigenmode $v_n(x,y)$ is repeated every $P$ unit cell.
 Therefore, if $P$ is large enough so that BT eigenmodes do not overlap over the distance $Pa_x$, the restriction of $v'_n(x,y)$ to $-Pa_x/2 \leq x \leq Pa_x/2$ gives the exact eigenmode $v_n(x,y)$.

The only difficulty in the above method is to find the values of the normalization constants $K_{p,n}$. We did not manage to find a normalization scheme that would give access to them. However, one can easily find their values numerically by treating them as unknown quantities and solving Eq. \eqref{eq:eigenbis_superposition} as an optimization problem (typically, the optimization criterion is to cancel $v'_n(x,y)$ in as many unit cells as possible). This is how we obtained the eigenmodes presented in Fig. \ref{fig:eigenmodes} (discussed below).

\subsection{Asymptotic behavior at low and large gradients}

At large gradient strength, the gradient length $\ell_g$ becomes much smaller than any other geometrical length in the medium.
Therefore, there is no difference between a bounded and an unbounded domain because in both cases the eigenmode is localized over the length $\ell_g$ and its properties depend only on the local properties of the obstacle's boundaries such as curvature \cite{Grebenkov2018b,Almog2018a,Almog2019a,Moutal2019b}.

At low gradient strength, the diffusion effect becomes predominant over the gradient effect so that the $p$-pseudo-periodic numerical eigenmodes $v'_{p,n}$ are close to the $p$-pseudo-periodic Laplacian eigenmodes $u_{p,n}$. In the matrix product \eqref{eq:matrix_magn_bis} that \cc{represents} $\exp(-\tau\mathcal{B})$, the main effect of the matrices $G_{p\to p+q_0}$ is to ``move'' along the Bloch bands of the medium by projecting $u_{p,n}$ onto $u_{p+q_0,n}$, whereas the decay of the eigenmode is mainly caused by the diffusion matrices \cc{$\exp(-\tau_j \Lambda_{p_j})$}. Thus, as the gradient becomes infinitely small and the sampling of $q_x(t)$ becomes infinitely fine, the numerical eigenvalues of $\exp(-\tau \mathcal{B})$ tend to (we keep the notations of Eq. \eqref{eq:matrix_magn_bis}):
\begin{eqnarray}
\exp(-\tau &\mu'_n)\underset{g_x\to 0}{\approx} \cc{\exp(-\tau_N\lambda_{0,n}) \dots\exp(-\tau_2\lambda_{p_2,n}) \exp(-\tau_1\lambda_{p_1,n})}
\\
&\cc{\;\underset{\mathrm{fine~sampling}}{\rightarrow} \exp\left(- \int_0^\tau \lambda_{p(t)}\,\mathrm{d}t\right)}\\
&\cc{\quad \;\;\; = \exp\left(\frac{-\tau a_x}{2\pi}\int_0^{2\pi/a_x} \lambda_{p,n}\,\mathrm{d}p\right)\;,}
\end{eqnarray}
so that the true eigenvalues of the BT operator are given by (see Eq. \eqref{eq:true_eigenvalues}):
\begin{equation}\cc{
\mu_n(g_x=0^+) = \langle  \lambda_{p,n} \rangle\;,
\label{eq:asymptot_low_g}}
\end{equation}
where $\langle  \lambda_{p,n} \rangle$ is the average value of $\lambda_{p,n}$ over $p$:
\begin{equation}
\langle  \lambda_{p,n} \rangle = \frac{a_x}{2\pi} \int_0^{2\pi/a_x} \lambda_{p,n}\,\mathrm{d}p\;.
\end{equation}
\cc{Remarkably,} the continuous bands $\lambda_{p,n}$ of the Laplace operator (i.e., BT operator with $g_x=0$) are collapsed into their average values $\langle  \lambda_{p,n} \rangle$, $n=0,1,\ldots$ when $g_x$ is very small but non zero. Thus, the gradient term of the BT operator cannot be treated as a small perturbation of the Laplace operator because the limit $g_x \to 0$ is singular. This peculiar behavior is shown in Figs. \ref{fig:spectrum} and \ref{fig:spectrum_long} where the bands of the Laplace operator are drawn as vertical segments at $g_x=0$ and the asymptotic formula \eqref{eq:asymptot_low_g} is plotted as horizontal dashed lines. One can see that these dashed lines naturally extend the solid curves beyond our computational limit shown by thick black line (see Sec. \ref{section:numerical_eigenmodes}).
\cc{This effect is similar to Wannier-Stark localization for electrons in a crystal under a weak electric field \cite{Wannier1959a,Emin1987a}. In that case the linear potential term is real so that the spectrum is real. Energy states have the general form $\mu_{n,k}=\langle \lambda_{p,n}\rangle + k g_x a_x$ and form a dense ``double ladder'' on the real axis. In contrast, the imaginary potential that we study here produces a spectrum of the form $\mu_{n,k}=\langle \lambda_{p,n}\rangle + ik g_x a_x$. The indices $n$ and $k$ produce a ``ladder'' pattern along the real and imaginary axis, respectively, which results in a discrete spectrum.}

\begin{figure}[ht]
\centering
\includegraphics[width=0.49\linewidth]{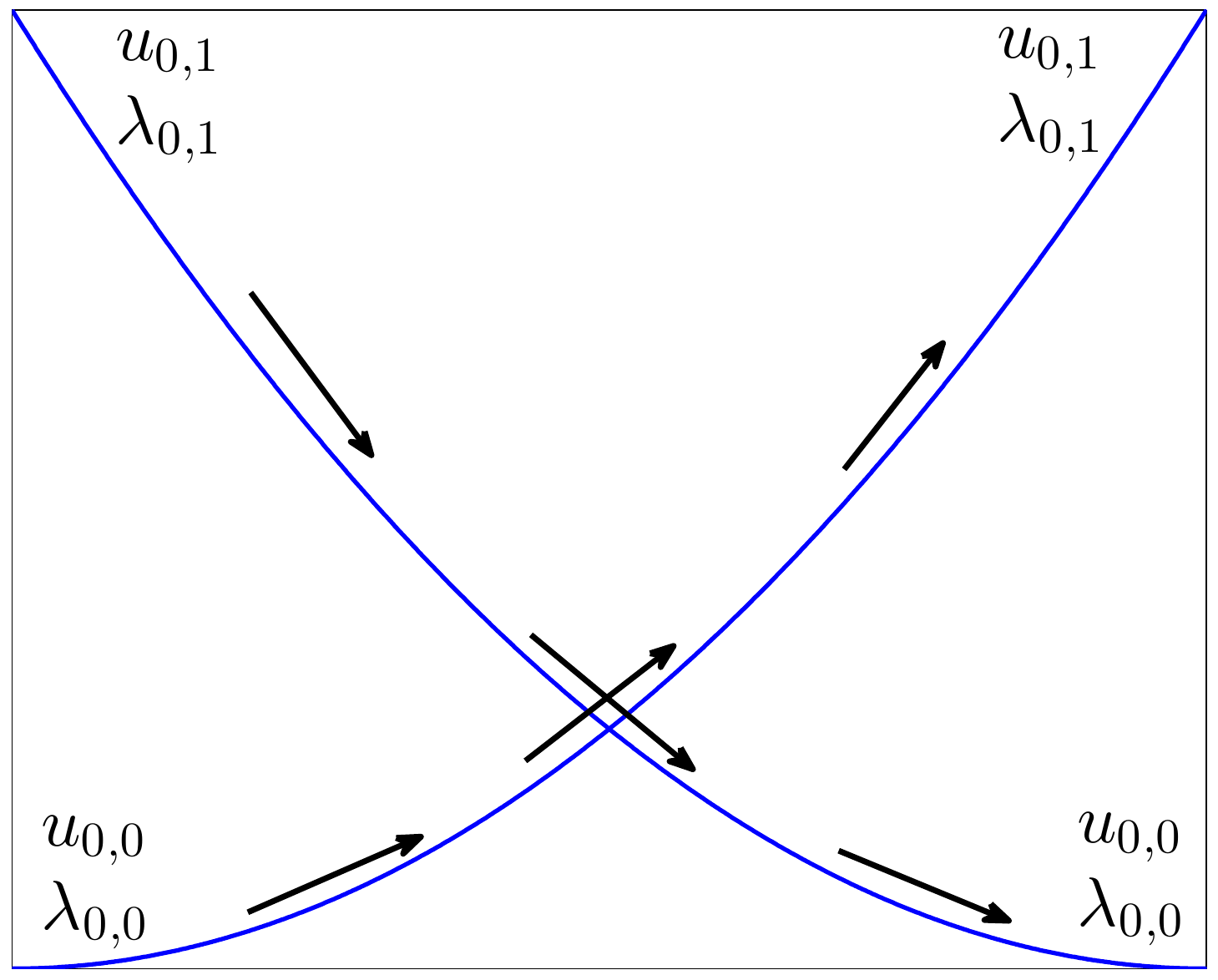}
\includegraphics[width=0.49\linewidth]{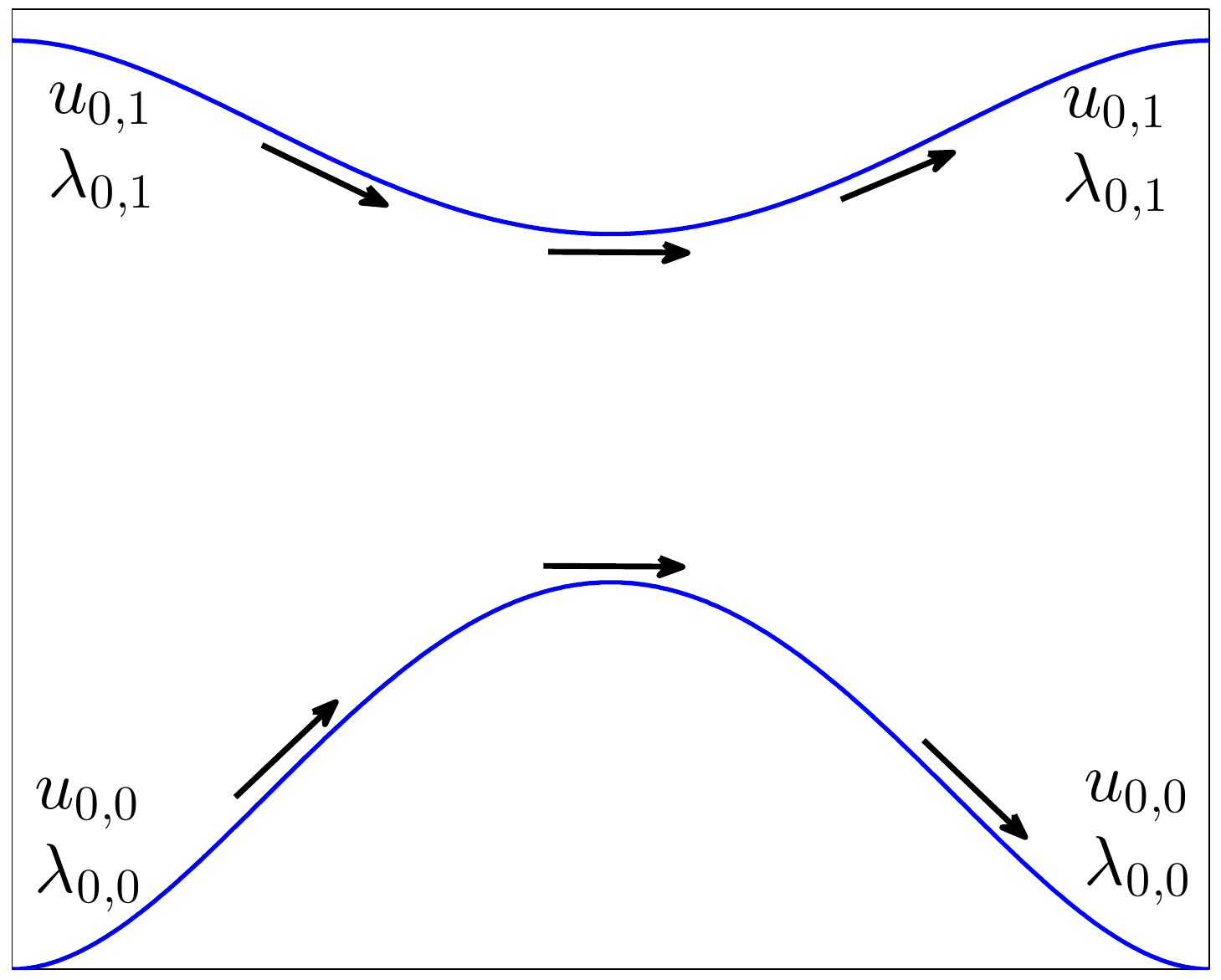}
\caption{The first two Laplacian bands for: (left) free space, where the bands cross each other; (right) a domain with obstacles such as the one considered throughout the text, with no crossing between bands. The arrows help to visualize the ``motion'' along bands created by small pulses $e^{iq_0 x}$ and show that $u_{p,n}$ cannot be an eigenmode of $\exp(-\tau \mathcal{B})$ for $g_x\to 0$ if bands $n$ and $n+1$ cross each other.}
\label{fig:band_free_obstacles}
\end{figure}

The above reasoning implicitly assumes that the Bloch bands of the medium are isolated, i.e. that by continuously increasing the wavenumber $p$, each band $u_{p,n}, \lambda_{p,n}$ continuously evolves without crossing any other bands and that the limit $p\to 2\pi/a_x$ yields the ``initial point'' $u_{0,n},\lambda_{0,n}$. The isolated versus non-isolated bands situations is illustrated on Fig. \ref{fig:band_free_obstacles}.
One can see that Eq. \eqref{eq:asymptot_low_g} is not applicable to the case where bands cross each other because the eigenmode $u_{0,0}$ is continuously transformed into $u_{0,1}$ by the successive narrow pulses.
As a consequence of the  ``avoided crossing'' theorem of von Neumann and Wigner \cite{Neumann1929a}, the Bloch bands cannot cross if the unit cell $\Omega_1$ is irreducible along $x$. In contrast, there does not exist any irreducible unit cell for free space so that the formula \eqref{eq:asymptot_low_g} is not applicable, which is consistent with emptiness of the spectrum of the BT operator \cite{Grebenkov2017a}.

\subsection{Bifurcation points}

The plot of the eigenvalues $\mu_n$ as functions of the gradient reveals some bifurcation (or ``branching'') points, where two eigenvalues with the same imaginary part and different real parts branch into two eigenvalues with the same real part and different imaginary parts. This mathematical phenomenon was first shown by Stoller \textit{et al.} for the BT operator in an interval with Neumann boundary condition \cite{Stoller1991a}. Here, we shall explain how these bifurcation points are related to the localization of eigenmodes.

If the unit cell $\Omega_1$ is symmetric under the parity transformation $x\to -x$, then the BT operator is invariant under parity and conjugation:
\begin{equation}
\left(-\nabla^2-ig(-x)\right)^* = -\nabla^2 - igx\;.
\end{equation}
Therefore, if $v_n(x,y,z)$ is an eigenmode of $\mathcal{B}$ with eigenvalue $\mu_n$, then $v_n^*(-x,y,z)$ is an eigenmode of $\mathcal{B}$ with eigenvalue $\mu_n^*$. 
This leads to two different situations.

(i) When $\mathrm{Im}(\mu_n)=kg_xa_x/2$ for a given integer $k$, then $\mu_n^*=\mu_n - ikg_xa_x$ so that it is actually the same eigenvalue but translated to another unit cell (see Eq. \eqref{eq:true_eigenvalues}). In general the eigenvalue $\mu_n$ is simple so that \cc{$v_n^*(-x,-y,-z)=v_n(x-ka_x,y,z)$}, which means that the eigenmode $v_n$ has a symmetric shape (its real part is symmetric and imaginary part antisymmetric), and is centered on the middle of the unit cell (if $k$ is even) or on the boundary between two unit cells (if $k$ is odd). From the spectrum and the corresponding eigenmodes shown in Figs. \ref{fig:spectrum} and \ref{fig:eigenmodes}, one can see that this corresponds to (a1), (b1), (a4), (b4) where $k$ is odd (the eigenmodes are centered on the spacing between two obstacles at $x=a_x/2$) and to (a2), (a3) where $k$ is even (the eigenmodes are centered on the obstacle at $x=0$).

(ii) When $\mathrm{Im}(\mu_n)\neq kg_xa_x/2$, then $\mu_n^*$ does not belong to the same family of eigenvalues as $\mu_n$, i.e., there is an integer $n'$ such that $\mu_{n'}=\mu_n^*$.
Then one has $v_{n'}(x,y,z)=v^*_n(-x,-y,-z)$: the shape of the eigenmode $v_{n'}$ is the same as that of the eigenmode $v_n$ after parity transformation. One can see that this situation corresponds to eigenmodes (b2) and (b3) on Fig. \ref{fig:eigenmodes}, where $\mu_2$ and $\mu_3$ form a complex conjugate pair and the eigenmodes  $v_2$ and $v_3$ are localized on the left and right sides of the obstacle, respectively.

The transition between these two situations creates a branching point where two eigenvalues coalesce to form a complex conjugate pair. Note that, in contrast with Hermitian operators, the corresponding eigenmodes also coalesce at the branching point. This is supported by the fact that the eigenmodes (a2) and (a3) are very close to each other in Fig. \ref{fig:eigenmodes}, as they were plotted not far from their branching point (see vertical dashed lines on Fig. \ref{fig:spectrum}). This behavior and the comparison with Hermitian operators are illustrated with a simple matrix model in \ref{section:bifurcation_simple}.

If the domain is not invariant by parity symmetry, then the eigenmodes still localize at large gradients but there is no longer a sharp transition between ``delocalized'' and ``localized''. Note, however, that there are still branching points in the spectrum if one considers complex values of the gradient (not shown). The branching point with the smallest absolute value defines a convergence radius outside of which low-gradients asymptotic expansions would fail because of the non-analyticity of the bifurcation. The finite radius of convergence of the cumulant expansion in terms of $bD_0$ was investigated in \cite{Froehlich2006a} for a one-dimensional model in the limit of short gradient pulses (infinite gradient $g_x$ and $\delta\to 0$). In that case, the gradient pulse effectively applies a $e^{iq_x x}$ phase pattern across the domain and the decay of the magnetization is caused by the ``blurring'' of this pattern due to diffusion. In this regime, the signal can be described by a formula similar to Eq. \eqref{eq:structure_factor_pseudo} and is an analytic function of $bD_0=q_x^2 D_0 \Delta$. As the authors explain, the finite convergence radius of the cumulant expansion is merely caused by the Taylor series of the logarithm function and related to the smallest (in absolute value) complex value of $bD_0$ for which the signal is zero. In contrast, we argue that the non-analyticity of the BT spectrum at finite gradient strength should intrinsically restrict the range of applicability of low-gradient expansions in all non-trivial domains.

\section{Conclusion}
 \label{section:conclusion}

The aims of this paper were twofold. One one hand, we have developed a numerical method to solve efficiently the BT equation in periodic media. By replacing the continuous integrated gradient profile $(q_x(t),q_y(t),q_z(t))$ by a step function, this equation can be solved in a single unit cell by spectral methods, allowing for very fast and accurate computations, especially at high gradients. This is of significant practical importance for numerical simulations in dMRI as periodic media can describe a wide range of unbounded media if the spin-bearing particles visit at most a few unit cells during the duration of the gradient sequence. Numerical simulations in a simple model (array of circular obstacles) reveal diverse regimes (effective free diffusion, motional narrowing, localization, diffusion-diffraction) for the transverse magnetization and the signal. The spacing between obstacles along the gradient direction was shown to be a crucial parameter by comparing results for the gradient in the horizontal direction and in the diagonal direction. In particular, the competition between this spacing and the gradient length controls the emergence of the localization regime at high gradient strength.

On the other hand, this numerical method allowed us for the first time to compute the eigenmodes and eigenvalues of the BT operator in periodic media. The non-Hermitian character of the BT operator led to several qualitatively interesting phenomena. The most spectacular one is that its spectrum is discrete even though periodic domains are infinite. More precisely, even a very small gradient term causes the continuous Bloch bands of the Laplace operator to collapse on their average values. 
One sees therefore that the low gradient limit is singular in periodic domains that urges for re-thinking conventional perturbative results that are still dominant in the field of dMRI (see the review \cite{Grebenkov2007a}).
As the gradient increases, the BT eigenmodes start to localize near the obstacles of the domain and we have shown that this localization is associated to bifurcation points in the spectrum. Moreover, the emergence of this localization regime  corresponds to a strong deviation in the measured signal compared to the freely diffusing case and related perturbation formulas. Mathematically, the bifurcation points create  non-analyticity of the spectrum that prevents the use of low-gradient asymptotic expansions beyond some critical value of the gradient, hence the sharp difference in signal decay between low gradients and high gradients. Several mathematical questions remain open, among which the existence of the eigenmodes of the BT operator in general (non trivial) domains and their completeness outside of the set of bifurcation points are probably the most important.

\section*{Acknowledgements}

The authors thank B.~Helffer for very fruitful discussions.

\clearpage

\appendix

\section{Numerical results for a short-gradient pulses sequence}
\label{section:NPA}

\begin{figure}[t]
\centering
\includegraphics[width=0.24\linewidth]{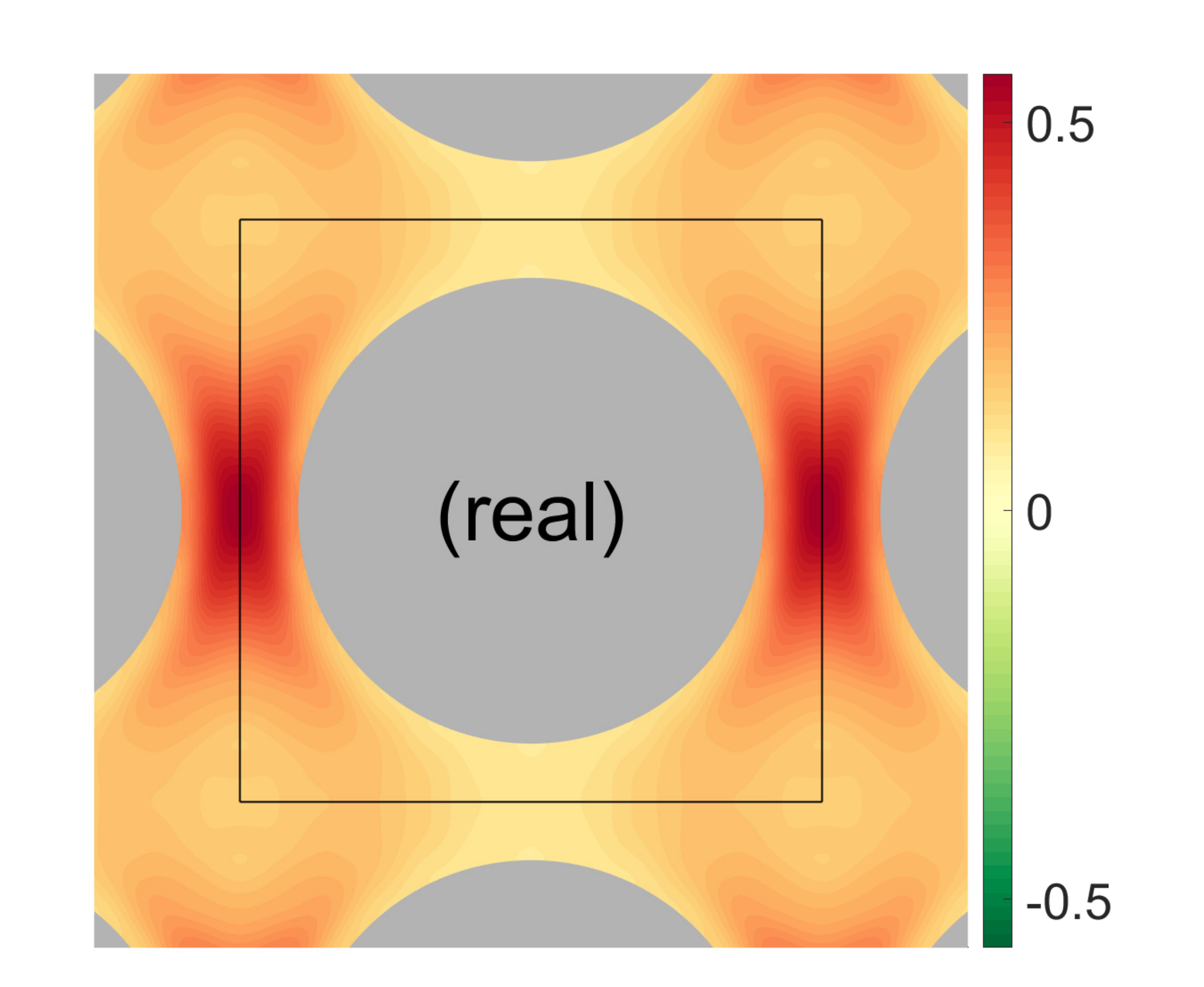}
\includegraphics[width=0.24\linewidth]{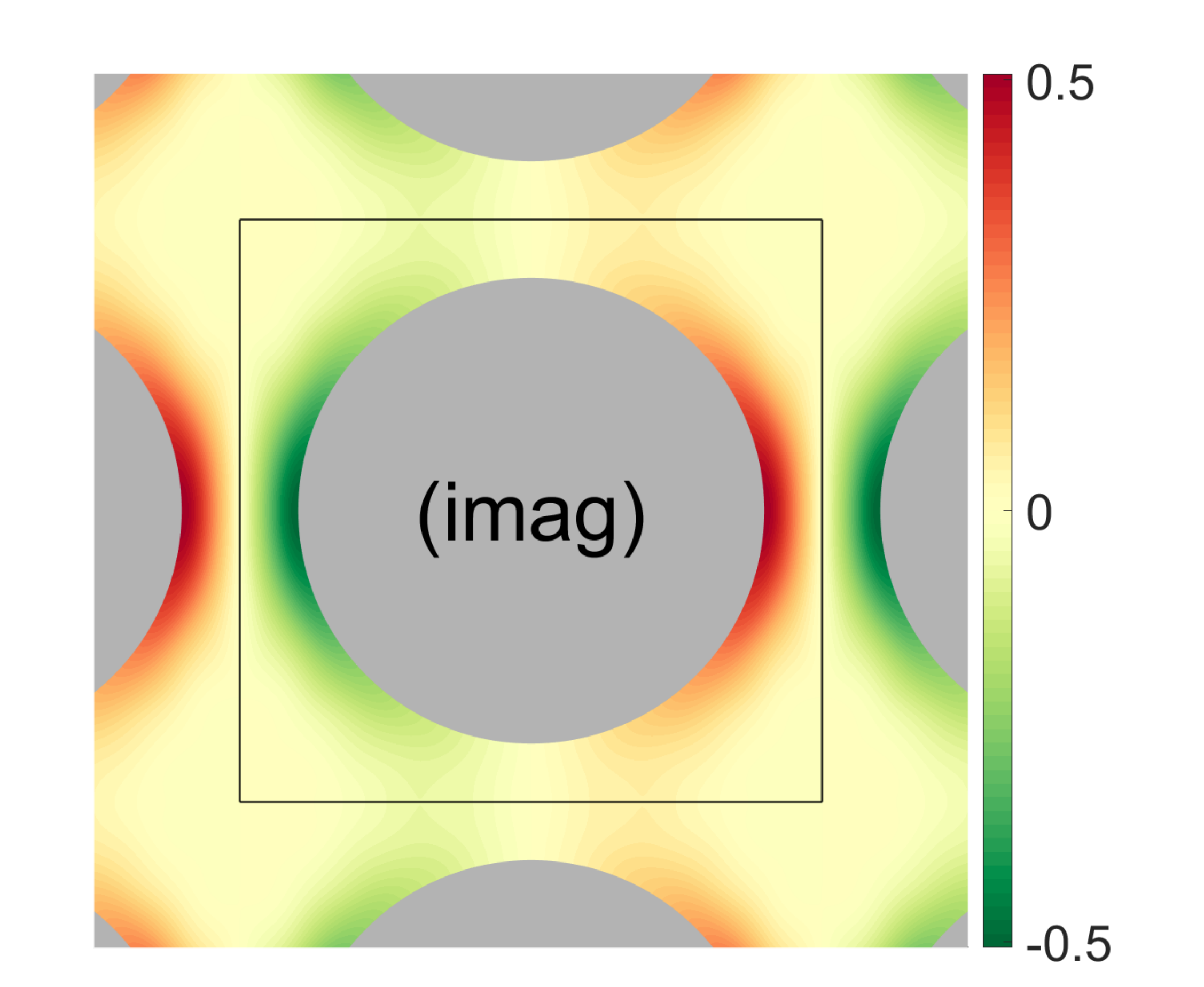}
\includegraphics[width=0.24\linewidth]{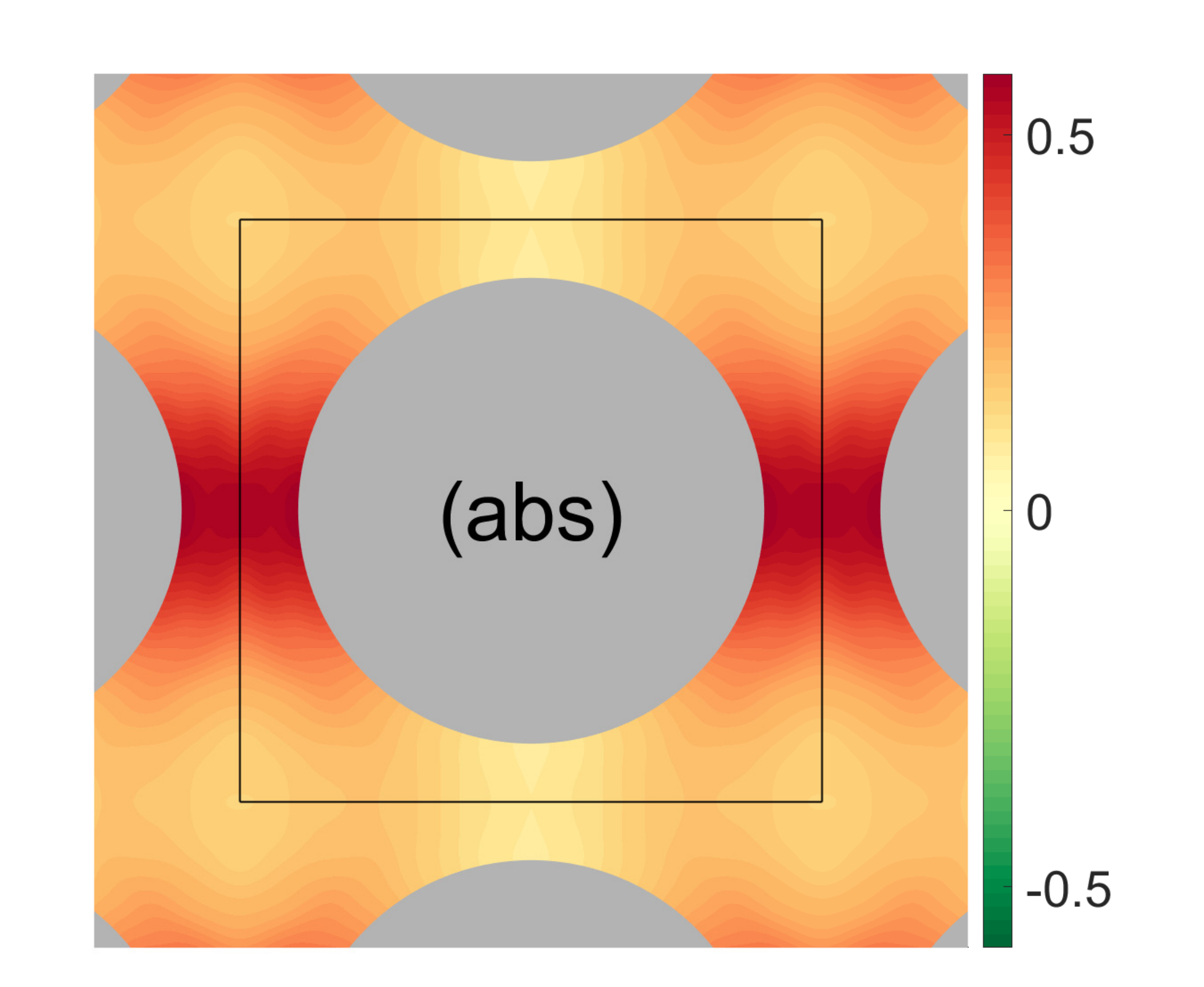}
\includegraphics[width=0.24\linewidth]{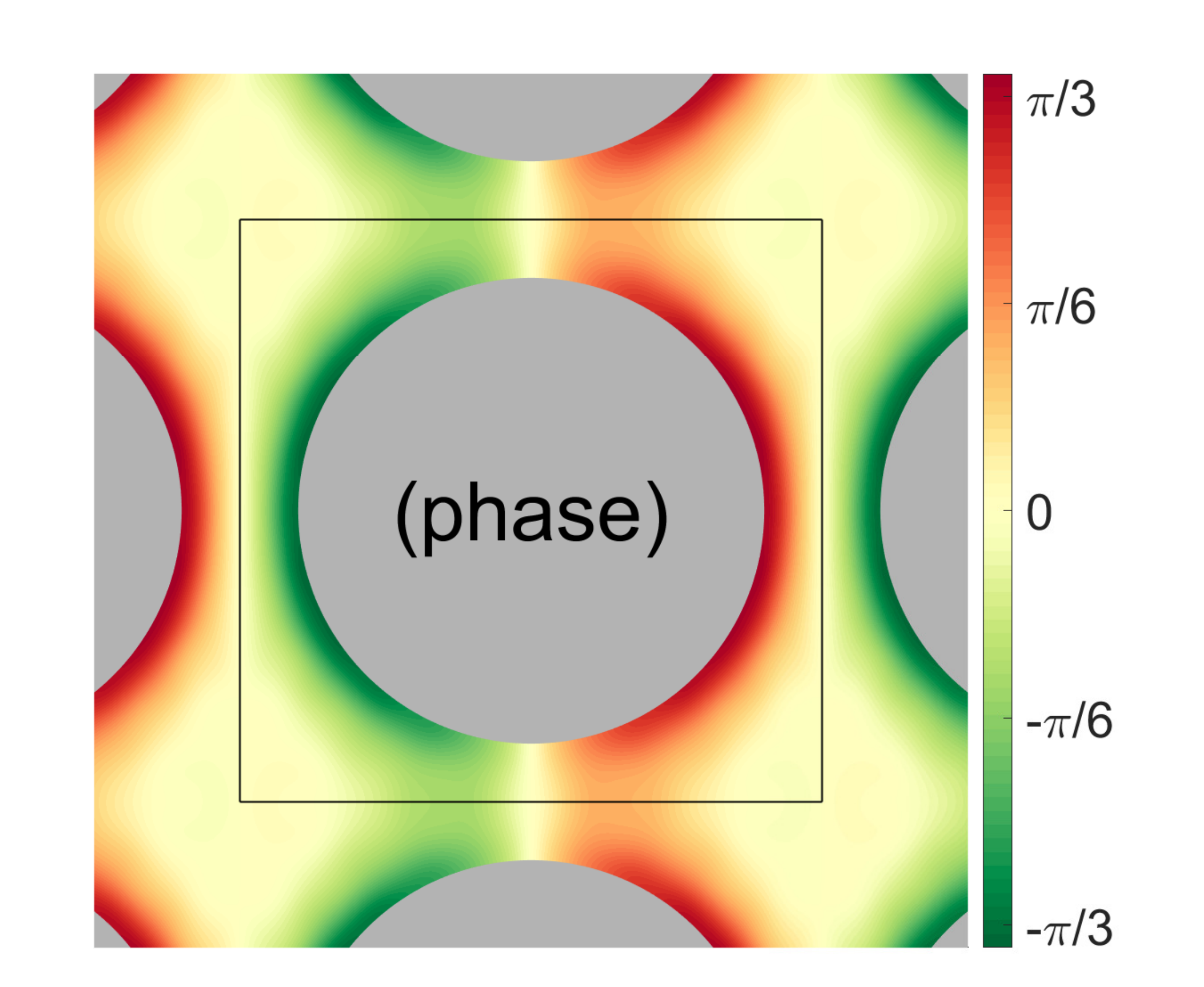}
\noindent\rule[5pt]{\linewidth}{0.4pt}
\includegraphics[width=0.24\linewidth]{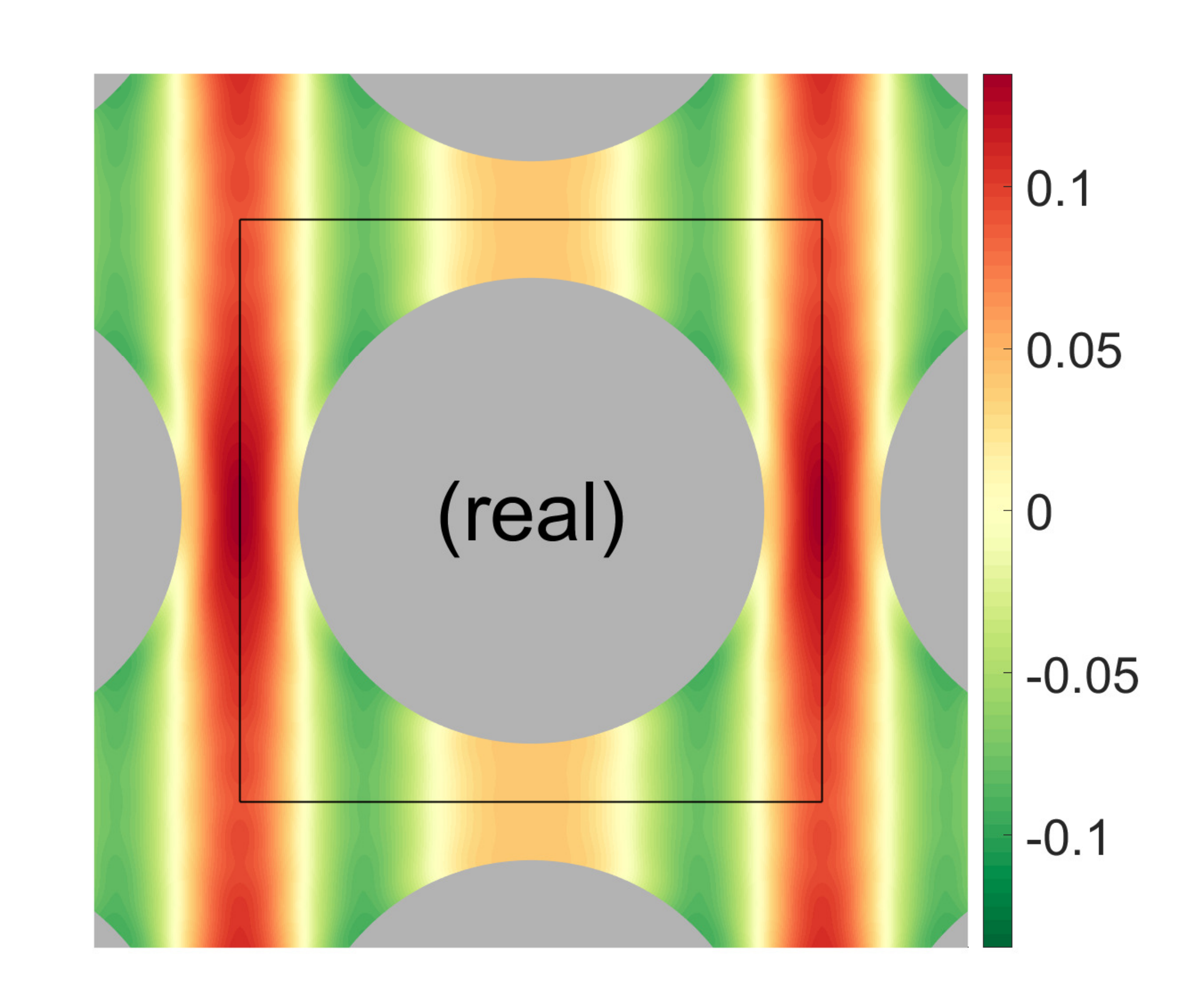}
\includegraphics[width=0.24\linewidth]{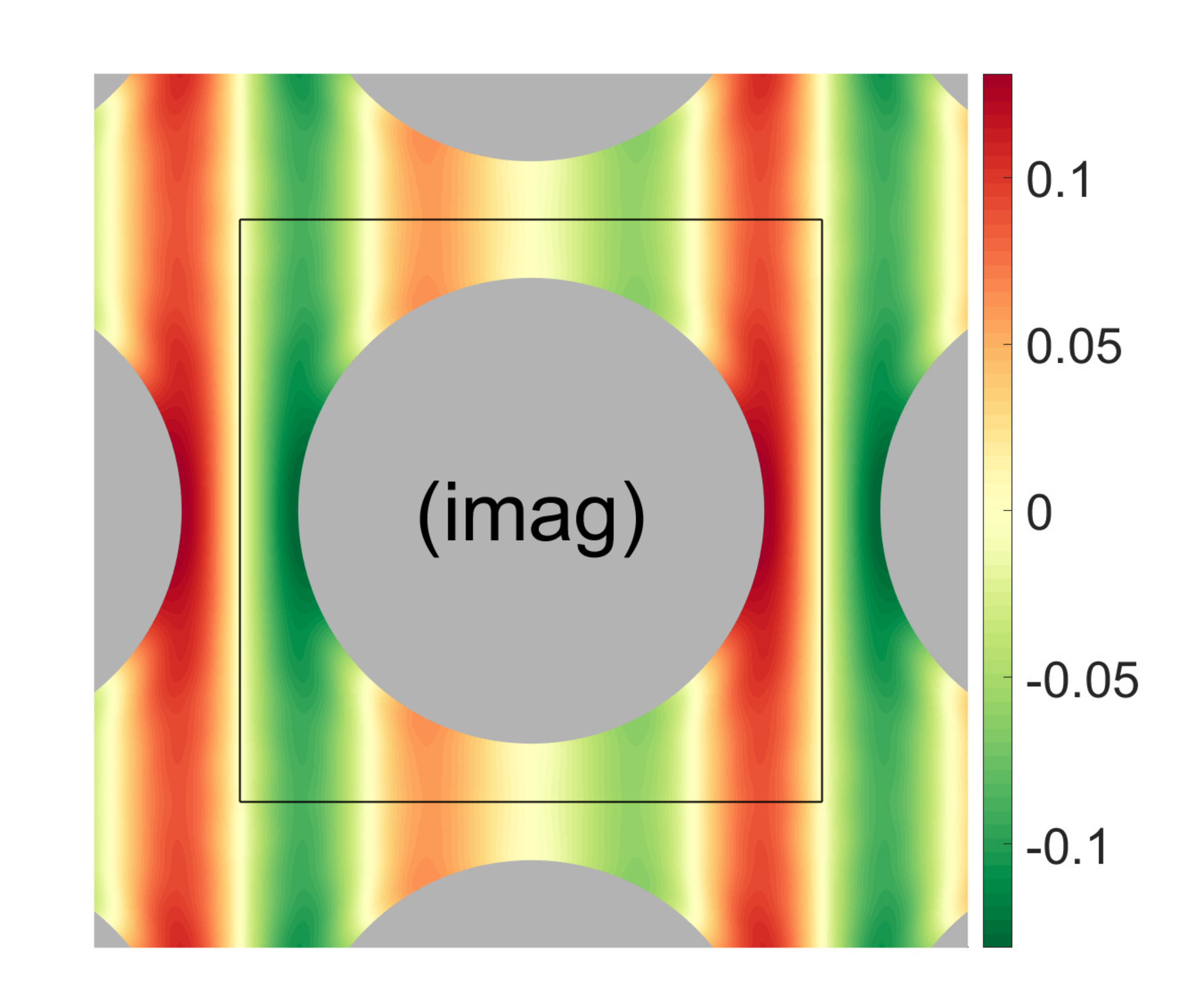}
\includegraphics[width=0.24\linewidth]{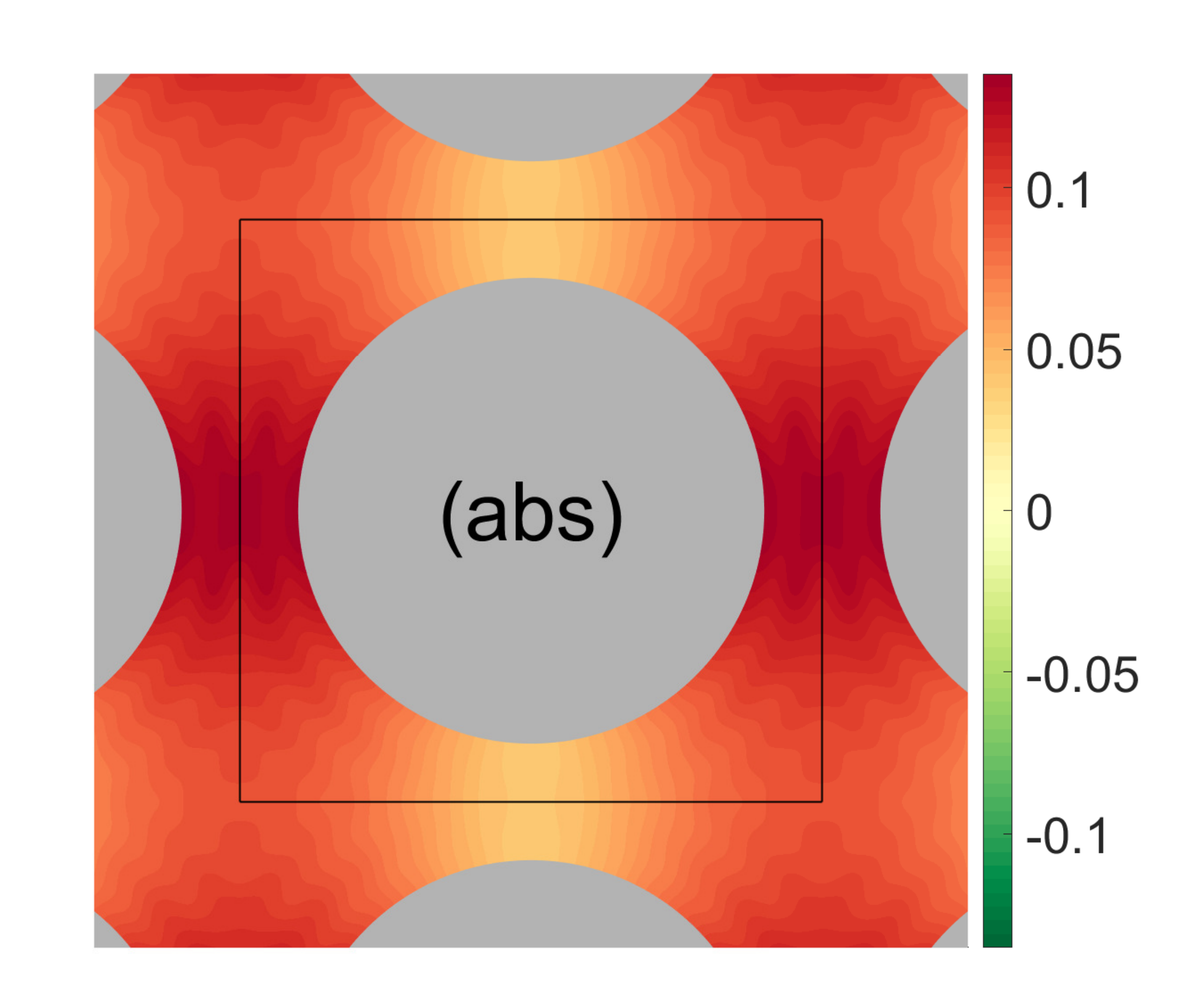}
\includegraphics[width=0.24\linewidth]{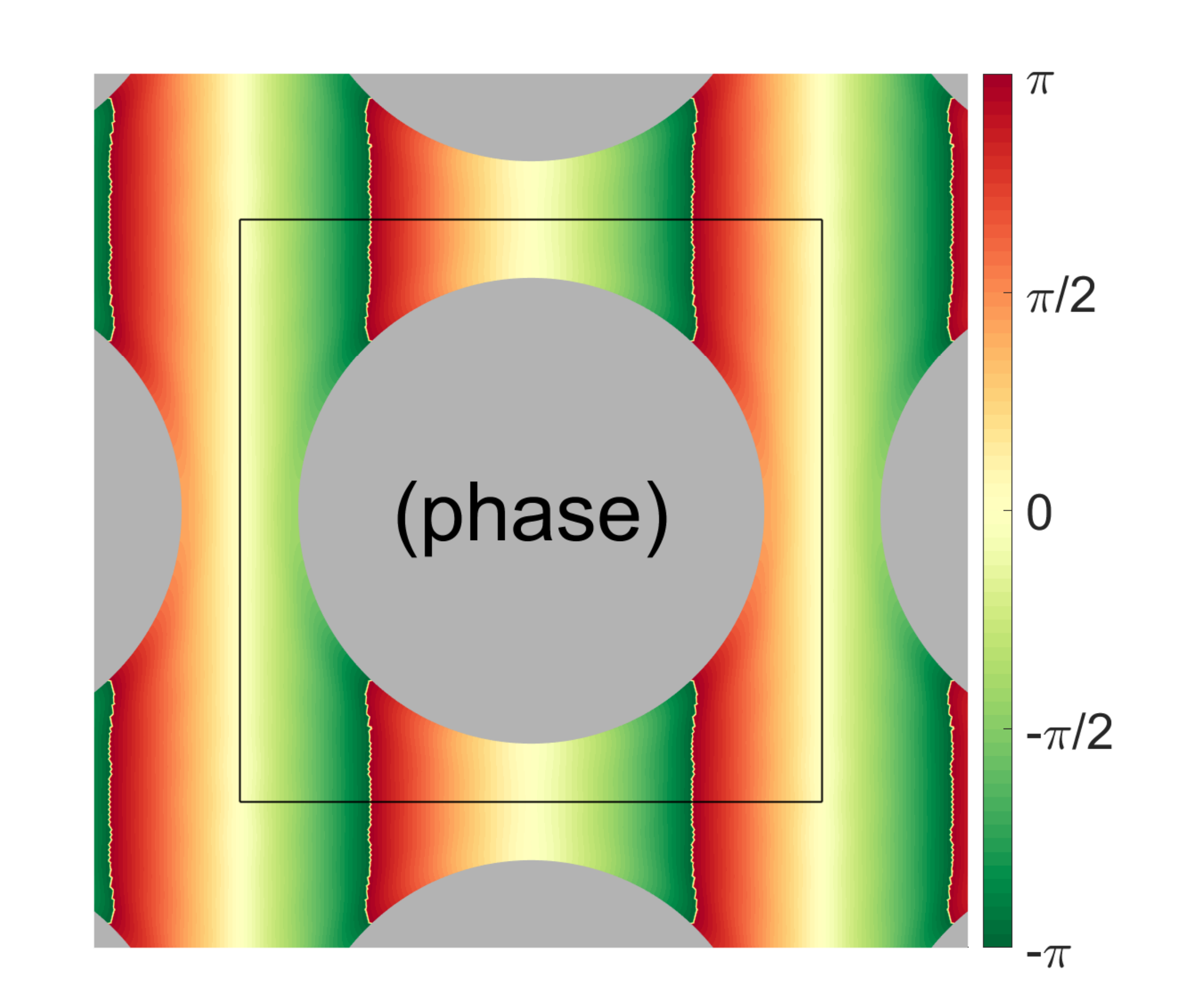}
\noindent\rule[5pt]{\linewidth}{0.4pt}
\includegraphics[width=0.24\linewidth]{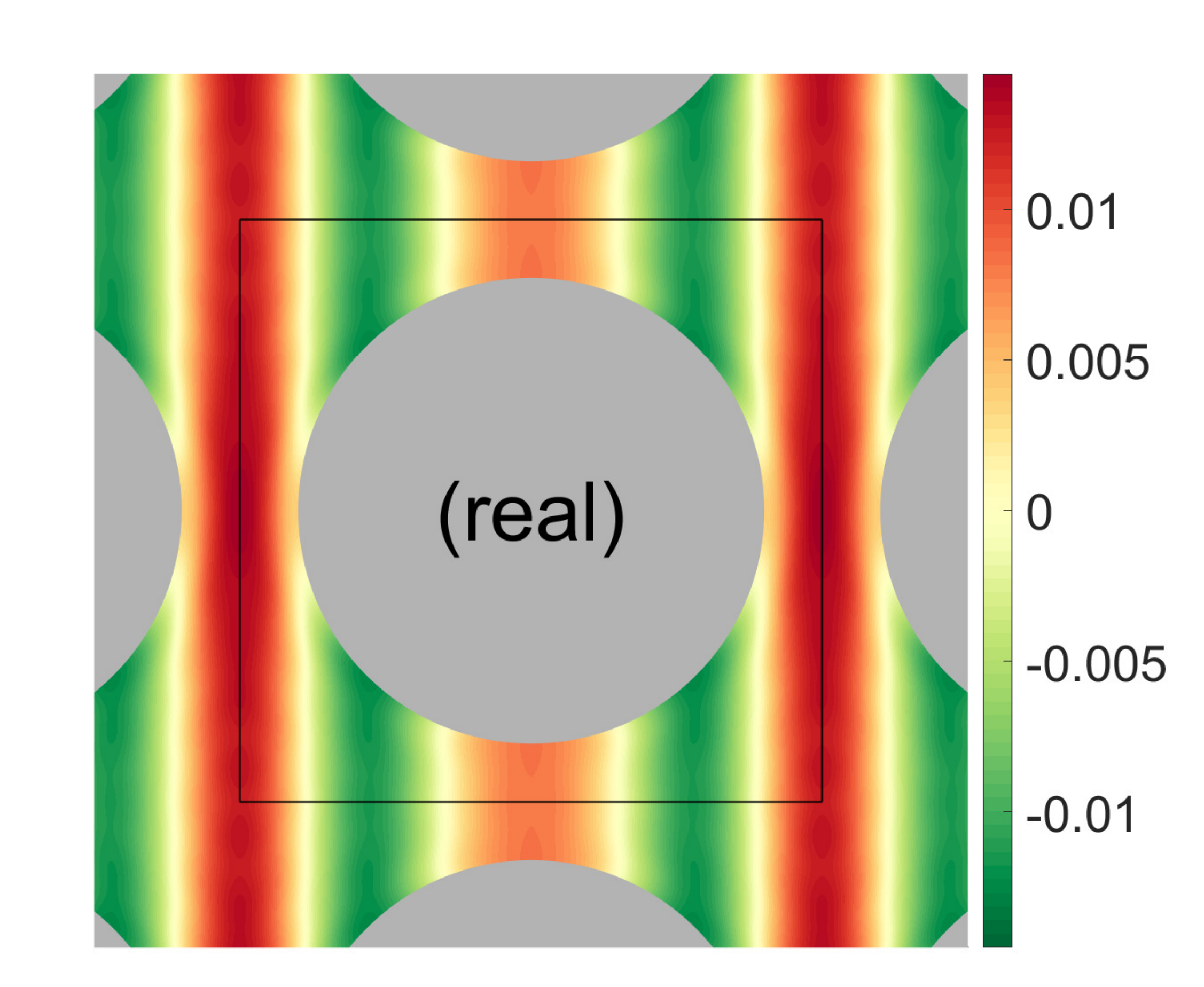}
\includegraphics[width=0.24\linewidth]{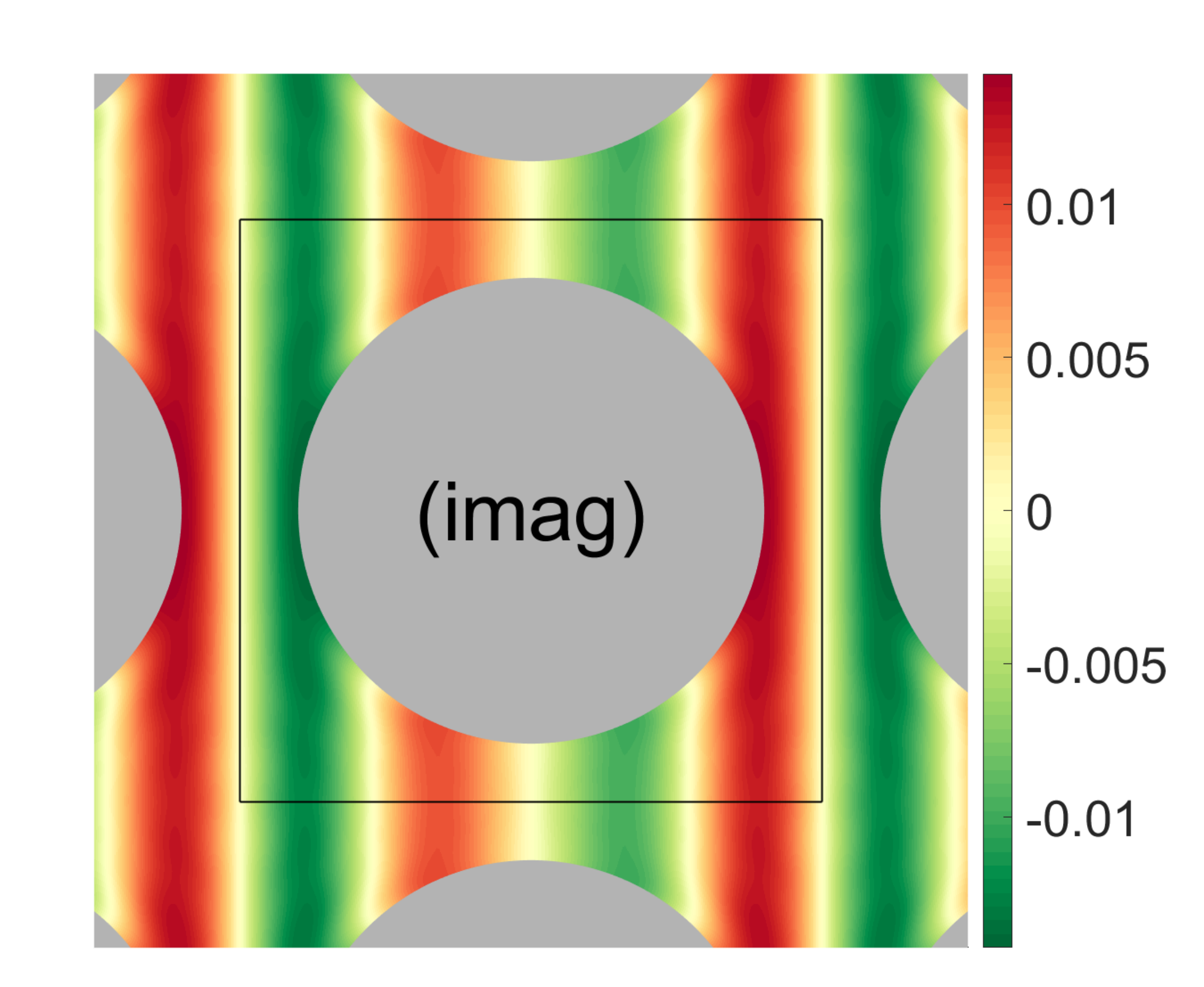}
\includegraphics[width=0.24\linewidth]{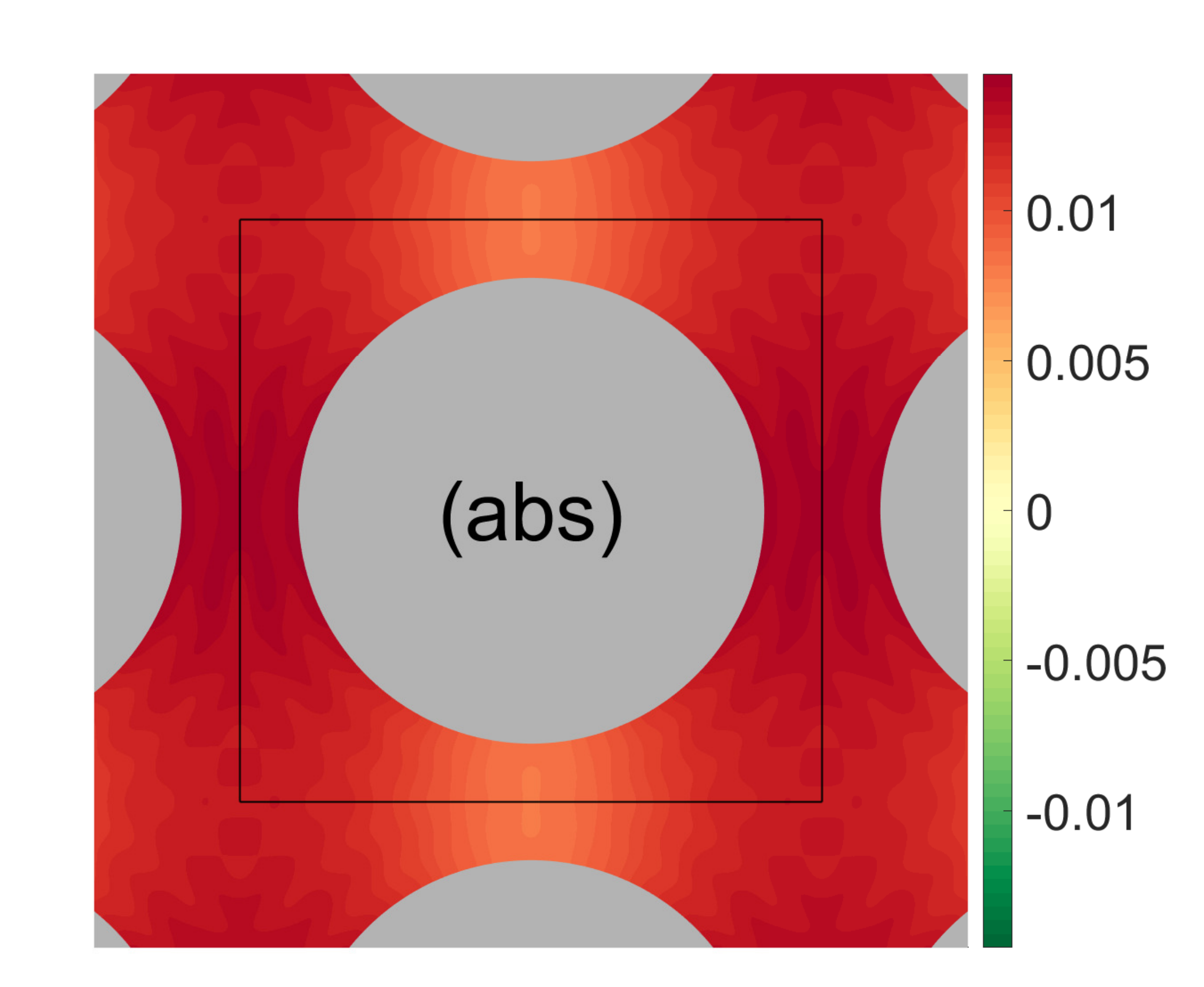}
\includegraphics[width=0.24\linewidth]{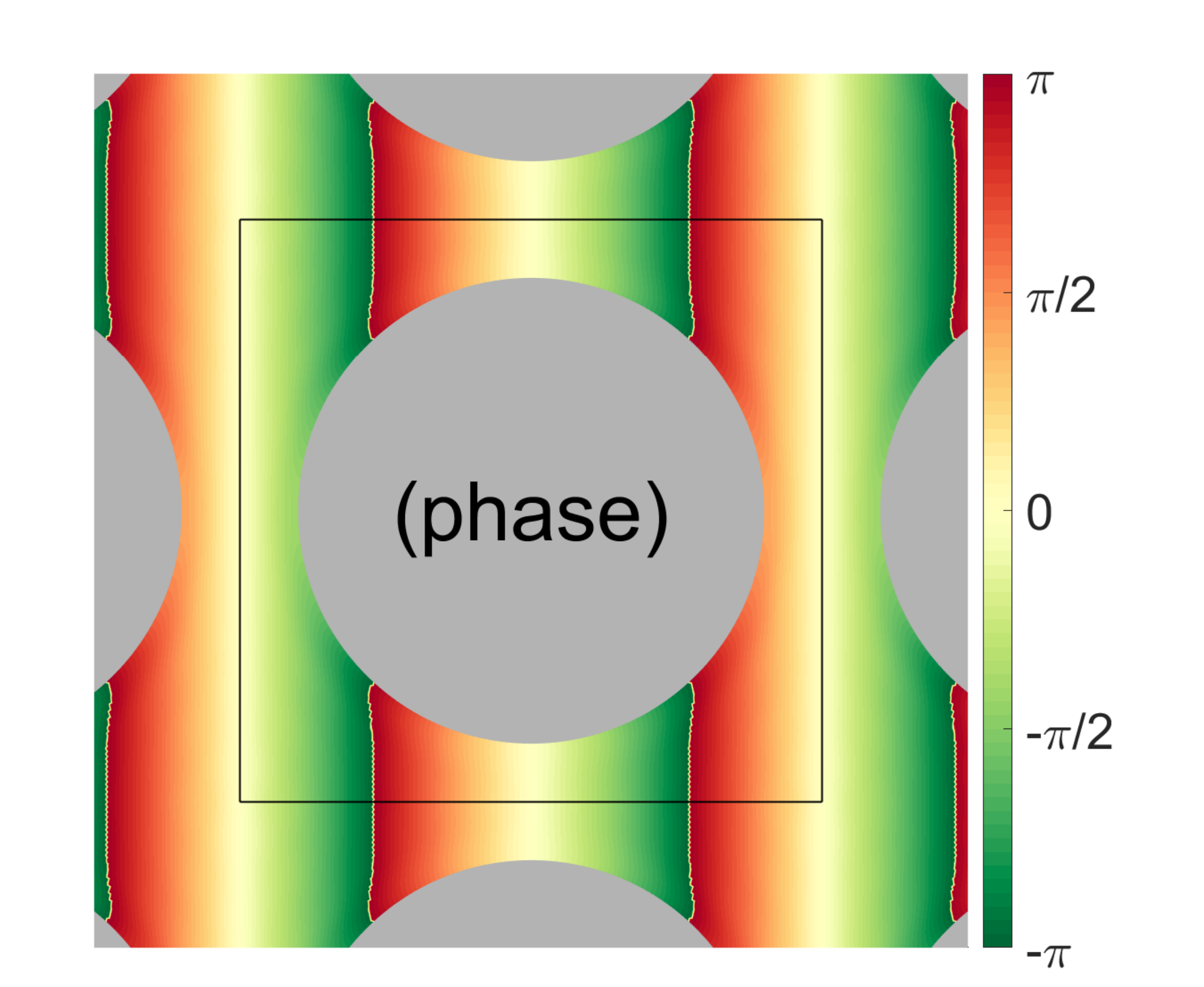}
\caption{Plot of the magnetization (real and imaginary part, absolute value and phase) after a narrow-pulse sequence. The gradient is in the left to right horizontal direction. The black square indicates the unit cell in which the computation was performed. For all figures, $R/a=0.4$, and we kept a fixed value $qa=14\pi/3$. The corresponding normalized signal is shown on the left panel of Fig. \ref{fig:signal_q}.
(top) $\ell_\Delta/a=0.1$; 
(middle) $\ell_\Delta/a=0.3$; 
(bottom) $\ell_\Delta/a=1.0$. 
}
\label{fig:NPA_magn_horiz}
\end{figure}

In this appendix, we present and discuss the behavior of the magnetization and the signal for a short gradient pulse sequence (see Fig. \ref{fig:sampling} with $\delta\to 0$).
As in the main text, we consider a 2D square lattice of impermeable circular obstacles with radius $R$ and lattice step $a$.
In that case, there are three relevant dimensionless quantities: $R/a$, $qa$, and $\ell_\Delta/a$, where $q$ is the weight of the narrow gradient pulses and $\ell_\Delta=\sqrt{D_0 \Delta}$ is the diffusion length traveled by spin-bearing particles during the time $\Delta$ between two pulses. Diffusion in free space would yield a uniform magnetization
\begin{equation}
m=\exp(-bD_0)=\exp\left(- q^2\ell_\Delta^2 \right)\;.
\end{equation}
Note that the short-gradient pulse limit corresponds to $\ell_g\to 0$, $\ell_\delta \to 0$ and $\ell_\delta/\ell_g \to 0$ so that the mechanism behind the attenuation of the signal is different from the extended-gradient pulse situation presented in the main text, that was controlled by the competition between $\ell_g/a$, $\ell_\delta/a$, and $R/a$. Correspondingly, the magnetization and the signal exhibit new behaviors as we shall now explain.
Let us first assume that the gradient is along $x$, i.e. in the horizontal direction. The first gradient pulse multiplies the magnetization in the medium by $e^{iqx}$, then diffusion ``blurs'' this pattern and the second pulse multiplies the magnetization by $e^{-iqx}$. As a consequence, the magnetization shows two very different patterns depending on the duration of the diffusion step.

(i) If the diffusion step duration is short so that $q \ell_\Delta \lesssim 1$, there is little ``blurring'' of the phase pattern by diffusion. Just before the second pulse, the magnetization is close to $e^{iqx}$ but with a lower amplitude, because spins with different phases are mixed by diffusion: the average phase at a given position remains the same but dephasing of spins causes attenuation of magnetization. Close to obstacles, the phase pattern $e^{iqx}$ is modified because it is ``cut'' by the boundaries. For this reason, the attenuation of the magnetization is less pronounced and the resulting phase of spins is modified as well. Thus, right after the second pulse, the magnetization is nearly uniform except for boundary regions where the magnetization is more intense (so-called ``edge enhancement'', see \cite{Swiet1995a}) and has a significative imaginary part (after integration, this imaginary part cancels so that the signal is real).

(ii) if the diffusion step duration is long so that $q\ell_\Delta \gg 1$, the phase pattern is completely blurred by diffusion. However, the magnetization is not uniform because of the $p$-pseudo-periodicity created by the gradient pulse, where $p=q\quad (\mathrm{mod}\; 2\pi/a)$. In terms of Laplacian eigenmodes, all $u_{n,p}$ with $n>1$ relax and the magnetization is close to $u_{0,p}$ (with attenuation) after the diffusion step (and before the second pulse). Therefore, after the second pulse, the magnetization is close to $u_{0,p}e^{-iqx}$, that is somewhat similar to $e^{-i\tilde{q}x}$, where $\tilde{q}$ denotes here the multiple of $2\pi/a$ that is the closest to $q$.

These two regimes are shown on Fig. \ref{fig:NPA_magn_horiz} for the gradient in the horizontal direction and $qa=14\pi/3$, where the top panel corresponds to $\ell_\Delta/a=0.1$, i.e. $q\ell_\Delta=1.5$ (case (i)), and the bottom panel corresponds to $\ell_\Delta/a=1.0$, i.e. $q\ell_\Delta=15$ (case (ii)). The middle panel corresponds to $\ell_\Delta/a=0.3$, i.e. $q\ell_\Delta=4.4$, that is an intermediate case between (i) and (ii).

\begin{figure}[t]
\centering
\includegraphics[width=0.24\linewidth]{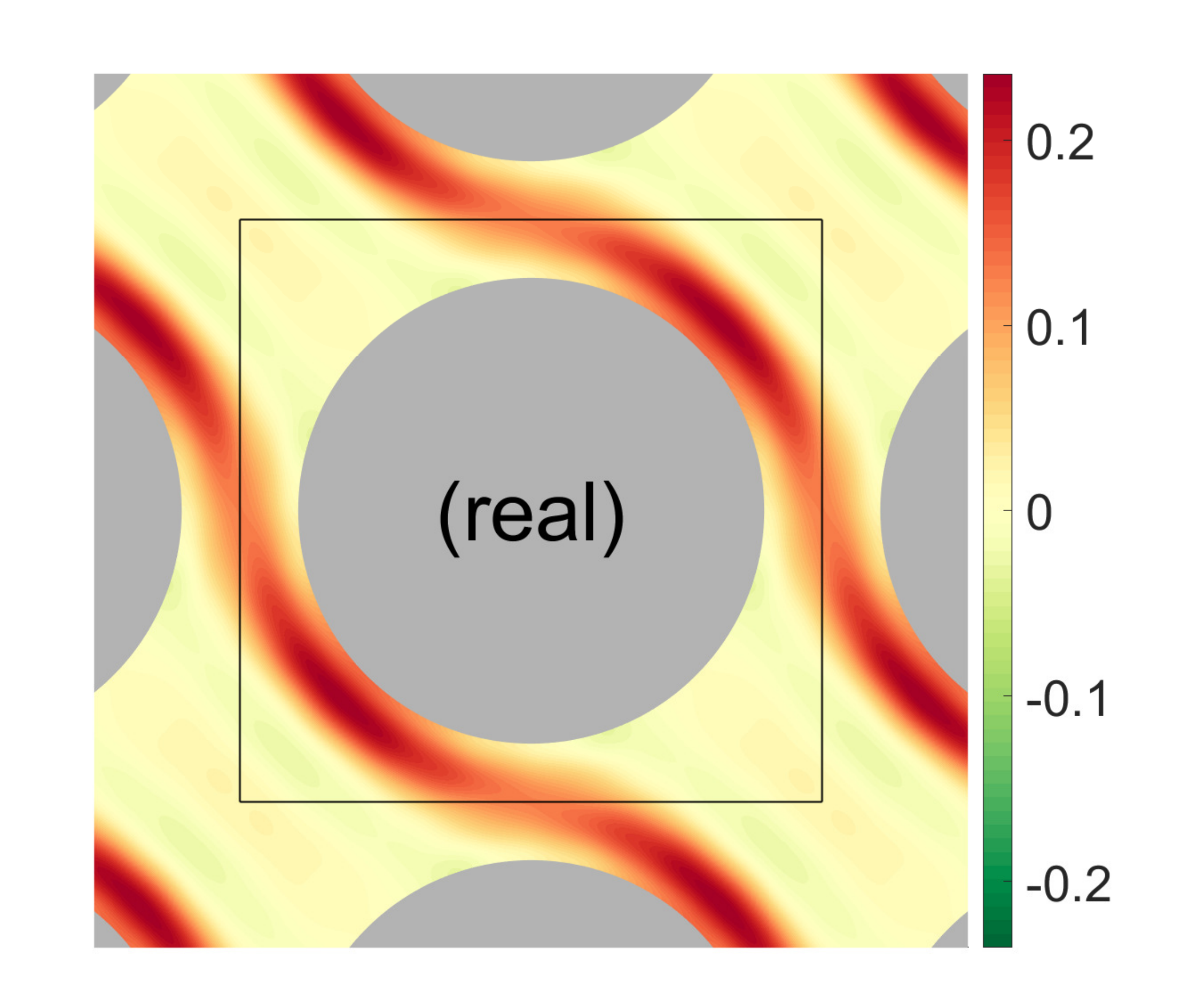}
\includegraphics[width=0.24\linewidth]{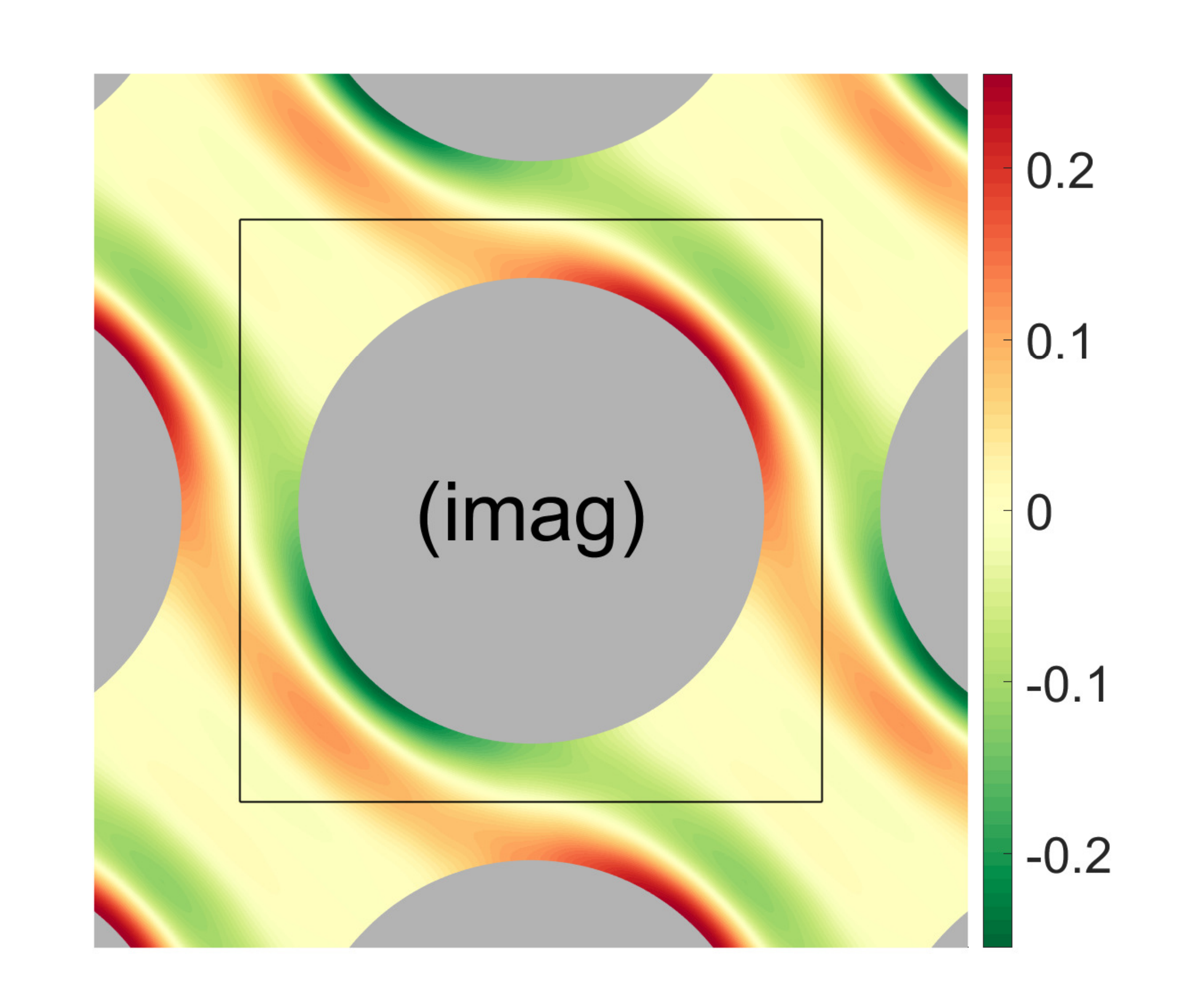}
\includegraphics[width=0.24\linewidth]{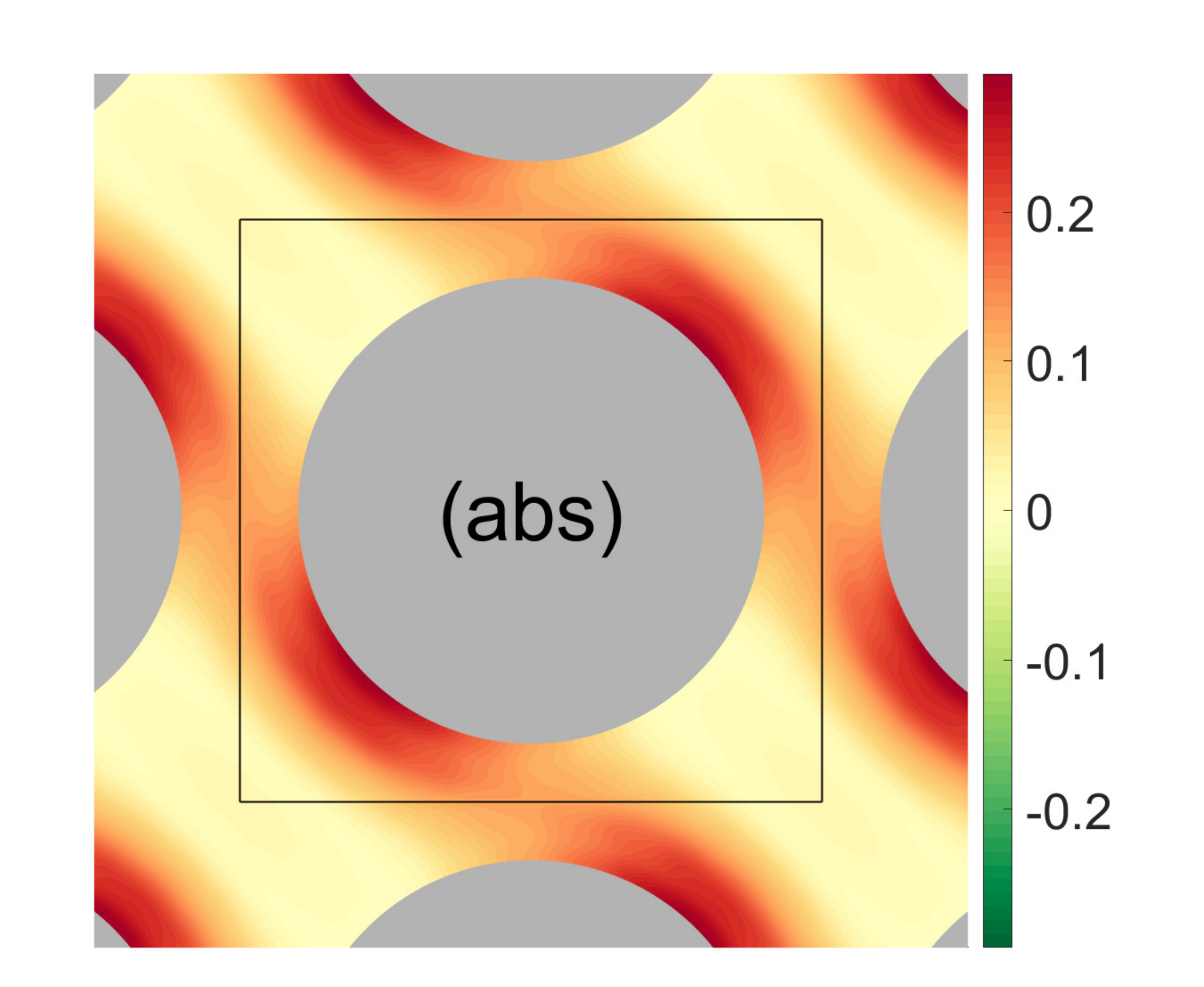}
\includegraphics[width=0.24\linewidth]{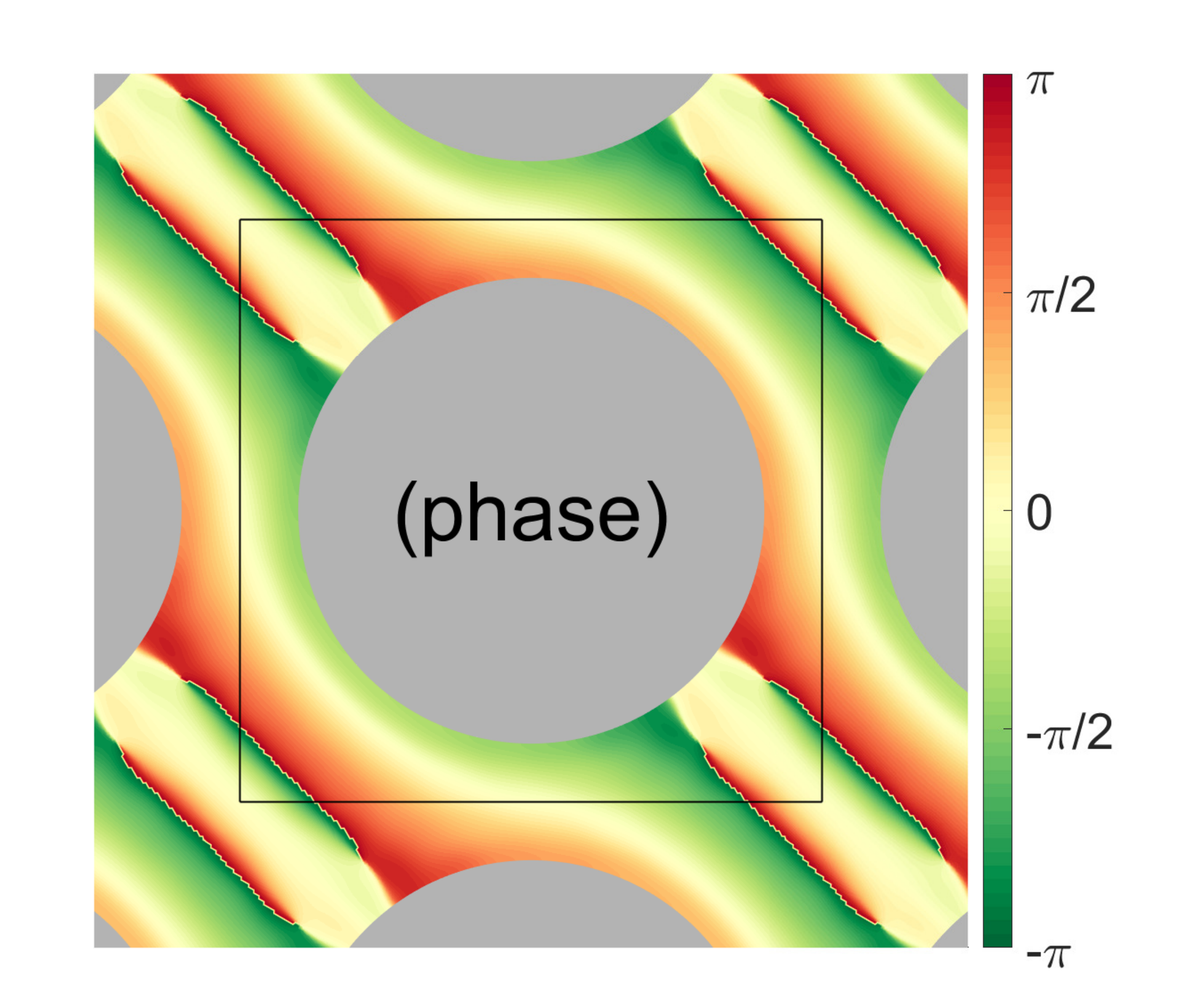}
\noindent\rule[5pt]{\linewidth}{0.4pt}
\includegraphics[width=0.24\linewidth]{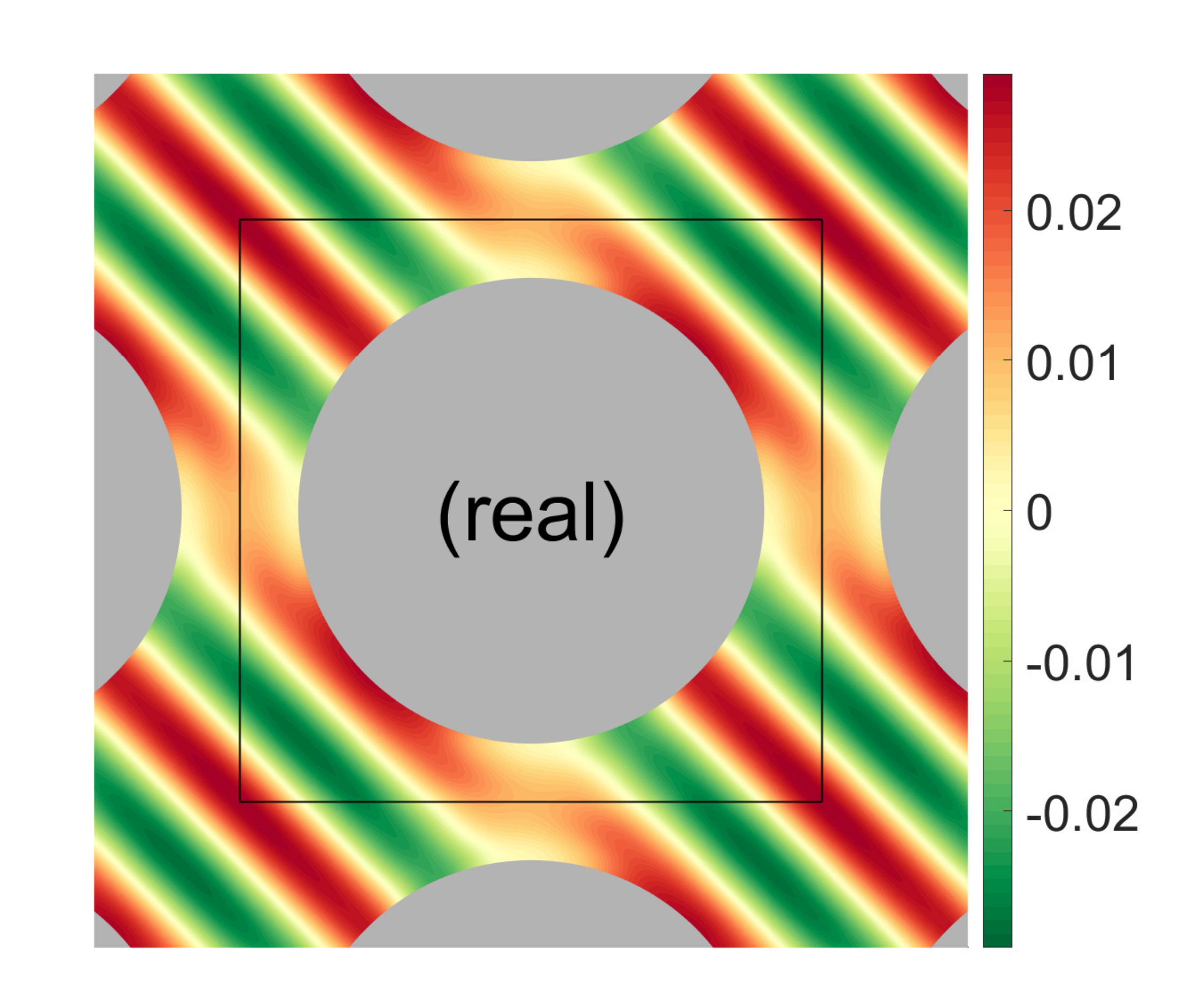}
\includegraphics[width=0.24\linewidth]{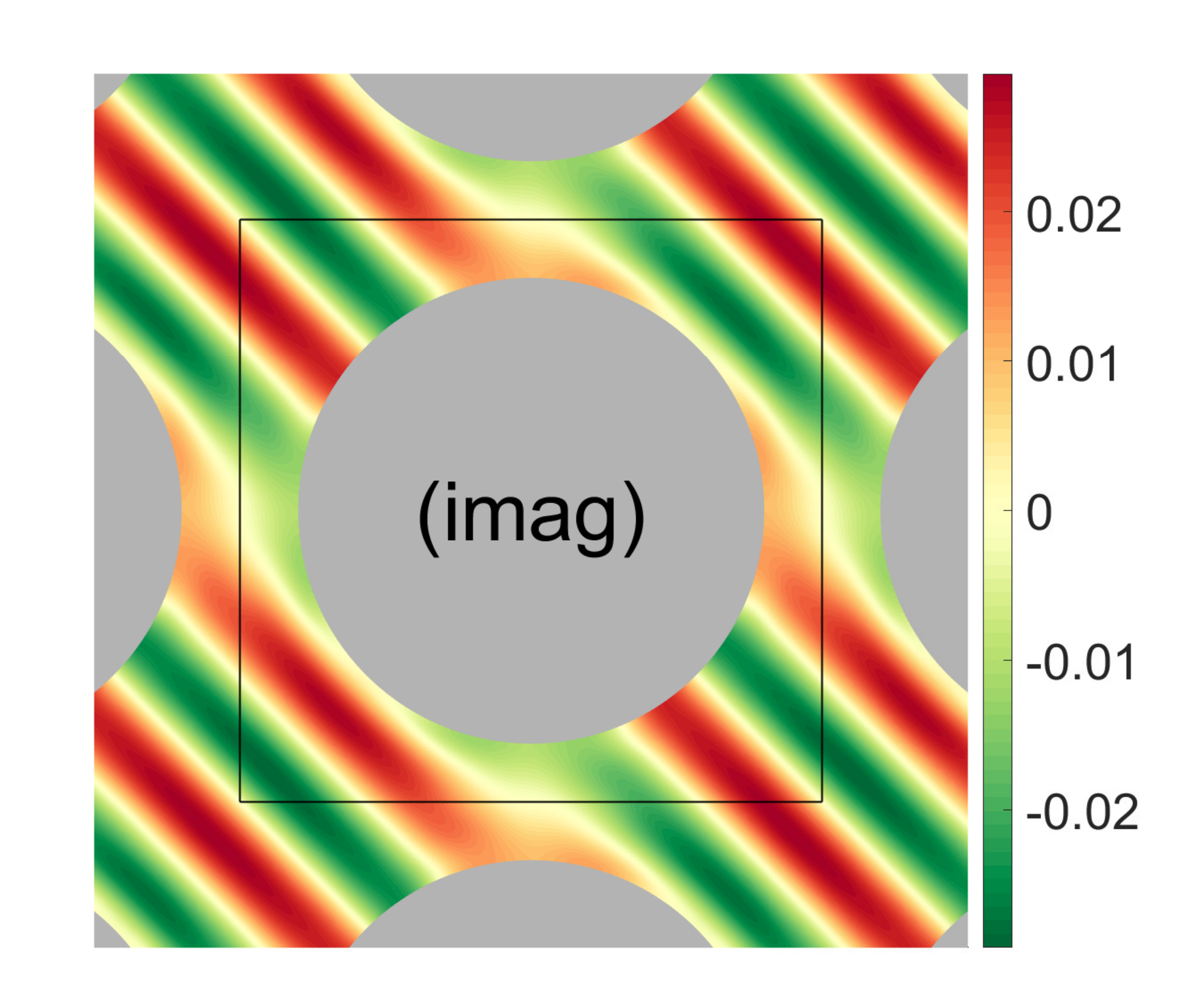}
\includegraphics[width=0.24\linewidth]{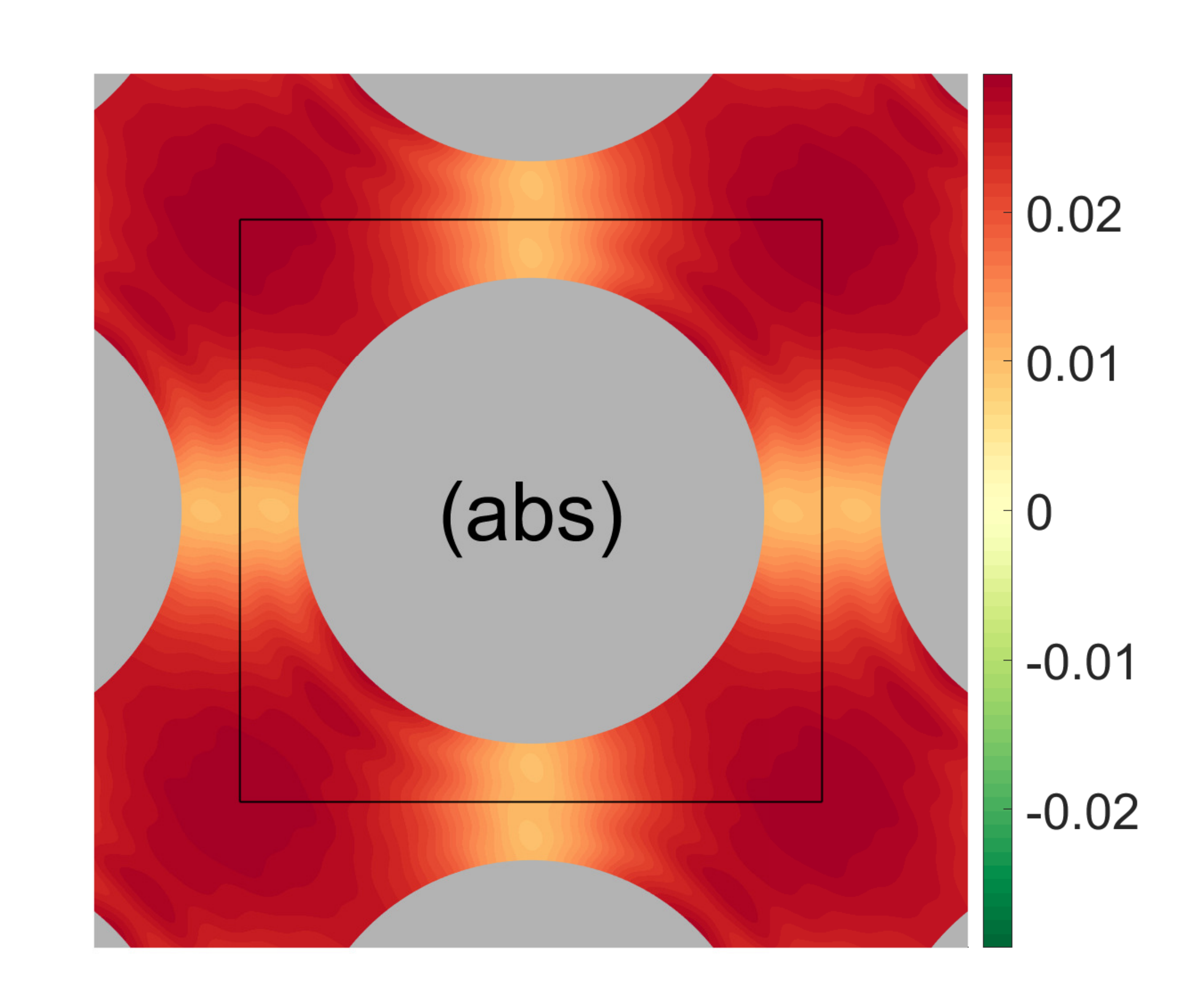}
\includegraphics[width=0.24\linewidth]{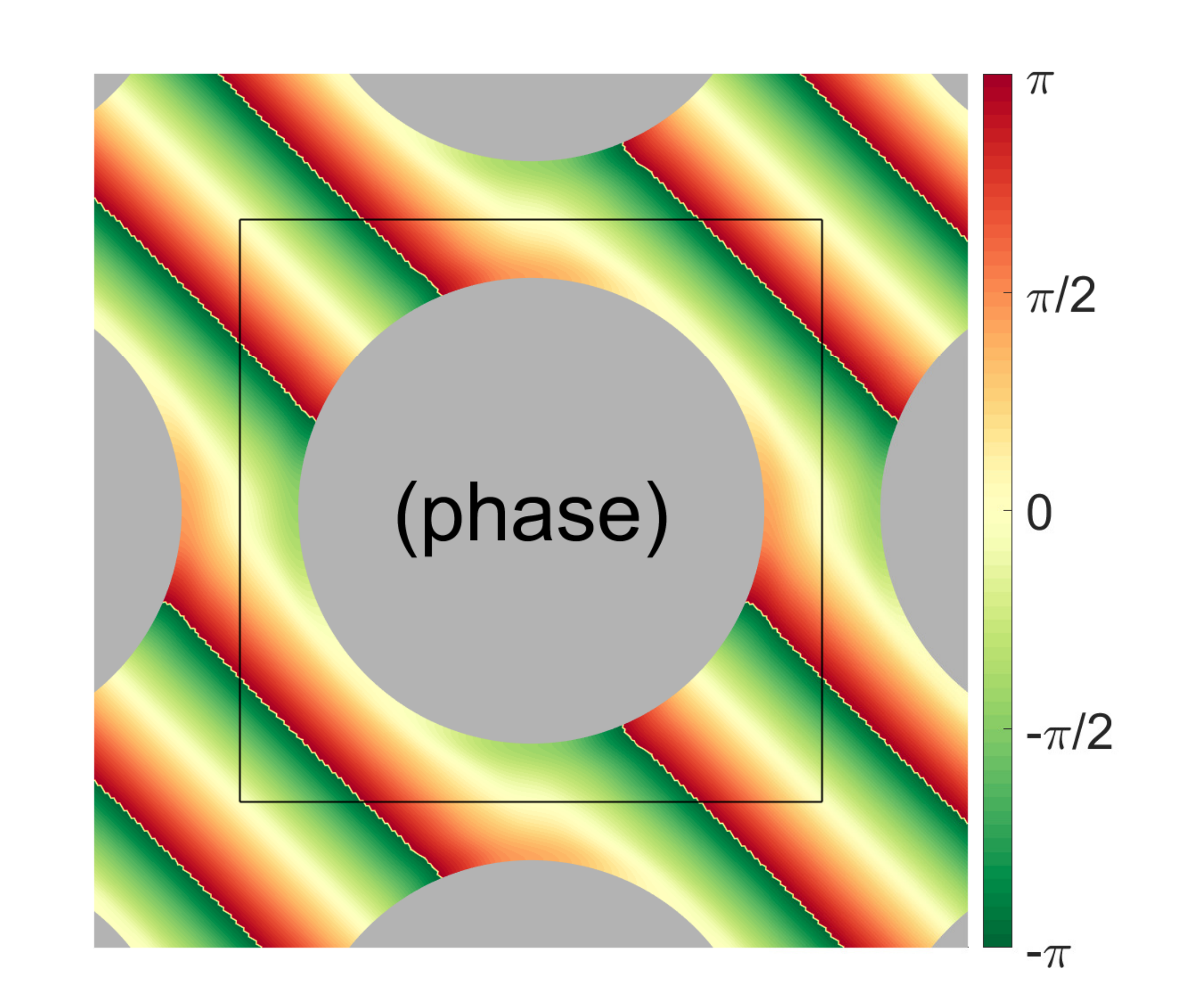}
\noindent\rule[5pt]{\linewidth}{0.4pt}
\includegraphics[width=0.24\linewidth]{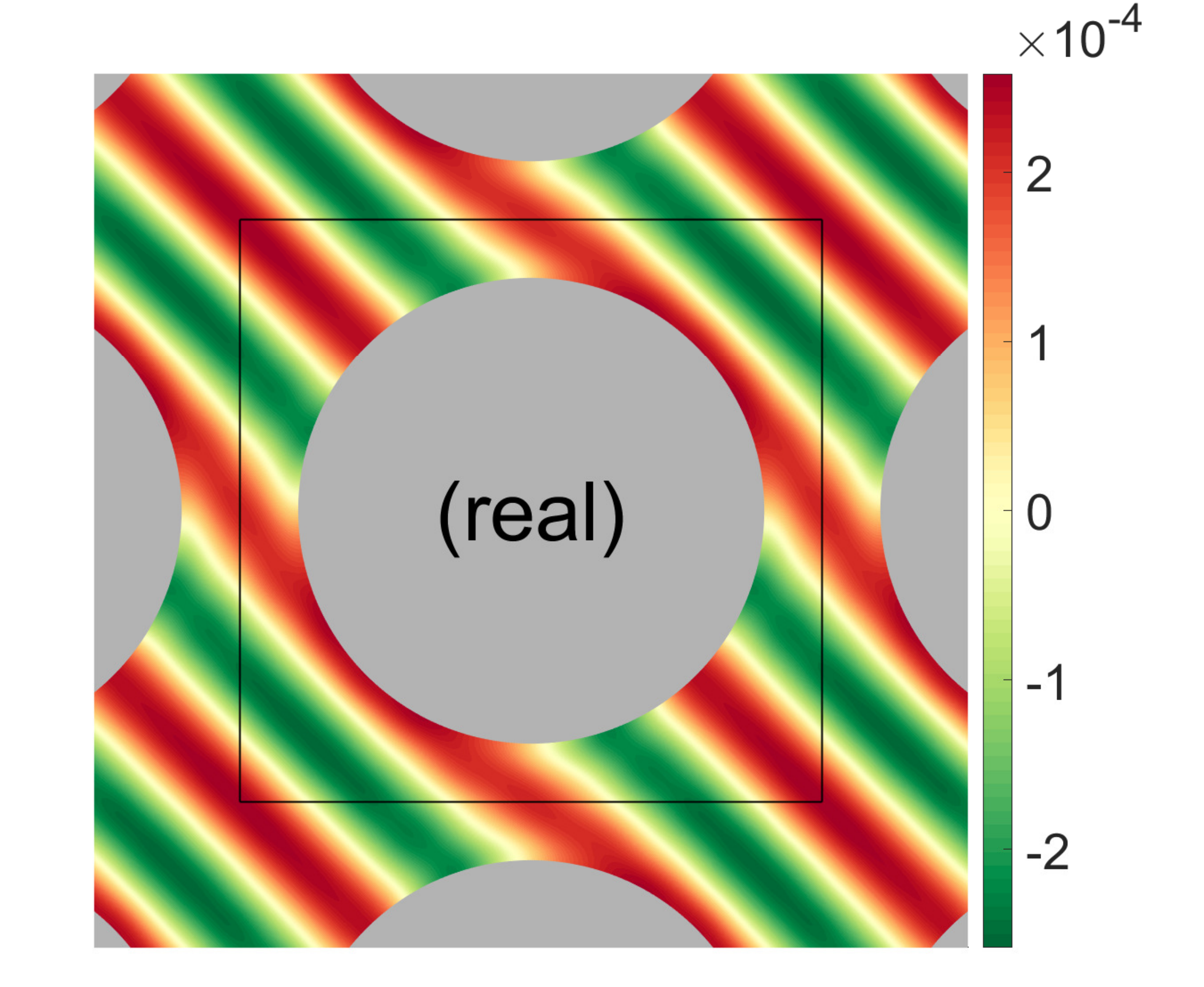}
\includegraphics[width=0.24\linewidth]{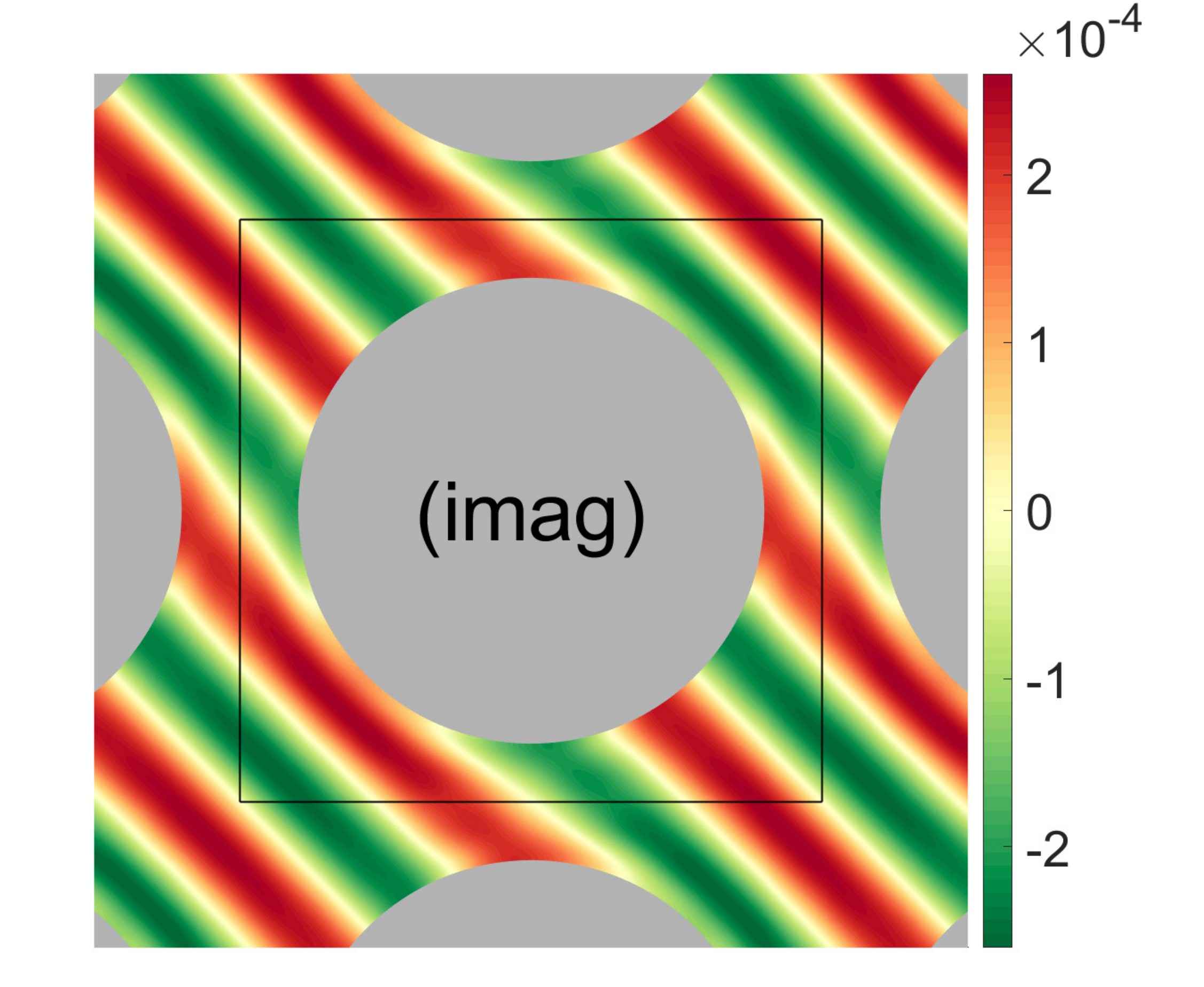}
\includegraphics[width=0.24\linewidth]{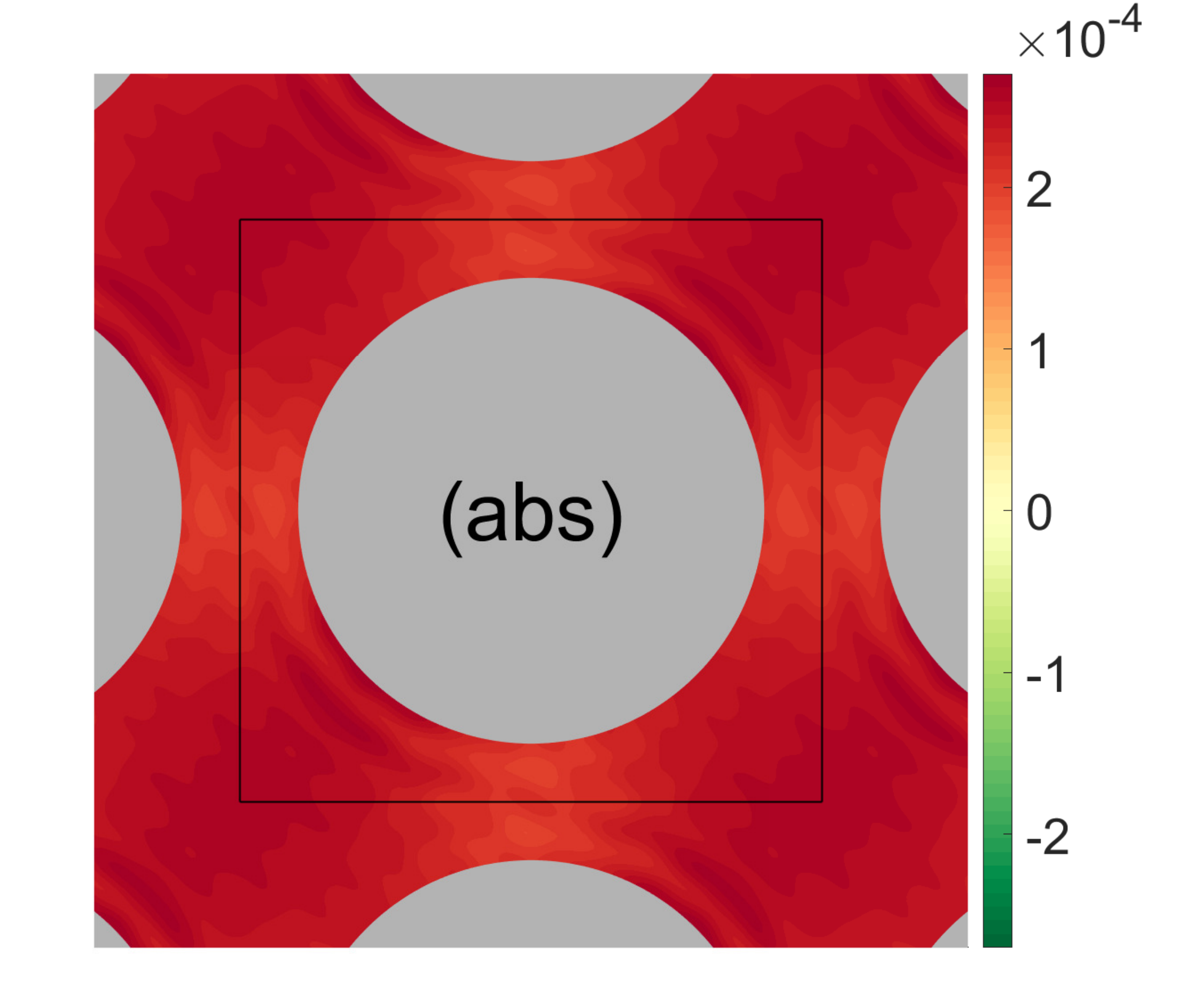}
\includegraphics[width=0.24\linewidth]{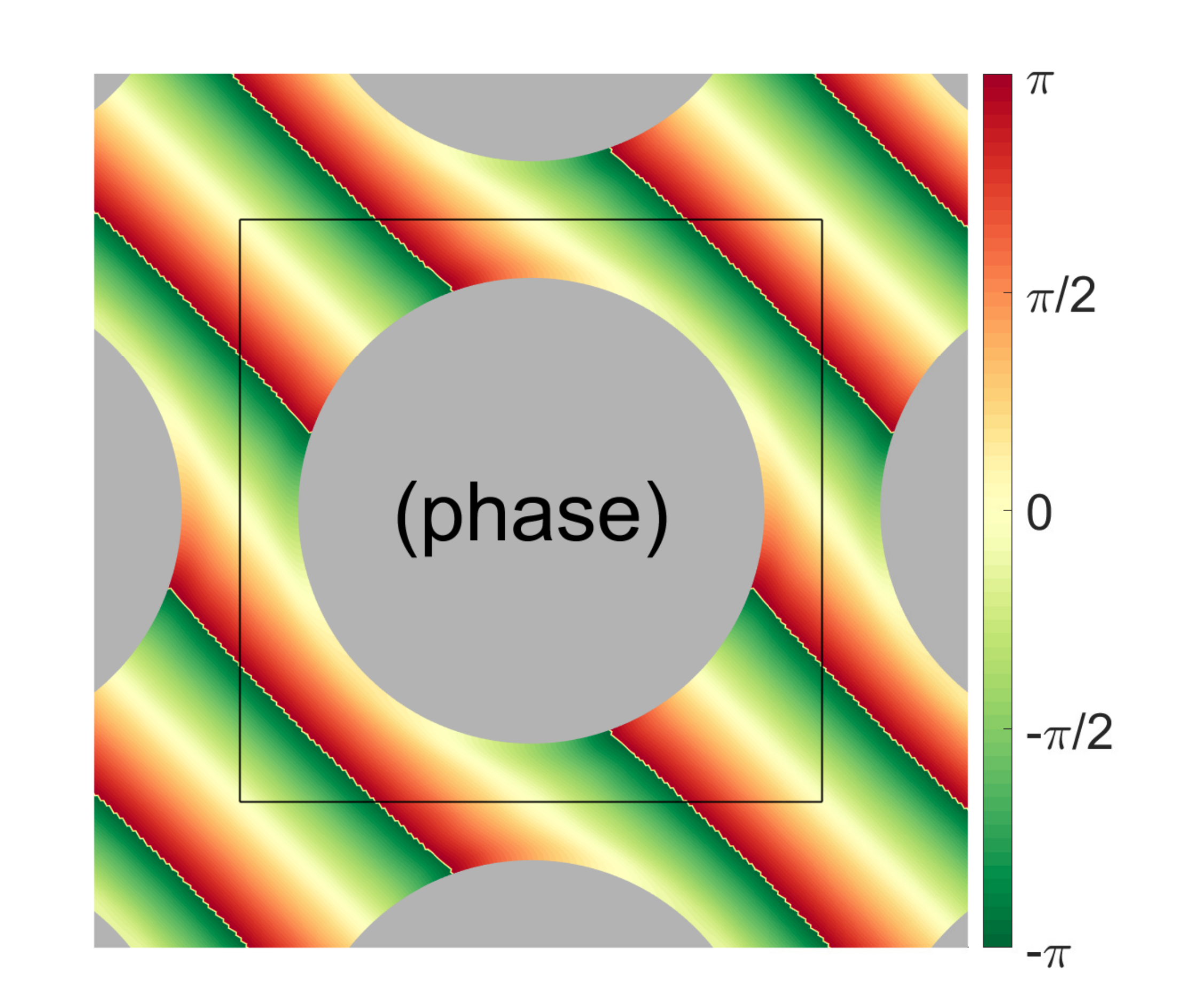}
\caption{Plot of the magnetization (real and imaginary part, absolute value and phase) after a narrow-pulse sequence. The gradient is in the bottom-left to top-right diagonal direction. The black square indicates the unit cell in which the computation was performed. For all figures, $R/a=0.4$, and we kept a fixed value $qa/\sqrt{2}=14\pi/3$. The corresponding normalized signal is shown on the right panel of Fig. \ref{fig:signal_q}.
(top) $\ell_\Delta/a=0.1$; (middle) $\ell_\Delta=0.3$; (bottom) $\ell_\Delta/a=1.0$.
}
\label{fig:NPA_magn_diag}
\end{figure}

The case of the gradient in the diagonal direction is very similar except that the length of the unit cell along the gradient direction is different. As it is shown in Fig. \ref{fig:periodic_2d}, although the diagonal of the unit cell is equal to $a\sqrt{2}$, one can reduce it further so that the actual period along the gradient direction is $a/\sqrt{2}$. Another way to see this is that the set $\lbrace \mathbf{g}\cdot \mathbf{e}\rbrace$, where $\mathbf{e}$ spans all vectors of the lattice, is equal to $(g a/\sqrt{2}) \,\mathbb{Z}$. Thus, the same discussion as that for the horizontal case holds if one replaces $a$ by $a/\sqrt{2}$. Following this conclusion, Fig. \ref{fig:NPA_magn_diag} was obtained with $qa/\sqrt{2}=14\pi/3$ and the gradient in the diagonal direction.

\begin{figure}[ht]
\centering
\includegraphics[width=0.49\linewidth]{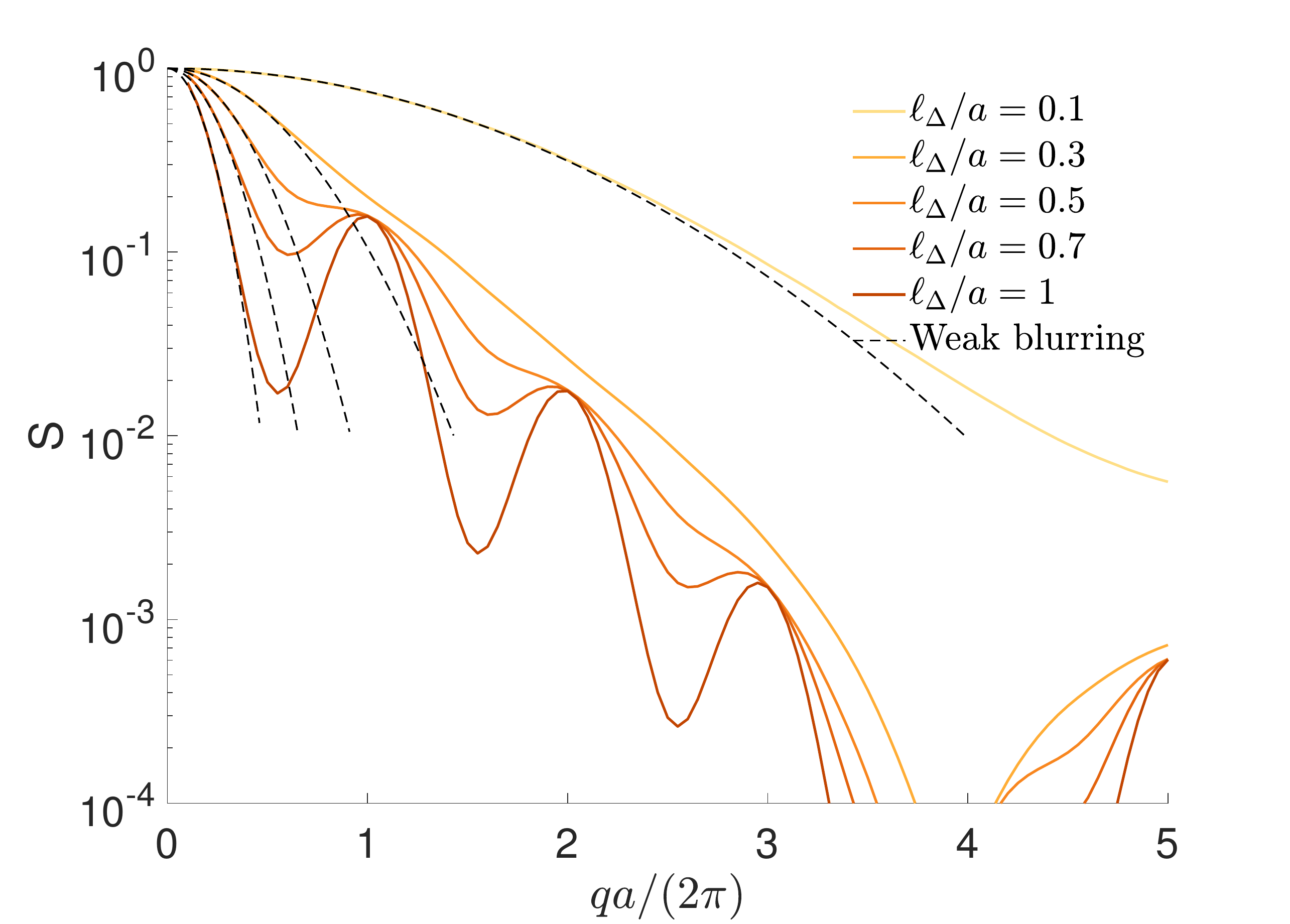}
\includegraphics[width=0.49\linewidth]{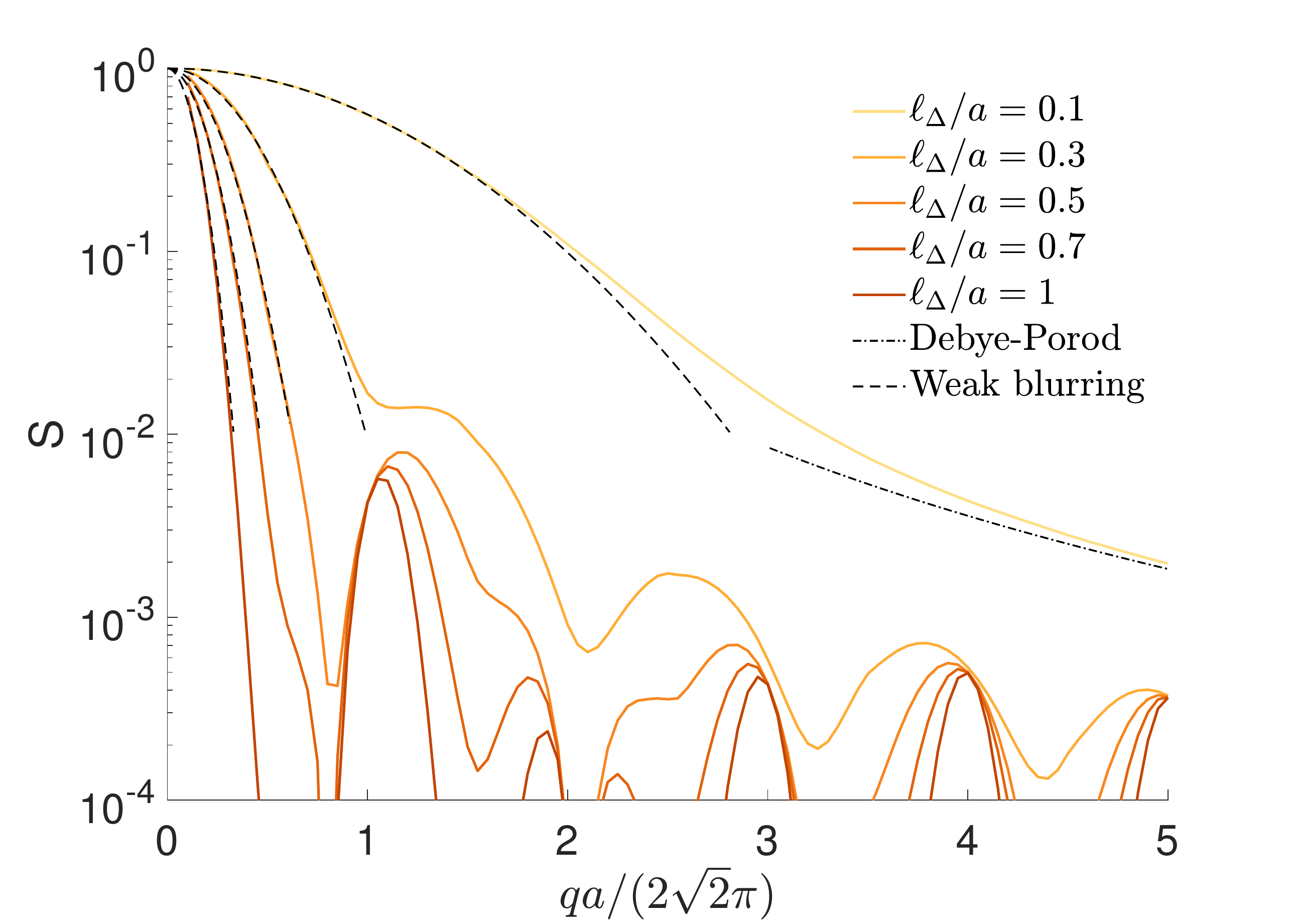}
\caption{Signal after a narrow-pulse sequence for different values of $\ell_\Delta/a$ and asymptotic formulas \eqref{eq:free_decay_bis_NPA} and \eqref{eq:Debye_Porod} for the shortest diffusion time considered here, $\ell_\Delta/a=0.1$.  (left) The gradient is in the horizontal direction. (right) The gradient is in the diagonal direction.
Refer to the text for discussion of the figure.
}
\label{fig:signal_q}
\end{figure}

The normalized signal is plotted on Fig. \ref{fig:signal_q} as a function of $qa/(2\pi)$ for the gradient in the horizontal direction and as a function of $qa/(2\sqrt{2}\pi)$ for the gradient in the diagonal direction. 
In the weak blurring regime (i.e., $q\ell_\Delta \lesssim 1$), the signal decays according to an expression similar to Eq. \eqref{eq:free_decay_bis}:
\begin{equation}
S\approx\exp(-\beta(\ell_\Delta/a) bD_0)=\exp(-\beta(\ell_\Delta/a)q^2\ell_\Delta^2)\;,
\label{eq:free_decay_bis_NPA}
\end{equation}
where $0<\beta(\ell_\delta/a)<1$ is the ratio of effective diffusion coefficient to intrinsic diffusion coefficient $D_0$ that accounts for the restriction by obstacles in the domain. Because the gradient sequence considered here is not the same as the one for which Eq. \eqref{eq:free_decay_bis} was written, the functions $\alpha$ and $\beta$ do not coincide  but share some common features \cite{Moutal2019c}: $\beta(0)=1$, $\beta$ is a linear function of $\sigma \ell_\Delta$ close to $0$ and $\beta(\infty)=\alpha(\infty)$ yields the universal tortuosity limit of the medium.
We have plotted Eq. \eqref{eq:free_decay_bis_NPA} on Fig. \ref{fig:signal_q} for different values of $\ell_\Delta/a$ (the parameter $\beta(\ell_\Delta/a)$ was obtained by fitting the low-$q$ part of each curve).
 In the strong blurring regime (i.e., $q\ell_\Delta \gg 1$), the signal exhibits different behaviors depending on the diffusion length, that can be interpreted with the help of Eq. \eqref{eq:structure_factor_pseudo} and related to the above discussion of the magnetization profile.

At short diffusion time (e.g. $\ell_\Delta/a\approx0.1$), nearly all eigenmodes contribute to the signal in \eqref{eq:structure_factor_pseudo} so that this expansion is not the best tool to understand the behavior of the signal. Because of the short diffusion time, one can treat the effect of the obstacle's boundary as a sum of independent contributions from small boundary regions (as in \cite{Froehlich2006a} where the signal in an interval is split into a sum ``left boundary + bulk + right boundary''). In particular, the strong blurring regime yields the following expression for the \emph{non-normalized} signal by a single barrier of length $\ell_b$ at angle $\theta$ with respect to the gradient (we recall that the initial magnetization is uniform and equal to $1$):
\begin{equation}
s_b \approx \frac{\ell_\Delta\ell_b \exp(-bD_0\sin^2\theta)}{\sqrt{\pi} bD_0 \cos^2\theta} =\frac{\ell_b\exp(-q^2\ell_\Delta^2\sin^2\theta)}{\sqrt{\pi}q^2\ell_\Delta\cos^2\theta} \;,
\end{equation}
where $\theta=0$ corresponds to a boundary perpendicular to the gradient and naturally yields the highest signal. Note that the above formula is valid only if $bD_0\cos^2\theta \gg 1$, otherwise the factor $1/(bD_0\cos^2\theta)$ should be modified. However, because of the attenuation by the exponential factor $\exp(-bD_0\sin^2\theta)$, this correction is of little importance in the following. Following \cite{Froehlich2006a}, we average over all boundary orientations (with a saddle-point approximation around $\theta=0$) and normalize the signal, \cc{that yields for a medium with isotropic boundary orientations (such as the one considered here)}:
\begin{equation}
S\approx\frac{\sigma \ell_\Delta}{\pi (bD_0)^{3/2}} =  \frac{\sigma}{\pi q^3 \ell_\Delta^2}\;,
\label{eq:Debye_Porod}
\end{equation}
where $\sigma = 2\pi R / (a^2- \pi R^2)$ is the surface-to-volume ratio of the domain. This is the two-dimensional Debye-Porod law where the signal is dominated by contributions from the boundaries in the medium. The three-dimensional case would yield
\begin{equation}
S\approx\frac{\sigma \ell_\Delta}{2\sqrt{\pi} (bD_0)^{2}} =  \frac{\sigma}{2\sqrt{\pi} q^4 \ell_\Delta^3}\;,
\label{eq:Debye_Porod_3D}
\end{equation}
for a medium with isotropic boundary orientations, i.e., isotropic (spherical) or statistically isotropic obstacles. \cc{We retrieved the formulas \eqref{eq:Debye_Porod}, \eqref{eq:Debye_Porod_3D} from Ref. \cite{Froehlich2006a} (note that a factor $2$ was missing in these formulas).}

At slightly longer diffusion time (e.g., $\ell_\Delta/a \approx 0.3$), high-order eigenmodes are almost fully attenuated and the signal is nearly equal to the first form factor $C_{p,0}(q)$ that depends on the structure of the unit cell. For example, the drop in \cc{the} signal at $qa/(2\pi)\approx 4$ for the gradient in the horizontal direction is characteristic of the particular value $R/a=0.4$ for which the computation was performed.
At even longer diffusion time, the exponential decay of the first eigenmode emerges and the signal is close to \cc{$C_{p,0}(q)\exp(-\lambda_{p,0} \Delta)$}. As we explained in Sec. \ref{section:blabla}, $\lambda_{p,0}=0$ for $p=0$ so that the signal exhibits ``diffusion-diffraction'' peaks that reveal the periodicity of the medium. The signal for the gradient in the diagonal direction shows peaks at integer values of $qa/(2\sqrt{2}\pi)$, that confirms the value of the period $a/\sqrt{2}$. Moreover, for $\ell_\Delta/a \gtrsim 1$, the decay of the signal at small values of $qa$ is mainly dictated by \cc{$\exp(-\lambda_{0,p}\Delta)$} and not by the form factor $C_{p,0}(q)$ that has a slower decay with $q$. Combined with Eq. \eqref{eq:free_decay_bis_NPA}, this observation yields the following low-$p$ asymptotic behavior:
\begin{equation}
\lambda_{0,p} \approx D_0\beta(\infty) p^2\;,
\end{equation}
i.e., the behavior of the first Laplacian band at low \cc{wavenumbers} is directly related to the tortuosity limit of the medium.

The comparison of Figs. \ref{fig:NPA_magn_horiz}, \ref{fig:NPA_magn_diag} and \ref{fig:signal_q} with Figs. \ref{fig:horiz_magn}, \ref{fig:diag_magn} and \ref{fig:signal_b} reveals important qualitative differences. 
First, one can note a visual similarity between the localized magnetization in the bottom panels of  Figs. \ref{fig:horiz_magn}, \ref{fig:diag_magn} and the edge enhancement effect that can be observed on the top panels of Figs. \ref{fig:NPA_magn_horiz} and \ref{fig:NPA_magn_diag}. However, we argue that these two regimes are vastly different. In fact, the localization regime arises when the motion encoding by the gradient is strong (i.e., $bD_0 \gg 1$) so that the transverse magnetization is strongly attenuated everywhere but in a small layer of thickness $\ell_g$ close to the obstacles, resulting in a weak signal. In contrast, the edge enhancement effect shown here appears even at weak gradient encoding (i.e., $bD_0 \lesssim 1$) so that the transverse magnetization is rather intense everywhere in the medium but enhanced near obstacles, resulting in a strong signal. Furthermore, a short-gradient pulse sequence with strong encoding (i.e., $bD_0 \gg 1$) gives rise to a peculiar striped pattern as shown on the bottom panels of Figs. \ref{fig:NPA_magn_horiz} and \ref{fig:NPA_magn_diag}. This delocalized pattern is in some sense the ``opposite'' of the localizated magnetization pockets shown on Figs. \ref{fig:horiz_magn} and \ref{fig:diag_magn}. This is especially apparent in the resulting signal: whereas the short-gradient pulse experiment probes the global structure of the domain that is revealed through the diffusion-diffraction pattern, the extended-gradient pulse experiment probes the local properties of obstacle's boundaries around  localization points.
Intuitively, the reason behind these differences is that the limit $\delta\to 0$, $g\to \infty$ with constant $g\delta=q$ yields $\ell_\delta/\ell_g=0$. In other words, there is no motion encoding during the narrow gradient pulse, and the attenuation of the transverse magnetization is caused by the subsequent diffusion step. This is in sharp contrast with extended-gradient pulses that continuously encode the random motion of spin-bearing particles.

\section{Another spectral method in 1D-periodic medium}
\label{section:spectral_method_bis}

In this appendix, we consider a 1D-periodic medium and show how to implement the effect of $g_y$ and $g_z$ gradients with an alternative spectral method to the one presented in Sec. \ref{section:periodic_bounded}. Instead of replacing $g_y$ and $g_z$ by a collection of narrow pulses, one can replace them by stepwise functions. In fact, in bounded domains the effect of a constant gradient can be computed exactly with matrix multiplications.

Between two narrow $g_x$ pulses, the magnetization is $p$-pseudo-periodic with a given wavenumber $p$ and one can compute two matrices $B_y$ and $B_z$:
\begin{eqnarray}
&\left[B_{y,p}\right]_{n,n'}=\int_{\Omega_1}y\, u^*_{p,n} u_{p,n'} \,\mathrm{d}x\,\mathrm{d}y\,\mathrm{d}z\\
&\left[B_{z,p}\right]_{n,n'}=\int_{\Omega_1} z\, u^*_{p,n} u_{p,n'} \,\mathrm{d}x\,\mathrm{d}y\,\mathrm{d}z\;.
\end{eqnarray}
These two matrices encode the $y$ and $z$ terms of BT equation, so that a constant $(y,z)$-gradient pulse of duration $\tau$ is represented by the left-multiplication by \cc{$\exp\left(-\tau \Lambda_p + i\tau g_y B_{y,p} + i \tau g_z B_{z,p}\right)$}. Note that one has to compute as many different $B_{y,p}$ and $B_{z,p}$ matrices as there are different values of $p$ involved in the sampling.


\section{Bifurcation for a $2\times2$ matrix}
\label{section:bifurcation_simple}

We shall illustrate the mathematical phenomenon of spectral bifurcation on the simplest case of a $2\times 2$ matrix. Although a differential operator acting on an infinite-dimension functional space cannot be reduced to a matrix, we argue that the coalescence of two eigenmodes and eigenvalues is essentially captured by a computation on a vector space of dimension $2$. We first consider the example of an Hermitian matrix, then we show how the general, non-Hermitian case, differs qualitatively.
Without loss of generality, we consider a trace-free matrix of the general form
\begin{equation}
A=\left[\begin{array}{cc}a& b\\c &-a \end{array}\right]\;.
\end{equation}
One can easily compute its eigenvalues $\lambda_{\pm}$ and eigenvectors $X_{\pm}$:
\begin{eqnarray}
\lambda_{\pm}=\pm\sqrt{d}\;,\\
X_{\pm}=\left[\begin{array}{c} b\\ \pm\sqrt{d}-a \end{array}\right]\;,\\
d=bc+a^2\;,
\end{eqnarray}
if $d\neq 0$.
Let $a,b,c$ be smooth functions of a parameter $g$ (that represents the gradient strength in the BT operator) and that $d=0$ at $g=g_0$. Thus, both eigenvalues coalesce at the critical value $g=g_0$: $\lambda_\pm(g_0)=0$.

\subsection{Hermitian case}

If $A$ is Hermitian, then $a\in\mathbb{R}$ and $c=b^*$, so that $d=|b|^2+a^2$ is real and non-negative. Furthermore, $a(g_0)=b(g_0)=c(g_0)=0$. In fact, this simply derives from the fact that a diagonalizable matrix with all eigenvalues equal to zero is the null matrix. This also implies that $d'(g_0)=0$ and $d''(g_0)\neq0$ in general, so that close to $g_0$ the eigenvalues are approximately equal to 
\begin{equation}
\lambda_\pm = \pm \sqrt{d''(g_0)/2}\, |g-g_0|\;.
\end{equation}
Thus we can draw two main conclusions: (i) the spectrum does not present non-analytical bifurcation points, the eigenvalues merely cross each other at $g=g_0$; (ii) the dimension of the eigenspace $E_{\lambda=0}$ at $g=g_0$ is $2$.

\subsection{Non-Hermitian case}

Now we consider the general, non-Hermitian case. The function $d(g)$ takes complex values and crosses $0$ at $g=g_0$ with a non-zero derivative $d'(g_0)$. The phases of $\lambda_{\pm}$ undergo a $\pi/2$ jump when $g$ increases through the critical value $g_0$ and the absolute values of $\lambda_{\pm}$ have a typical $\sqrt{\left|d'(g_0)(g-g_0)\right|}$ shape for $g$ close to $g_0$. In particular, if $d(g)$ is real, positive for $g<g_0$ and negative for $g>g_0$, one obtains close to the critical value $g_0$:
\begin{eqnarray}
g<g_0 \qquad \left\lbrace
\begin{array}{l}
\mathrm{Re}(\lambda_{\pm})\approx\pm \sqrt{d'(g_0)(g_0-g)} \\ \mathrm{Im}(\lambda_{\pm})=0
\end{array}\right.  \vspace{6pt}\\
g>g_0 \qquad 
\left\lbrace\begin{array}{l}
\mathrm{Re}(\lambda_{\pm})=0\\ \mathrm{Im}(\lambda_{\pm})\approx\pm \sqrt{d'(g_0)(g-g_0)}
\end{array}\right.
\end{eqnarray}
This behavior reproduces the bifurcations shown on Figs. \ref{fig:spectrum} and \ref{fig:spectrum_long}.

At the critical value $g=g_0$, the matrix $A$ is in general not diagonalizable. Without loss of generality, let us assume that $b(g_0)\neq 0$.
The matrix $A$ can then be reduced to a Jordan block with an eigenvector $X_0$ and a generalized eigenvector $Y_0$:
\begin{eqnarray}
AX_0=0 \;, \qquad X_0=\left[\begin{array}{c}b(g_0)\\-a(g_0)\end{array}\right]\;, \\
AY_0=X_0 \;, \qquad \cc{Y_0}=\left[\begin{array}{c}0\\1\end{array}\right]\;.
\end{eqnarray}
Note that since the derivative of $\sqrt{d(g)}$ is infinite at $g=g_0$, one has
\begin{equation}
Y_0=\restr{\frac{d X_\pm}{d \lambda_\pm}}{g=g_0}\;,
\end{equation}
where the derivative yields the same result for $(X_+,\lambda_+)$ and $(X_-,\lambda_-)$.

In comparison to the Hermitian case, our main conclusions are: (i) the spectrum is non-analytical at $g=g_0$; (ii) the eigenvectors $X_\pm$ of $A$ collapse onto one single eigenvector $X_0$, the matrix $A$ can be reduced to a Jordan block with a generalized eigenvector $Y_0$ given by the rate of change of the eigenvectors $X_\pm$ with their corresponding eigenvalues $\lambda_\pm$, evaluated at the critical point $g=g_0$.

We emphasize that the dichotomy ``Hermitian, no bifurcation'' versus ``non-Hermitian, bifurcation'' is specific to two-dimensional matrices. In fact, if one considers a $4\times4$ matrice made of two $2\times2$ blocks where one is Hermitian and the other is non-Hermitian, then the eigenvalues of the Hermitian block will not display any bifurcation point when they cross even if the whole operator is not Hermitian. This somewhat artificial example shows that there is no general relation between bifurcation points and non-Hermitianity except that the spectrum of an Hermitian operator never bifurcates. By reducing the full operator to a low-dimensionality matrix on the subspace associated to the coalescing point, one can make precise statements about bifurcation and Hermitianity, as we did in this two-dimensional example.

\section*{References}
\end{document}